\documentclass[10pt]{iopart}

\DeclareMathAlphabet\mathbfcal{OMS}{cmsy}{b}{n}
\usepackage{mathrsfs}
\usepackage{bm}
\usepackage{iopams}
\usepackage{epsfig}
\usepackage{amssymb}
\usepackage{graphicx}
\usepackage{color}
\usepackage{cite}
\usepackage{xcolor}

\def\I#1{\!#1\!}
\def\vecb#1{\boldsymbol{#1}}
\def\ket#1{|#1\rangle}
\def\bra#1{\langle#1|}
\def\scal#1#2{\langle#1|#2\rangle}
\def\bscal#1#2{\bigl\langle#1\bigr|#2\bigr\rangle}

\def\matr#1#2#3{\langle#1|#2|#3\rangle}
\def\bmatr#1#2#3{\bigl\langle#1\bigl|#2\bigr|#3\bigr\rangle}
\def\abs#1{\left\lvert#1\right\rvert}

\def\ave#1{\langle#1\rangle}
\def\biave#1{\bigl\langle#1\bigr\rangle}

\def\dis#1{\langle\!\langle#1\rangle\!\rangle}
\def\bidis#1{\bigl\langle\!\!\hspace{1pt}\bigl\langle#1\bigr\rangle\!\!\hspace{1pt}\bigr\rangle}

\def\abs#1{\left|#1\right|}
\def\Hi{\mathscr{H}}
\def\Ph{\mathscr{P}}
\def\Gr{\mathscr{G}}
\def\psiin{\psi_{\rm in}}
\def\lamin{\lambda_{\rm in}}
\def\lamfi{\lambda_{\rm fi}}
\def\ii{{\rm i}}

\def\eqref#1{\eref{#1}}
\def\uvo#1{\lq\lq #1\rq\rq}
\def\ESQPT{\textsc{esqpt}}
\def\ESQPTs{\textsc{esqpt}s}
\def\QPT{\textsc{qpt}}
\def\QPTs{\textsc{qpt}s}
\def\TPT{\textsc{tpt}}
\def\TPTs{\textsc{tpt}s}
\def\DoF{\textsc{dof}}
\def\DoFs{\textsc{dof}s}
\def\TC{\textsc{tc}}
\def\D{\textsc{d}}
\def\N{\textsc{n}}
\def\S{\textsc{s}}

\mathchardef\mhyphen="2D
\def\most#1{[#1]}

\begin{document}

\pagestyle{plain}

\review[Excited-state quantum phase transitions]{Excited-state quantum phase transitions}
\author{Pavel Cejnar$^1$, Pavel Str{\'a}nsk{\'y}$^1$, \\ Michal Macek$^2$, Michal Kloc$^{1,3}$}
\address{$^1$ Institute of Particle and Nuclear Physics, Faculty of Mathematics and Physics, Charles University, 
  V~Hole{\v s}ovi{\v c}k{\' a}ch 2, 180\,00 Prague, Czechia}
\address{$^2$ Institute of Scientific Instruments, The Czech Academy of Sciences, Kr{\' a}lovopolsk{\' a} 147, 612\,64 Brno, Czechia}
\address{$^3$ Department of Physics, University of Basel, Klingelbergstrasse 82, 4056 Basel, Switzerland}
\ead{cejnar@ipnp.mff.cuni.cz}

\begin{abstract}
We review the effects of excited-state quantum phase transitions (\ESQPTs) in interacting many-body systems with finite numbers of collective degrees of freedom.
We classify typical {\ESQPT} signatures in the spectra of energy eigenstates with respect to the underlying classical dynamics and outline a variety of quantum systems in which they occur.
We describe thermodynamic and dynamic consequences of {\ESQPTs}, like those in microcanonical thermodynamics, quantum quench dynamics, and in the response to nearly adiabatic or periodic driving.
We hint at some generalizations of the {\ESQPT} concept in periodic lattices and in resonant tunneling systems.  
\end{abstract}



\tableofcontents


\section{Introduction}
\label{S:intro}

The term excited-state quantum phase transition ({\ESQPT}) denotes various kinds of criticality in discrete spectra of excited states of some isolated quantum systems.
While the standard quantum phase transition ({\QPT}) represents an irregular evolution of the system's ground state with a suitable control parameter $\lambda$ (an external field intensity or an internal coupling strength) \cite{Carr10,Vojta03,Sachd99,Gilmo81}, the {\ESQPT} typically impacts a significant amount of excited states (not just a few above the ground state) and implies unsmooth variations of their energy and structure with both the control parameter $\lambda$ and energy~$E$.
In contrast to thermal phase transitions ({\TPTs}), which also affect excited states but with energy smeared by thermal fluctuations, the {\ESQPTs} influence excited states (possibly a subset of them) at a sharp value of energy.
An evidence has grown that such spectral singularities are relevant in numerous simple quantum mechanical models as well as in diverse many-body systems of condensed-matter physics.
These include qubits with long-range interactions, coupled atom-field systems, interacting boson systems describing collective motions of molecules, atomic nuclei and cold atoms, and interacting Fermi and Bose-Fermi systems.

The {\ESQPT} shows up in non-analyticities of the level density and level flow in the plane ${\lambda\times E}$, where we consider a single control parameter~$\lambda$ of the system (usually chosen from a larger set). 
These non-analyticities appear along some critical borderlines, i.e., curves $E=E_{\rm c}(\lambda)$ in this plane, and are accompanied by anomalous changes of the averaged form of the Hamiltonian eigenvectors.
The critical borderlines divide the spectrum at a given $\lambda$ into several excited, or dynamic \uvo{quantum phases} such that some features of the corresponding eigenstates fail to smoothly continue across the borderline.
The point ${\lambda=\lambda_{\rm c}}$ where the {\ESQPT} critical energy $E_{\rm c}(\lambda)$ eventually reaches the ground-state energy $E_{\rm gs}(\lambda)$ becomes a {\QPT} critical point.
In this sense, the {\ESQPTs} extend the ground-state phase diagrams to the excited domain ${\lambda\times E}$, similarly as {\TPTs} extend these diagrams to $\lambda\times T$ (with $T$ denoting temperature). 
Let us stress, however, that {\ESQPTs} can be observed also in models with no {\QPT} in their parameter set.
As any type of criticality, the {\ESQPT} becomes nonanalytic only in the limit of infinite size of the system, ${\aleph\to\infty}$. Nevertheless, since finite-size precursors of the true singularity are often meaningful already for moderate size parameter $\aleph$, it makes sense to study these effects in finite systems.

Most of the known {\ESQPTs} are connected with stationary points of the associated classical-limit Hamiltonians, or with other singularities of the underlying classical dynamics.
Manifestations of such singularities in quantal spectra strongly depend on the number~$f$ of the system's degrees of freedom ({\DoFs}), the visibility of the related effects in typical situations being much stronger for lower values of $f$.
Let us preview (before the detailed explanation given in the forthcoming sections) an example of a non-degenerate stationary point in the $2f$-dimensional phase space.
At such a point the Hessian matrix of the Hamiltonian second derivatives must have only non-zero eigenvalues.
So the stationary point is locally quadratic, its type (minimum, maximum or saddle point) being determined by the the number of negative eigenvalues of the Hessian matrix, the so-called index.
The singularity associated with a non-degenerate stationary point appears in the $(f\I{-}1)$st energy derivative of the $\aleph\I{\to}\infty$ level density $\overline{\rho}(E)$, which in a vicinity of the stationary-point energy $E_{\rm c}(\lambda)$ obeys the formula
\begin{equation}
\frac{\partial^{f-1}}{\partial E^{f-1}}\ \overline{\rho}(\lambda,E)\propto h(E)\pm
\left\{\begin{array}{ll}
\theta[E-E_{\rm c}(\lambda)] & {\rm for\ even\ index},\\
\ln|E-E_{\rm c}(\lambda)| & {\rm for\ odd\ index}.
\end{array}\right.
\label{clas}
\end{equation}
Here, $h(E)$ is an unspecified smooth function and $\theta(x)$ denotes the step function ($\theta\I{=}0$ for $x\I{<}0$ and $\theta\I{=}1$ for $x\I{\geq}0$).
So for $f\I{=}1$ system the {\ESQPT} shows up directly in the level density, for $f\I{=}2$ in its first derivative, etc. 
Qualitatively the same type of $\aleph\I{\to}\infty$ singularity appears in the smoothed slopes $\frac{\partial}{\partial\lambda}E_i$ of individual energy levels $E_i(\lambda)$.

The notion of {\ESQPT} has many predecessors (see below), but it was for the first time explicitly considered by Cejnar, Macek, Heinze {\it et al.} in Refs.\,\cite{Cejna06,Cejna07}, closely following the results of Refs.\,\cite{Heinz06,Macek06}.
The idea was subsequently established on a firmer theoretical ground by Caprio, Cejnar and Iachello in Ref.\,\cite{Capri08}, which gave several examples of {\ESQPTs} in models with ${f=1}$.
The {\ESQPT} effects in systems with ${f=2}$ were first mentioned by Cejnar and Str{\'a}nsk{\'y} \cite{Cejna08} and then systematically investigated by the same authors together with Macek and Leviatan in the series of papers \cite{Stran14,Stran15,Macek19}.
The classification of {\ESQPTs} caused by non-degenerate stationary points of classical-limit Hamiltonians for any $f$ was given in Ref.\,\cite{Stran16}.
Today, there exists an extensive literature studying {\ESQPT} signatures in discrete energy spectra of various interacting many-body systems that exhibit some kinds of {\em collective behavior\/} with moderate numbers of effective {\DoFs}.
Effects connected with {\ESQPTs} are reported in quantum quench dynamics, driven dynamics and thermodynamics.
Analogies of  {\ESQPTs} exist in Floquet dynamics and in quantum tunneling. 
Particularly the following authors and groups, except the above-mentioned ones, made essential contribution to the {\ESQPT} field: Brandes, Bastidas, Kopylov {\it et al.}, P{\'e}rez-Bernal, Santos {\it et al.}, Rela{\~n}o, P{\'e}rez-Fern{\'a}ndez, Dukelsky {\it et al.}, Hirsch, Bastarrachea-Magnani, Lerma-Hern{\'a}ndez {\it et al.},  and Puebla {\it et al.}
The references will be given below.

Before proceeding to these results, we briefly mention the closely related, though usually more specialized concepts preceding the notion of {\ESQPT}.
In systems with $f\I{=}1$, the impact of phase-space separatrix (a trajectory crossing in asymptotic time a stationary point) on quantum spectra was studied in the early 1990s by Cary and Rusu \cite{Cary92,Cary93}.
Continuation of such analyses in the many-body context \cite{Leyvr05,Reis05,Zhang05,Riber07,Riber09} roughly coincides with the dawn of the {\ESQPT} physics.
We stress that crossing of a phase-space separatrix can have rather conspicuous quantum consequences, such as the diverging oscillation periods experimentally detected in a spinor condensate \cite{Kawag12,Zhao14}.
Even an earlier predecessor of the {\ESQPT}-focused research  was the study of monodromy in integrable $f\I{=}2$ systems, introduced in the 1980s by Duistermaat, Cushman and V{\~u} Ng{\d o}c \cite{Duist80,Cushm88,Ngoc99}.
Monodromy is a topological singularity which disables global introduction of action-angle variables in the phase space. 
On the quantum level it shows up as a defect in the lattices of conserved quantum numbers associated with individual energy eigenstates, which was incorporated into the {\ESQPT} phenomenology already in the first papers \cite{Heinz06,Macek06,Cejna06}.
Note that these effects were verified in experimental spectra of some molecules, including ${\rm H}_2{\rm O}$ \cite{Winne05,Zobov05}, which up to now remain the clearest experimental evidence of {\ESQPTs} \cite{Lares13}. 
Another very close antecedent of the {\ESQPT} is the microcanonical thermodynamic singularity caused by a stationary point of classical dynamics, investigated in the 2000s by Franzosi and Pettini \cite{Franz04} and Kastner {\it et al.} \cite{Kastn07,Kastn08,Kastn08b}.
In low-$f$ systems, these singularities are indeed equivalent to {\ESQPTs}, but the cited works aimed at identification of an underlying mechanism of standard {\TPTs} in $\aleph\I{\sim}f\I{\to}\infty$ systems through some bulk properties of the singularities. 

The {\ESQPTs} are most commonly recognized in entirely interacting, so-called fully connected many-body systems, for which the size parameter $\aleph$ grows independently of the effective number of {\DoFs} associated with some collective modes of motions.
However, various forms of criticality in excited states were found in many-body systems with local interactions, for which non-collective {\DoFs} play an important role and $\aleph\I{\to}\infty$ is equivalent to $f\I{\to}\infty$.
One example is the so-called metastable {\QPT} introduced by Carr {\it et al.} \cite{Kanam09,Kanam10} in a model of bosons on a ring.
Similarly, the eigenstate phase transition, described by Huse {\it et al.} in the context of spin and fermion lattices \cite{Pal10,Huse13}, characterizes the onset of many-body localization in individual excited states with varying Hamiltonian parameters.
Another {\ESQPT}-like critical effect, known as the van Hove singularity \cite{Hove53}, appears in periodic lattice systems with local interactions.
The energy spectrum of a particle moving through such a lattice is determined via the dispersion relation in the quasimomentum space. 
Its extremes and saddles produce singularities observed in electronic and vibrational spectra of two-dimensional (2D) crystals like graphene \cite{Dietz13,Iache15,Dietz17}.
Finally, let us mention various forms of the dynamical quantum phase transition \cite{Sciol10,Heyl13,Heyl18,Zunko18}, which in general represents critical behavior of quantum many-body systems out of equilibrium in the time variable.
Also this kind of criticality is, in some cases, related to the {\ESQPT} singularities in excited spectra.    

Recent work on {\ESQPTs} can be sorted to the following categories: 
recognition of {\ESQPTs} in various specific models and identification of their signatures in quantum spectra and classical dynamics 
\cite{Berna08,Morei08,Shche09,Cejna09,Figue10,Berna10,Capri11,Lopez11,Lares11,Peres11b,Brand13,Puebl13,Lares13,Basta14a,
Basta14b,Relan14,Engel15,Graef15,Chuma15,Zhang16,Graef16,Lobez16,Chave16,Puebl16,Relan16,Relan16b,Sinde17,
Eckle17,Romer17,Basta17,Garci17,Zhang17,Karam17,Kloc17a,Kloc17b,Opatr18,Bychek18,Rodrig18,Nyawo18,Macek19,
Zhu19,Khalo19,Khalo20,Nitsc20,Monda20,Feldm20}, 
analysis of thermodynamic consequences of {\ESQPTs} 
\cite{Basta16,Peres17,Cejna17,Webst18,Relan18,Garcia18,Kloc19}, 
search for dynamic effects caused by {\ESQPTs}
\cite{Relan08,Peres09,Peres11a,Yuan12,Puebl15,Kopyl15b,Santo15,Santo16,Kopyl17,Wang17,Perez17,Furtm17,Zimme18,Kloc18,
Wang19,Wang19b,Humme19,Pilat19,Tian20,Puebl20,Kloc20}, 
and extension of the {\ESQPT} concept out of its home territory
\cite{Dietz13,Basti14a,Basti14b,Bondy15,Gessn16,Stran20}.
This division of literature roughly determines the structure of our review:
Section~\ref{Open} offers an opening example of a quantum system with several types of {\ESQPTs}.
Section~\ref{Finite} outlines a variety of models to which the notion of {\ESQPT} has already been applied.
Section~\ref{Singul} analyzes the model-independent typology of {\ESQPTs} and describes their manifestation in energy spectra and eigenstates.
Section~\ref{Thermo} summarizes known thermodynamic consequences of {\ESQPTs}.
Section~\ref{Dynam} describes various signatures of {\ESQPTs} in driven dynamics. 
Section~\ref{Exten} reports on generalizations of {\ESQPTs} to spatially extended, periodically driven and scattering systems. 
Section~\ref{Conc} sketches an outlook.

We chose a \uvo{talkative} style, trying to explain a wider context of the presented subjects and make the review accessible to students and non-specialists.
Technical details are outsourced to the cited literature. 
Known results are illustrated by original new figures, which often extend the reported subject to different situations.
We note that all quantities are taken dimensionless, i.e., expressed in suitable implicit units.

\section{Opening example: Dicke superradiance models}
\label{Open}

As an invitation to the {\ESQPT} field, we outline in this section the spectral signatures of {\ESQPTs} in the Dicke model of atomic physics and quantum optics, and in its various extensions.
The model we consider represents a slight modification \cite{Brand13,Sorien18} of the original Dicke Hamiltonian \cite{Dicke54,Jayne63,Tavis68}, known for its superradiant {\TPT} and {\QPT}. 
These transitions are nowadays accessible to experimental investigation by means of advanced quantum simulators \cite{Bauma10,Bauma11,Baden14,Klind15,Lewis19,Li20}. 
The {\ESQPTs}, whose structure is richer in the extended version of the model, may be detected by similar techniques in future.
Studies of {\ESQPTs} in the Dicke and related models can be found, e.g, in Refs.\,\cite{Peres11b,Brand13,Puebl13,Basta14a,Basta14b,Engel15,Chuma15,Lobez16,Chave16,Puebl16,Relan16b,Eckle17,Basta17,
Kloc17a,Kloc17b,Rodrig18,Zhu19} (classical and spectral properties), \cite{Basta16,Peres17} (thermodynamics) and \cite{Peres11a,Furtm17,Kloc18,Pilat19,Puebl20,Kloc20} (dynamic consequences).
We note that numerous other extension of the original model exist, see, e.g., Refs.\,\cite{Hayn17,Rodrig18,Zhu19}.

\subsection{The model and its thermal and quantum phase transitions}
\label{EDM1}

At first we describe the model and its standard phase transitions.
The model was proposed in the 1950s to illustrate the effect of dynamic superradiance, i.e., collectivity-enhanced coherent radiation of atoms in a cavity \cite{Dicke54}. 
It can be cast as a schematic description of the  interaction of one-mode electromagnetic field with a dense ensemble of $N$ two-level atoms.
Photons with energy $\omega$ are created and annihilated by operators $\hat{b}^{\dag}$ and $\hat{b}$.
The ensemble of atoms is described by operators 
\begin{equation}
\hat{J}_{0}=\frac{1}{2}\sum_{i=1}^{N}\hat{\sigma}^{(i)}_{0}\,,\quad
\hat{J}_{+}=\frac{1}{2}\sum_{i=1}^{N}\hat{\sigma}^{(i)}_{+}\,,\quad
\hat{J}_{-}=\frac{1}{2}\sum_{i=1}^{N}\hat{\sigma}^{(i)}_{-}\,,
\label{quasis}
\end{equation}
where $\hat{\sigma}_{0}^{(i)}\I{=}\hat{\sigma}_{z}^{(i)}$ and ${\hat{\sigma}_{\pm}^{(i)}=\hat{\sigma}_{x}^{(i)}\pm\ii\hat{\sigma}_{y}^{(i)}}$ are Pauli matrices in the two-dimensional Hilbert space of the $i$th atom with level energies $\pm\omega_0/2$.
The total Hilbert space of the system is the direct product ${\Hi=\Hi_{\rm A}\otimes\Hi_{\rm B}}$, where $\Hi_{\rm A}$ is the $2^N$-dimensional space of the atomic subsystem spanned by vectors $\ket{\updownarrow^{(1)}\updownarrow^{(2)}...\updownarrow^{(N)}}$, with $\ket{\updownarrow^{(i)}}=\ket{\uparrow^{(i)}}$ or $\ket{\downarrow^{(i)}}$ denoting the up or down eigenvectors of $\hat{\sigma}_z^{(i)}$, and $\Hi_{\rm B}$ is an infinite-dimensional space of the boson field spanned by basis vectors with $N_b\I{=}0,1,2...$ field quanta.

The model represents an example of a {\em coupled system}.
A general discussion of such systems, including proper definition of their size parameters, will be presented in Sec.\,\ref{Fisie}.
In the following, we assume that elementary excitation energies $\omega$ and $\omega_0$ of both subsystems are roughly the same.
Arbitrary redistribution of energy between the field and atoms then implies approximately the same overall numbers of excitation quanta---photons and excited atoms.
In this situation the number of atoms $N$ is a good measure of the system's total size, so we set $\aleph\I{=}N$.
Asymmetric regimes with $\omega\I{\ll}\omega_0$ or $\omega\I{\gg}\omega_0$ need to be treated differently, see Sec.\,\ref{EDM3}. 

The Hamiltonian of the extended Dicke model reads \cite{Brand13}
\begin{equation}
\hspace{-10mm}
\hat{H}(\lambda,\delta)=\underbrace{\omega\,\hat{b}^{\dag}\hat{b}+\omega_0\hat{J}_0}_{\hat{H}_0}+
\lambda\underbrace{\frac{1}{\sqrt{N}} \left(\hat{b}^{\dag}\hat{J}_-+\hat{b}\hat{J}_++\delta\,\hat{b}^{\dag}\hat{J}_++\delta\,\hat{b}\hat{J}_- \right)}_{\hat{H}'(\delta)}
\label{HDic}
,
\end{equation}
where $\lambda$ and $\delta$ are two interaction parameters.
The principal control parameter ${\lambda\in[0,\infty)}$ represents the overall atom-field coupling strength in front of the interaction Hamiltonian $\hat{H}'(\delta)$. 
We instanly introduce two special values of this coupling,
\begin{equation}
\lambda_{\rm c}(\delta)=\frac{\sqrt{\omega\omega_0}}{(1+\delta)}
\,,\qquad
\lambda_{0}(\delta)=\frac{\sqrt{\omega\omega_0}}{(1-\delta)}
\,,
\label{lac}
\end{equation}
whose meaning for the quantum phase structure of the model will be explained later.
The second parameter $\delta$ weights the so-called counter-rotating terms in $\hat{H}'(\delta)$  \cite{Jayne63}.
The variation of this parameter within the interval $\delta\I{\in}[0,1]$ realizes a crossover between the Tavis-Cummings ({\TC}) regime with $\delta\I{=}0$ \cite{Tavis68} and the original Dicke ({\D}) regime with $\delta\I{=}1$ \cite{Dicke54}.
The difference between these regimes follows from their symmetries.
Since the general Hamiltonian \eqref{HDic} with arbitrary $\delta$ changes the total number of field and atomic excitations only by 0 or 2, it has an additional integral of motion represented by a parity operator ${\hat{\Pi}=(-1)^{\hat{M}}}$, where ${\hat{M}=\hat{N}_{b}\I{+}\hat{N}_*}$ is a sum of the number of field bosons ${\hat{N}_{b}=\hat{b}^{\dag}\hat{b}}$ and the number of excited atoms ${\hat{N}_*=\hat{J}_z\I{+}\frac{1}{2}N}$.
From the form of the ${\delta=0}$  Hamiltonian it is however evident that the quantity $\hat{M}$ itself becomes an integral of motion in the {\TC} regime. 
This makes the ${\delta=0}$ limit integrable, hence fully regular, in contrast to partially chaotic regimes with ${\delta\neq 0}$.

In the weak-coupling case, for $\lambda$ small, the atomic subsystem demonstrates the phenomenon of dynamic superradiance, i.e., a non-exponential (flash-like) decay of the atomic ensemble from its fully excited to fully de-excited state.
This is a consequence of uniformity of the atom-field interaction \cite{Dicke54}.
However, the model exhibits also another type of superradiance---the one encoded in its equilibrium properties at low temperatures for large values of the coupling strength $\lambda$ \cite{Wang73,Hepp73}.
In particular, for $\lambda$ above a critical value $\lambda_{\rm c}(\delta)$ from Eq.\,\eqref{lac}, there exists a certain critical temperature $T_{\rm c}(\lambda,\delta)$ such that the equilibrium phase below $T_{\rm c}$ (the so-called {\em superradiant phase}) generates macroscopic excitation of both atomic and field subsystems while the phase above $T_{\rm c}$ ({\em normal phase}) resembles the non-interacting regime.
The transition has a critical character and is classified as a second-order {\TPT}.
Soon after the pioneering analyses \cite{Wang73,Hepp73} it was pointed out \cite{Rzaze75} that the inclusion of the physical term proportional to the squared electromagnetic vector potential into the Hamiltonian of the type \eqref{HDic} destroys its critical properties.
However, an effective Hamiltonian without this term is approximately valid in the setup of confined quantum optical systems (see, e.g., Ref.\,\cite{Keeli14}), so the above-described critical properties are realistic in the present-day laboratory realizations. 
Thermal properties of the extended Dicke model will be further discussed in Sec.\,\ref{Esther}.

The parameter value $\lambda_{\rm c}(\delta)$ from Eq.\,\eqref{lac} represents the {\em quantum critical point} where the critical temperature $T_{\rm c}(\lambda,\delta)$ is zero and the system therefore exhibits a {\QPT} from the normal to superradiant phase in its {\em ground state}.
This continuous (second-order) phase transitions and its quantum signatures were for the first time studied theoretically in Refs.\,\cite{Emary03,Lambe04} and later verified experimentally using the Bose gas in an optical cavity \cite{Bauma10,Bauma11}.

It is known that {\TPTs} and {\QPTs} of the continuous type commonly originate in the general phenomenon of spontaneous symmetry breaking \cite{Carr10,Vojta03,Sachd99}.
For the present system, the symmetry that is spontaneously broken in the superradiant phase can be identified with the above-defined parity symmetry.
In the {\QPT} case, for instance, we observe that in the normal phase the ground state has positive parity, while in the superradiant phase  the lowest energy states with both parities become degenerate in the ${N\I{\to}\infty}$ limit, allowing thus for parity-violating ground-state solutions.
The spontaneous symmetry breaking mechanism generally selects a specific order parameter characterizing the phase transition, i.e., a quantity whose expectation value is zero in one phase and non-zero in the other phase.
In the present case, the order parameter exists for both subsystems: It is the field intensity ${(\hat{b}+\hat{b}^\dag)}$, also referred to as a \uvo{quadrature}, for the bosonic subsystem, and the quasispin component~$\hat{J}_x$ for the atomic subsystem.
Indeed, it is easy to show that the expectation values of both these quantities vanish in any fixed-parity state of the normal phase while they can generally take non-zero values in parity-violating states of the superradiant phase.
In the {\TC} regime with ${\delta=0}$, the spontaneously broken symmetry is a continuous \uvo{gauge} symmetry under the U(1) transformations $e^{\ii \hat{M}\varphi}$ with ${\varphi\in[0,2\pi)}$ \cite{Kloc17b}.
In this case, the superradiant ground state is a doublet containing $M$ and $M\I{+}1$ eigenvalues [increasing with $\lambda\I{>}\lambda_{\rm c}(0)$] of the operator $\hat{M}$ \cite{Buzek05}.

The Hamiltonian (\ref{HDic}) obviously conserves the quantity ${\hat{\vecb{J}}^2=\hat{J}_x^2\I{+}\hat{J}_y^2\I{+}\hat{J}_z^2}$ whose spin-like quantum number $j$ can be set to any integer or half-integer value from ${j_{\rm min}=0}$ or $\frac{1}{2}$ (for $N$ even or odd, respectively) to ${j_{\rm max}=\frac{1}{2}N}$.
The physical meaning of~$j$ (in Ref.\,\cite{Dicke54} called the \uvo{cooperation number} of atoms) will be discussed in Sec.\,\ref{Quasi}, for now we just note that its most usual choice is ${j=j_{\rm max}}$, which
selects a subspace $\Hi_{j_{\rm max}}\I{\subset}\Hi_{\rm A}$ that is fully symmetric with respect to all permutations of atoms.
This subspace is invariant under the action of the evolution operator and enables one to constitute a representation in terms of coordinate and momentum operators $\hat{\phi}$ and $\hat{\zeta}$ in the following way: 
\begin{equation}
\hspace{-13mm}
\left(\begin{array}{c}
\hat{J}_x \\ \hat{J}_y \\ \hat{J}_z
\end{array}\right)
=\sqrt{j(j\I{+}1)}
\left(\begin{array}{c}
\sqrt{1-\hat{\zeta}^2}\cos\hat{\phi} \\ \sqrt{1-\hat{\zeta}^2}\sin\hat{\phi} \\ \hat{\zeta}
\end{array}\right)
\quad {\rm with\ }
[\hat{\phi},\hat{\zeta}]=\frac{\ii}{\sqrt{j(j\I{+}1)}}
\label{PhiPi}
.
\end{equation}
The number $[j(j\I{+}1)]^{-1/2}$, which for ${j=j_{\rm max}}$ and ${N\gg 1}$ roughly equals to $2/N$, plays the role of an effective Planck constant of the model.
The system therefore becomes classical in the ${N\to\infty}$ limit with ${[\hat{\phi},\hat{\zeta}]\to 0}$.
In this limit, the coordinate and momentum are ordinary numbers ${\phi\in[0,2\pi)}$ and ${\zeta\in[-1,+1]}$, and the sphere (the so-called Bloch sphere) of radius $\sqrt{j(j\!+\!1)}\approx j$ with spherical angles ${\theta\equiv\arccos\zeta}$ and $\phi$ represents the classical phase space of the atomic subsystem.

The classical limit of the field subsystem can be constructed via the coordinate-momentum representation of the boson operators
\begin{equation}
\left(\begin{array}{c}
\hat{b}^{\dag} \\ \hat{b}
\end{array}\right)
=\sqrt{\frac{\eta}{2}}
\left(\begin{array}{c}
\hat{q}-\ii\hat{p} \\ \hat{q}+\ii\hat{p}
\end{array}\right)
\quad {\rm with\ }
[\hat{q},\hat{p}]=\frac{\ii}{\eta}
\label{QP}
,
\end{equation}
where $\eta$ is an adjustable parameter quantifying classicality of the field subsystem. 
The choice ${\eta=\sqrt{j(j\I{+}1)}}$ yields the same value of the effective Planck constant as above (this symmetric choice is justified for balanced systems with ${\omega\sim\omega_0}$).
So the number of {\DoFs} of our atom-field system is ${f=2}$, the corresponding phase space $\Ph$ being identified with the Cartesian product ${\Ph_{\rm A}\times\Ph_{\rm B}}$, where the finite phase space $\Ph_{\rm A}$ of the atomic ensemble with a fixed $j$ is the Bloch sphere $(\phi,\zeta)$ and the infinite phase space $\Ph_{\rm B}$ of the field is the plane of variables $(q,p)$.
However, as shown in Refs.\,\cite{Kloc17a,Kloc17b}, the conservation of quantity $\hat{M}$ in the {\TC} regime with ${\delta=0}$ enables one to construct for any fixed value of classical observable $M$ a canonical transformation from variables $(\phi,\zeta)$ and $(q,p)$ to a single pair of coupled atom-field variables.
The model in this regime can therefore be treated as a system with ${f=1}$ effective {\DoF}.

\subsection{Singularities in the spectra of excited states}
\label{EDM2}

\begin{figure}[t!]
\begin{flushright}
\includegraphics[width=0.85\textwidth]{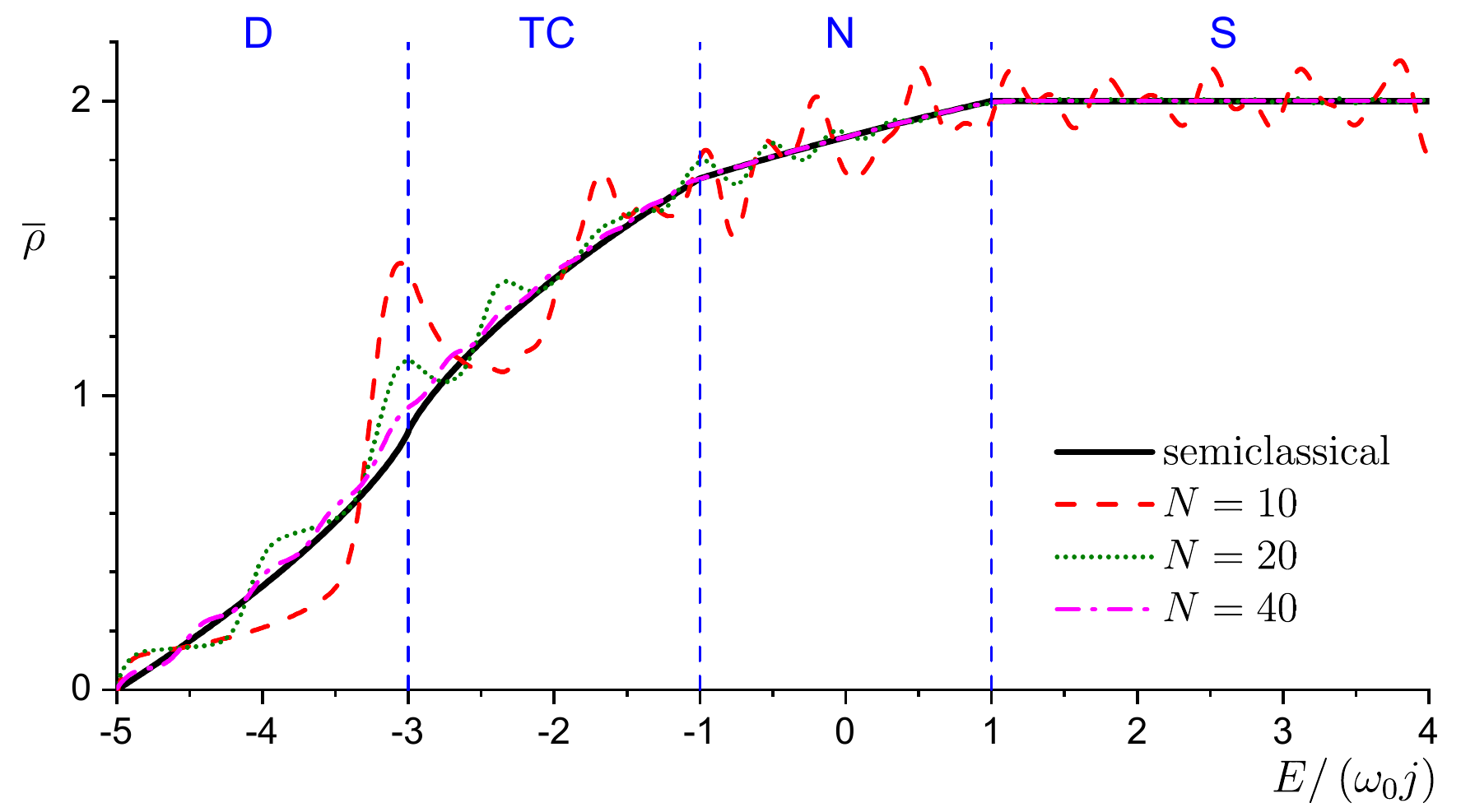}
\end{flushright}
\caption{
The energy dependence of the smoothed level density $\overline{\rho}(E)$ of Hamiltonian (\ref{HDic}) with ${\omega=\omega_0=1}$, ${\lambda=2.78}$, ${\delta=0.132}$ in the ${j=\frac{1}{2}N}$ subspace for several values of the size parameter $N$.
The finite-$N$ curves are  obtained from Eq.\,\eqref{leds} where the smoothening function $\overline{\delta}$ is a Gaussian with $\sigma=0.1$.
The semiclassical $N\I{\to}\infty$ curve is calculated from Eq.\,\eqref{smole} below.
Vertical lines indicate {\ESQPT} critical energies where $\frac{\partial}{\partial E}\overline{\rho}(E)$ for $N\I{\to}\infty$ has singularities (see Fig.\,\ref{dede}).
Acronyms {\D}, {\TC}, {\N} and {\S} mark quantum phases of the model.
}
\label{deda}
\end{figure}

Besides the above-outlined {\TPT} and {\QPT} kinds of criticality, the extended Dicke model exhibits also several types of {\ESQPTs}. 
They were first identified in Refs.\,\cite{Peres11a,Peres11b,Brand13} and then studied from the spectroscopic perspective in Refs.\,\cite{Basta14a,Basta14b,Kloc17a,Kloc17b}.
The key quantity allowing us to detect these effects is the density of energy eigenstates, or shortly the {\em level density}.
This is defined as
\begin{equation}
\rho(E)=\sum_{i}\delta(E-E_i)=-\frac{1}{\pi}\lim\limits_{\epsilon\to 0}\,{\rm Im}\,{\rm Tr}\,\frac{1}{E\I{+}\ii\epsilon\I{-}\hat{H}}
\label{led}\ ,
\end{equation}
where $E_i$ represents discrete eigenvalues of a general Hamiltonian of any bound quantum system.  
The expression on the right-hand side of Eq.\,\eqref{led} follows from the formula $\lim_{\epsilon\to 0}\epsilon/[(E\I{-}E_i)^2\I{+}\epsilon^2]=\pi\delta(E\I{-}E_i)$ and will be used later.
For the present moment, we skip the dependence of $\hat{H}$ (hence also $E_i$ and $\rho$) on control parameters.
The level density as a chain of $\delta$-functions is usualy smoothed by folding with a convenient normalized distribution function $\overline{\delta}(E)$, typically the Gaussian with zero mean and variance $\sigma^2$ such that the width $\sigma$ is larger than an average distance between neighboring levels.
In this way we obtain a smoothed level density
\begin{equation}
\overline{\rho}(E)=\int dE'\ \overline{\delta}(E-E')\,\rho(E')=\sum_{i}\overline{\delta}(E-E_i)
\label{leds}
.
\end{equation}
Note that $\overline{\rho}(E)$ for $\overline{\delta}(E)$ given by the Cauchy (Breit-Wigner) distribution of width $\gamma$ coincides with the right-hand side of Eq.\,\eqref{led} with $\epsilon$ kept at the fixed value $\epsilon\I{=}\gamma/2$ . 
 
Figure~\ref{deda} depicts examples of the smoothed level density of the extended Dicke Hamiltonian \eqref{HDic} in the subspace with $j\I{=}j_{\rm max}$ for several values of $N$.
We observe that with increasing $N$ the level density converges to an asymptotic dependence shown by the full curve.
It will be explained in Sec.\,\ref{Leden} that this ${N\to\infty}$ curve can be obtained independently of smoothing parameter $\sigma$ by means of semiclassical techniques involving the phase-space integration.
This asymptotic curve has three points of non-analyticity at energies indicated by the vertical dashed lines.
At the second and third lines, the curve $\overline{\rho}(E)$ shows a break, so the slope $\frac{\partial}{\partial E}\overline{\rho}(E)$ jumps.
At the first line, $\overline{\rho}(E)$ has a vertical tangent, so $\frac{\partial}{\partial E}\overline{\rho}(E)$ locally diverges.
In view of Eq.\,\eqref{clas} we can say that these singularities represent typical signatures of {\ESQPTs} in systems with $f=2$. 

\begin{figure}[t!]
\begin{flushright}
\includegraphics[width=0.85\textwidth]{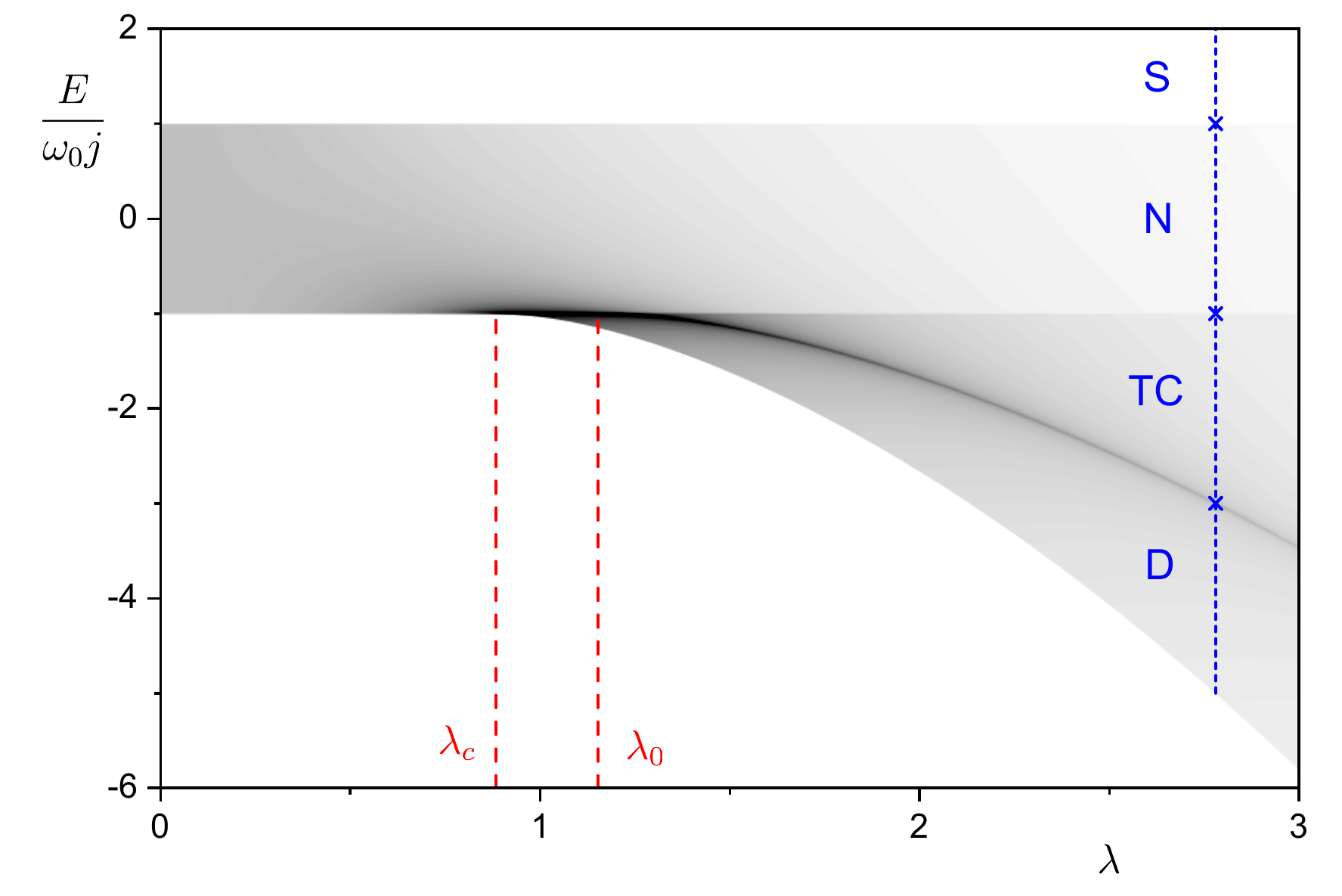}
\end{flushright}
\caption{
The derivative $\frac{\partial}{\partial E}\overline{\rho}(\lambda,E)$ of the $N\I{\to}\infty$ level density for Hamiltonian (\ref{HDic}) with ${\omega=\omega_0=1}$, ${\delta=0.132}$ and ${j=\frac{1}{2}N}$ in the plane $\lambda\times E$.
Darker shades mean larger values.
The {\ESQPTs} show up as jumps or divergences of the dependence. 
The value $\lambda\I{=}\lambda_{\rm c}$ marks the ground-state {\QPT} (drop of the lowest energy) while $\lambda_0$ corresponds to splitting of the {\ESQPT} borderlines. 
The vertical line with crosses at the {\ESQPT} critical energies and acronyms denoting quantum phases marks the value ${\lambda=2.78}$ used in Figs.\,\ref{deda} and \ref{nipere}.
Adapted from Ref.\,\cite{Cejna19}.}
\label{dede}
\end{figure}

Level densities $\rho(E)$ and $\overline{\rho}(E)$ depend on the control parameters of the system.  
The behavior of the first derivative $\frac{\partial}{\partial E}\overline{\rho}(E)$ in the plane ${\lambda\times E}$ is presented (for other model parameters fixed at the values indicated) in Fig.\,\ref{dede}.
We see that the non-analyticities from Fig.\,\ref{deda}, namely the jumps and divergences of the level-density derivative, exist within the whole ${\lambda\times E}$ plane and join to the {\QPT} non-analyticity affecting the ground state.
The curves where the level-density derivative is singular define {\em critical  borderlines\/} of the {\ESQPTs} in our model.
These curves separate domains in the spectrum of excited energy eigenstates that will be identified with different quantum phases distinguished by acronyms {\D}, {\TC}, {\N} and {\S}, see below.
For the critical borderlines in Fig.\,\ref{dede} there exist simple analytic expressions \cite{Kloc17a}.
We point out the role of the coupling parameter value $\lambda_0(\delta)$ from Eq.\,\eqref{lac}, where critical borderline between the {\D} and {\N} quantum phases splits and gives rise to the {\TC} phase.

\begin{figure}[t!]
\includegraphics[width=\textwidth]{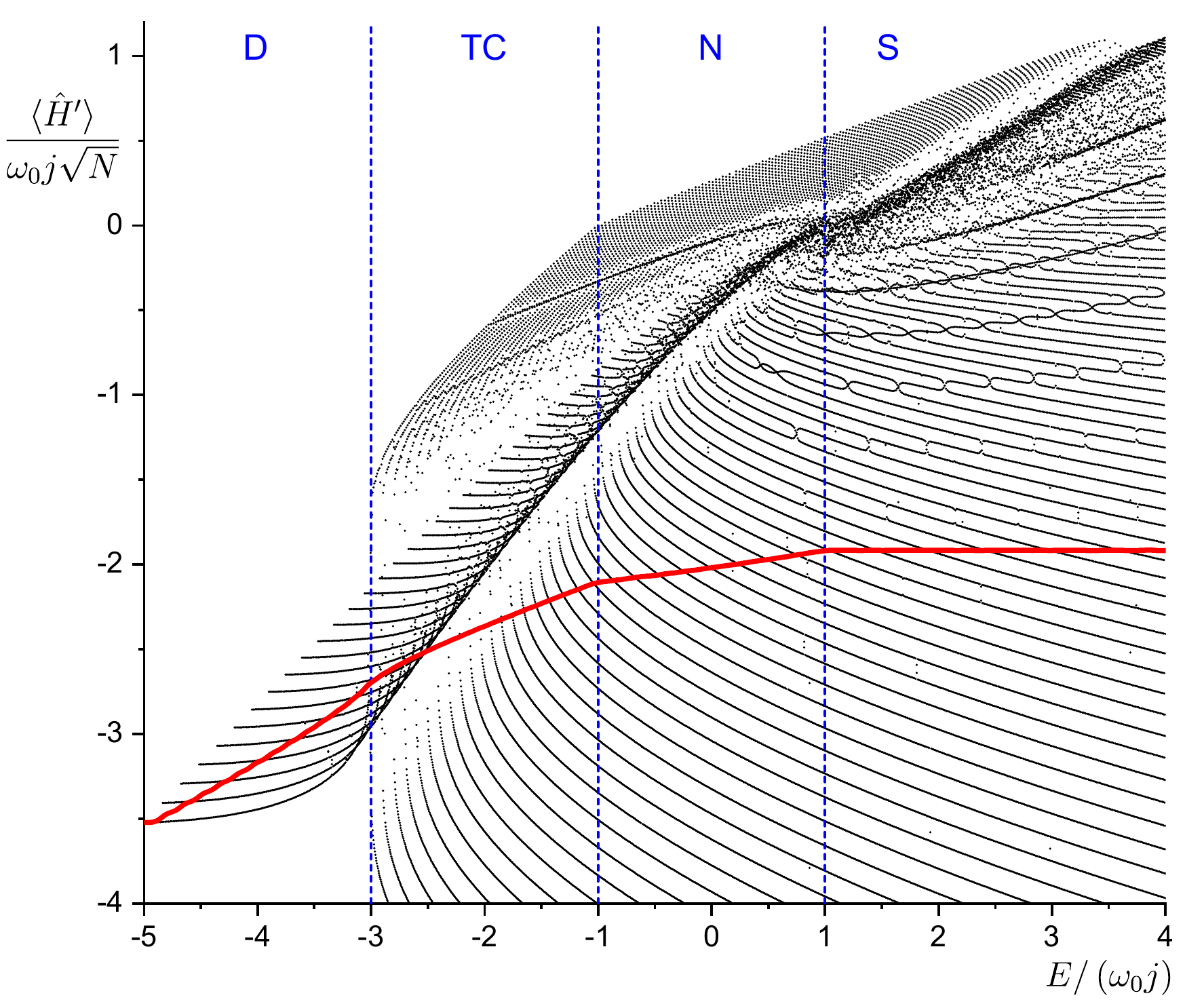}
\caption{Peres lattice for the observable $\propto\hat{H}'(\delta)$ of the model (\ref{HDic}) with ${\lambda=2.78}$ (see the vertical line in Fig.\,\ref{dede}), ${\delta=0.132}$, ${\omega=\omega_0=1}$ and ${N=2j=120}$. 
The red curve shows the smoothed energy dependence (\ref{Asmodep}). 
Vertical lines mark the {\ESQPT} critical energies (the same as in Fig.\,\ref{deda}).
Adapted from Ref.\,\cite{Cejna19}.}
\label{nipere}
\end{figure}

To show the {\ESQPT} spectral singularities in the perspective that allows for an introduction and interpretation of {\em quantum phases}, we present in Fig.\,\ref{nipere} a so-called {\em Peres lattice\/} of the observable ${\hat{A}\propto\frac{\partial}{\partial\lambda}\hat{H}(\lambda,\delta)=\hat{H}'(\delta)}$, i.e., the collective atom-field interaction term of Hamiltonian \eqref{HDic}.
In general, the Peres lattice \cite{Peres84} displays expectation values ${\ave{\hat{A}}_i=\matr{\psi_i}{\hat{A}}{\psi_i}}$ of an arbitrary observable $\hat{A}$ in a discrete set of Hamiltonian eigenstates $\ket{\psi_i}$ as a mesh in the plane ${E_i\times\ave{\hat{A}}_i}$.
This kind of representation of discrete spectra was used to distinguish regular and chaotic subsets of states, or to identify other relevant dynamic properties (see, e.g, Ref.\,\cite{Stran09}).
In the present case, the Peres lattice visualizes effects of the {\ESQPTs}.
The choice of model parameters in Fig.\,\ref{nipere} is the same as in Fig.\,\ref{deda} (the vertical line in Fig.\,\ref{dede}).
We notice that the {\ESQPT} critical borderlines $E_{\rm c}(\lambda)$ split the lattice into domains showing partially different arrangements of points.
When crossing any of these borderlines, there exists a certain feature of the lattice that changes abruptly, while between the borderlines the lattice remains qualitatively the same. 
This lays the ground for identifying the domains between borderlines with distinct quantum phases of the system.
Although specific types of lattice arrangements depend on the model and observable selected, it turns out that {\ESQPTs} in a general system always imply some qualitative changes of patterns in Peres lattices of any generic observable.
In the present model we use the following abbreviated names of the quantum phases: 
Acronyms {\D} and {\TC} stand for Dicke and Tavis-Cummings types of superradiant phase (the corresponding fragments of the Peres lattice resemble the low-energy lattices in the ${\delta=1}$ and 0 limits, respectively), {\N}~denotes the normal, i.e., non-radiant phase (with ${E>-\omega_0j}$), and {\S} marks the \uvo{saturated} phase (with constant level density, see Figs.\,\ref{deda} and \ref{dede}).
For a~more detailed description of these phases see Ref.\,\cite{Kloc17a}.

Given a general observable $\hat{A}$, we define its {\em energy density\/} in the discrete spectrum: $A(E)=\sum_i\ave{\hat{A}}_i\,\delta(E\I{-}E_i)$. 
The corresponding smoothed density of $\hat{A}$,
\begin{equation}
\overline{A}(E)=\sum_i\ave{\hat{A}}_i\,\overline{\delta}(E-E_i)
\label{Asmodep},
\end{equation}
can be also used for detection of {\ESQPTs}.
This quantity is obtained by a local averaging of values $\ave{\hat{A}}_i$ in the Peres lattice within a narrow energy interval around the given value $E$ and exhibits singularities at the {\ESQPT} critical energies.
In Sec.\,\ref{Leflo} we argue that the type of singularity in $\overline{A}(E)$ most likely coincides with the type of singularity in $\overline{\rho}(E)$.
This is illustrated in Fig.\,\ref{nipere}, where the smoothed density of observable $\propto\hat{H}'(\delta)$ is shown by the thick red curve.
We see breaks of the curve at the {\ESQPT} critical energies associated with {\TC}--{\N} and {\N}--{\S} phase boundaries, and an increased slope (precursor of the vertical tangent) of the curve at the {\D}-{\TC} phase boundary (cf.\,Figs.\,\ref{deda} and \ref{dede}).

\subsection{Singularities in the Tavis-Cummings and Rabi regimes}
\label{EDM3}

We have seen that the extended Dicke model shows {\ESQPTs} in the first derivative of the level density, as expected for the number of {\DoFs} $f\I{=}2$.
However, the model also presents special regimes in which exact or approximate integrals of motions enable us to apply  effective descriptions with $f\I{=}1$ to specific subsets of states.
Some of these subsets show abrupt changes of the lowest energy state, in a full analogy to the ground-state {\QPT}, while the spectra of states with higher energies exhibit {\ESQPT} singularities.
Due to the reduced {\DoF} number, these {\ESQPTs} affect the zeroth instead of first derivative of the level density and produce sharper signatures in Peres lattices.

\begin{figure}[t!]
\begin{flushright}
\includegraphics[width=\textwidth]{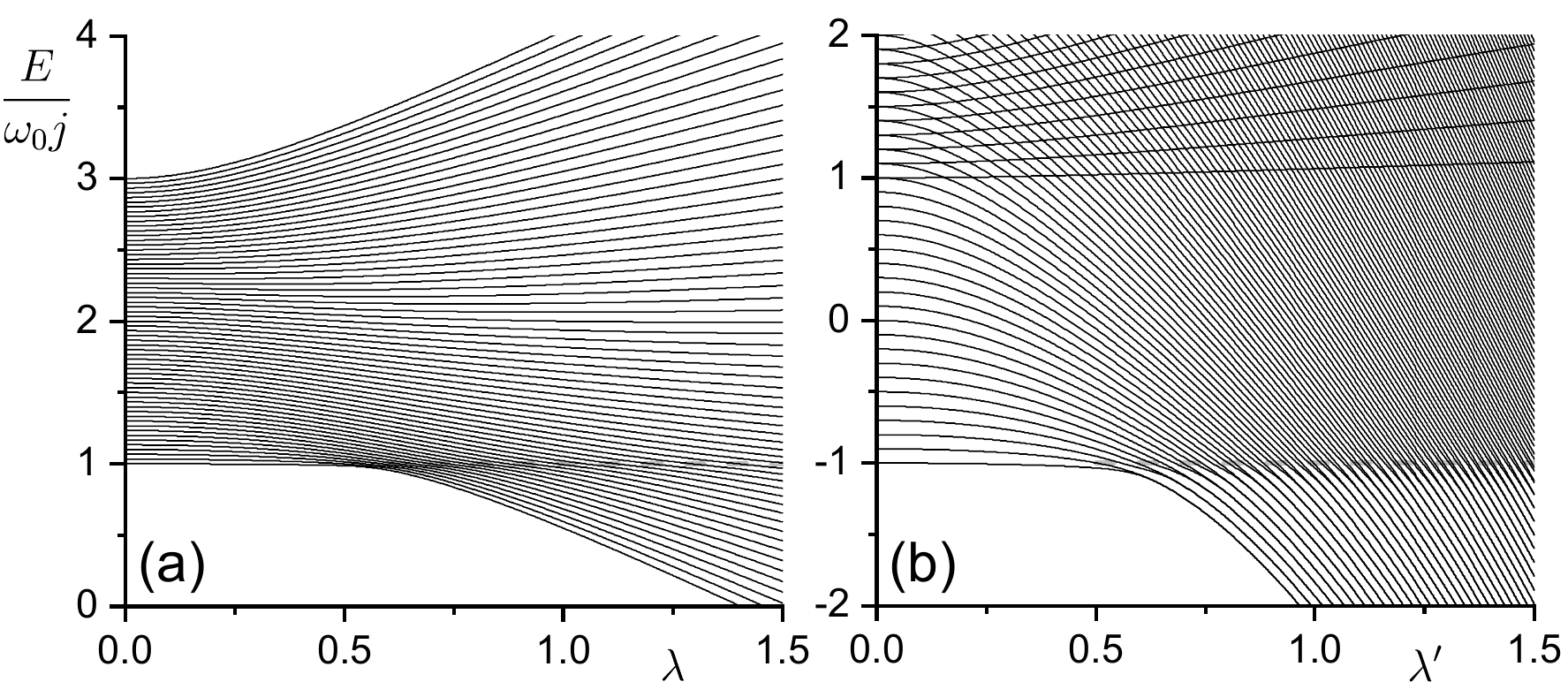}
\end{flushright}
\caption{The ${f=1}$ {\ESQPT} structures (marked by gray dashed lines) in (a) the Tavis-Cummings regime and (b) the Rabi regime of the Dicke model from Eq.\,\eqref{HDic}. Panel (a) depicts the spectrum of ${M=2j=N}$ states of the $\delta=0$ Hamiltonian with ${\omega=2},{\omega_0=1}$ for ${N=60}$. Panel (b) shows the spectrum of the ${N=1=2j}$ system with ${\omega=1},{\omega_0=R\omega=20}$ and ${\delta=1}$.}
\label{bunch}
\end{figure}

An immediate example of such a regime is the {\em Tavis-Cummings version\/} of the extended Dicke Hamiltonian \eqref{HDic} with $\delta\I{=}0$.
As explained above, the model in this case conserves the summed number of atomic and field excitations $\hat{M}$ and for each fixed value $M$ a canonical transformation maps the system to a reduced $f\I{=}1$ phase space \cite{Kloc17a,Kloc17b}.
The $\lambda$-dependent spectra of states in a given $M$-subspace can be analyzed independently from other $M$-subspaces.
The dimension of these subspaces is $M\I{+}1$ for ${M\leq 2j}$ and $2j\I{+}1$ for $M\I{>}2j$. 
The value $M\I{=}2j$ is critical not only because it is the highest $M$ number that permits states with $N_b\I{=}0$, but also because the corresponding spectrum exhibits both {\QPT} and {\ESQPT} effects in the limit $N,j\to\infty$ \cite{Peres11a,Peres11b,Kloc17a,Kloc17b}.
A finite-$N$ sample of the $M\I{=}2j$ spectrum is shown in panel (a) of Fig.\,\ref{bunch}.
We observe a precursor of a second-order {\QPT} at a critical value ${\lambda=\frac{1}{2}|\omega-\omega_0|\equiv\lambda_{\rm c}'}$, which differs from the critical coupling $\lambda_{\rm c}(\delta\I{=}0)$ associated with the global ground state of the whole model. 
The subsequent bunching of levels (sequence of avoided crossings behind the critical point) indicates an {\ESQPT} with diverging density of states in the ${N\to\infty}$ limit.
This is the most familiar form of {\ESQPT} present in numerous $f\I{=}1$ models, which initiated the study of {\ESQPTs} in the past \cite{Cejna06,Cejna07,Heinz06,Macek06,Capri08} and represents the case most discussed in the literature up to now (cf.\,Sec.\,\ref{Finite}).

\begin{figure}[t!]
\includegraphics[width=\textwidth]{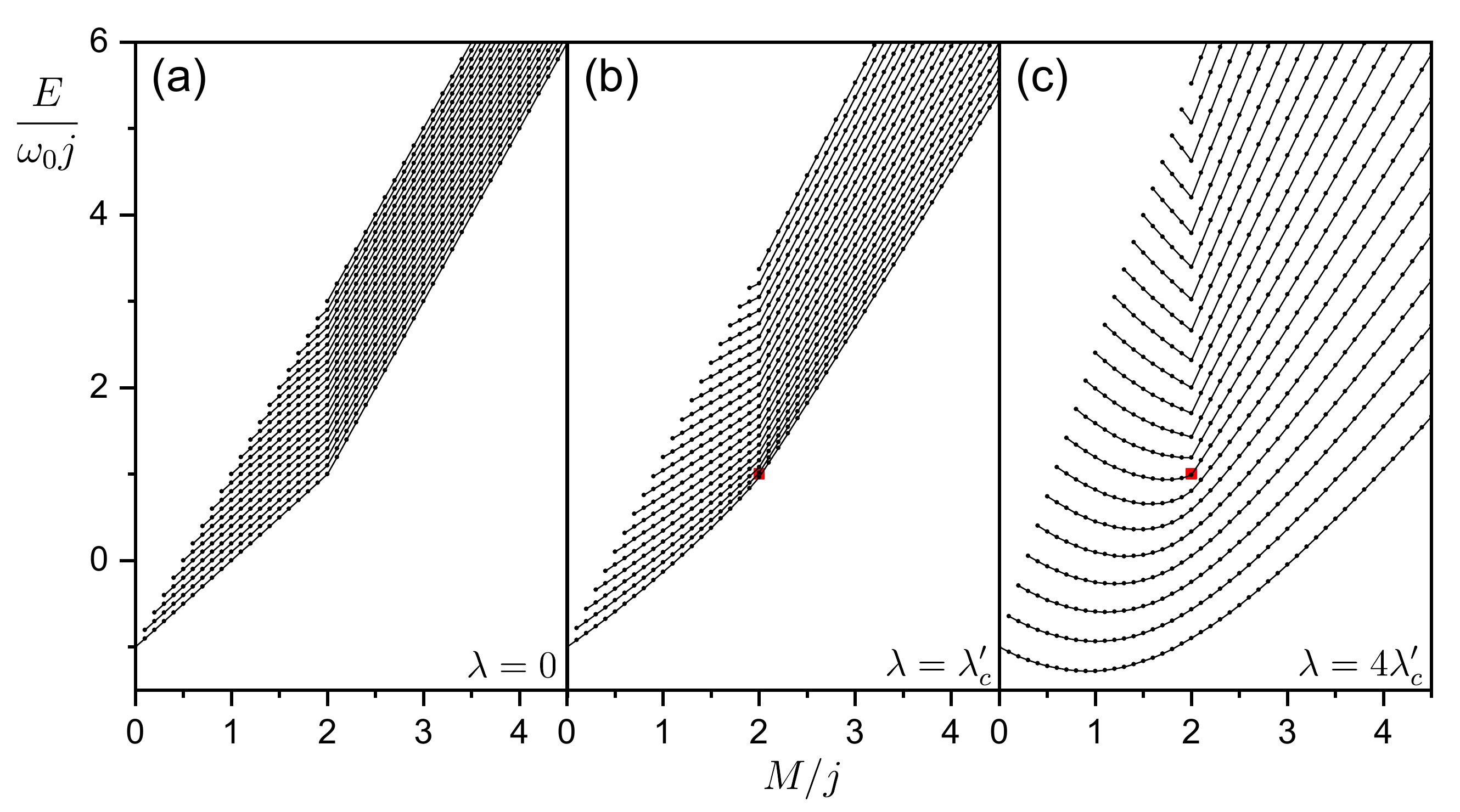}
\caption{Peres lattices of the Tavis-Cummings Hamiltonian in the plane $M$ {\it vs.} $E$
for three values of $\lambda$ indicated in panels (a)--(c).
The critical interaction strength is $\lambda’_{\rm c}=0.5$, see Fig.\,\ref{bunch}(a).
The other parameters in Eq.\,\eqref{HDic} are as follows: $\omega=2$, $\omega_0=1$ and $N=2j=20$.
The {\QPT} and {\ESQPT} points of the $M=2j$ spectrum are marked by the red rectangle.
In panel (c) this rectangle represents a point defect of the lattice indicating quantum monodromy. 
}
\label{inpere}
\end{figure}

Peres lattices of the {\TC} Hamiltonian with $\omega\I{=}2\omega_0$ for various values of parameter $\lambda$ are presented in Fig.\,\ref{inpere}.
The energy is on the vertical axis, while the horizontal axis shows values of $M$; so both axes now correspond to conserved quantities, in contrast to Fig.\,\ref{nipere}.
The vertical chain of points at $M/j\I{=}2$ represents the energy spectrum of the critical $M$-subspace.
Figure~\ref{inpere} in its three panels depicts lattices at (a) $\lambda\I{=}0$ (non-interacting case), (b) $\lambda\I{=}\lambda_{\rm c}'$ (the {\QPT} critical value for $M\I{=}2j$), and at $\lambda\I{=}4\lambda_{\rm c}'$ (the case with {\ESQPT}).
The red rectangle in panel (b) corresponds to the critical-point ground state of the $M\I{=}2j$ subset of states [cf.\,Fig.\,\ref{bunch}(a)] and that in panel (c) marks the critical {\ESQPT} energy $E'_c=\omega_0j$.

In panel (a) of Fig.\,\ref{inpere} we see that all lines connecting points with the same principal quantum number $i=0,1,...$ enumerating energy levels $E_{Mi}$ exhibit a break at $M\I{=}2j$.
This is so because above this value the increase of $M$ is possible only by adding photons whose energy $\omega$ differs from energy $\omega_0$ of the atomic excitation quantum  (the break angle depends on these energies).
The form of the lattice remains qualitatively the same for $\lambda\I{\leq}\lambda_{\rm c}'$ (panel b), but for $\lambda\I{>}\lambda_{\rm c}'$ (panel c) the lattice gets split into the upper part with the V-shaped breaks of the constant-$i$ lines, and the lower part with a~{U-shaped} bends of these lines.
The central point (the red rectangle) separating these structures at the {\ESQPT} energy $E'_{\rm c}$ of the $M\I{=}2j$ spectrum represents a point defect of the whole lattice.
It can be demonstrated that elementary cells of the lattice are getting distorted on a path around the defect so that after completing a full circle the final cell differs from the initial one. 
This is a manifestation of {\em quantum monodromy}---a specific anomaly in quantized spectra of some integrable systems associated with the existence of a singular bundle of classical trajectories in the phase space \cite{Duist80,Cushm88,Ngoc99}.  
Classical and quantum monodromy in the context of the {\TC} model was studied in Refs.\,\cite{Babel09,Kloc17b}.
We note that the connection of {\ESQPTs} to monodromy was described in a number of other integrable models with $f\I{=}2$, particularly in the molecular vibron models \cite{Lares11,Lares13} and in the nuclear interacting boson model along the transition between U(5) and SO(6) dynamical symmetries \cite{Heinz06,Macek06,Cejna06,Cejna07}, see Sec.\,\ref{Ibms}.

Another possibility to reduce the number of {\DoFs} in the Dicke model is connected with the {\em Rabi regime}, achieved under the condition of adiabatic separation of atomic and field variables \cite{Relan16b,Basta17}.
In particular, we assume that the atom to photon energy ratio $R=\omega_0/\omega$ is very large, $R\I{\gg}1$.
This violates the above-mentioned assumption of comparable energy costs for excitations of both subsystems.
As the expected number of photons $\ave{\hat{N}_b}_i$ in a typical eigenstate $\ket{\psi_i}$ at energies $E_i\sim N\omega_0$ fairly exceeds the expected number of atomic excitations $\ave{\hat{N}_*}_i$, the field and atomic subsystems need to be characterized by different size parameters.
It turns out (cf.\,Sec.\,\ref{Fisie}) that convenient size parameters are $N$ for the atomic subsystem and $NR$ for the field subsystem.
Using the coordinate-momentum representation of the field operators from Eq.\,\eqref{QP} with ${\eta=NR}$, we rewrite the Hamiltonian \eqref{HDic} in the form
\begin{equation}
\hspace{-15mm}
\frac{\hat{H}}{NR}=-\frac{\omega}{2NR}+\frac{\omega}{2}(\hat{q}^2+\hat{p}^2)+
\underbrace{\left(\sqrt{2}\,\lambda'(1\I{+}\delta)\,\hat{q},\sqrt{2}\,\lambda'(1\I{-}\delta)\,\hat{p},\frac{\omega}{N}\right)}_{\hat{\vecb{B}}}\cdot\ \hat{\vecb{J}}
\label{HRab}
\end{equation}
with $\hat{\vecb{J}}\equiv(\hat{J}_x,\hat{J}_y,\hat{J}_z)$ and $\lambda'=\lambda/N\sqrt{R}$.

For very large values of $R$ the field subsystem is close to the classical limit, so the field operators $(\hat{q},\hat{p})$ can be replaced by the corresponding classical variables $(q,p)$. 
The operators $\hat{\vecb{B}}$ in Eq.\,\eqref{HRab} turn into ordinary numbers $\vecb{B}$ and the Hamiltonian becomes a $(q,p)$ dependent operator acting solely in the atomic Hilbert space $\Hi_{\rm A}$.
Defining $\hat{J}_{z'}=\vecb{B}\cdot\hat{\vecb{J}}/|\vecb{B}|$ as the quasispin operator projected to a new $z'$ axis pointing along the $\vecb{B}/|\vecb{B}|$ direction, the diagonalization of the Hamiltonian \eqref{HRab} becomes trivial.
The eigenenergies $E_{m'}(q,p)$ are enumerated by the projection quantum number $m'=-j,...,+j$ of the operator $\hat{J}_{z'}$ and depend on the field variables. 
They can be interpreted as effective classical Hamiltonians $H_{m'}(q,p)$ controlling slow dynamics of the field subsystem for individual fixed quasispin projections $m'$.
Semiclassical quantization of the field states for each of these effective Hamiltonians yields a discrete energy spectrum enumerated by $i=1,2,...$.
The full spectrum resulting from this procedure is labeled by a pair of quantum numbers $m'$ and $i$, and consists of bands of densely spaced field excitations distinguished by $i$ built on sparsely spaced atomic excitations characterized by $m'$.
In this approximation, the energies $E_{m'i}$ of levels in different $m'$ bands can cross each other without repulsion, which is a consequence of the approximate integral of motion $\hat{J}_{z'}$.

As follows from Eq.\,\eqref{HRab}, some of the effective classical-limit Hamiltonians $H_{m'}(q,p)$ that determine the states within the same $m'$ band undergo a phase transitional change.
In particular, for ${m'<0}$, at a certain critical value of $\lambda'$, for $m'\I{=}-\frac{1}{2}N$ equal to ${\lambda_{\rm c}'=\omega/N(1\I{+}\delta)}$, the $(q,p)\I{=}(0,0)$ global minimum of $H_{m'}(q,p)$ switches into a local maximum. 
Hence the bandhead state (${i\I{=}1}$) shows a transition interpreted as a {\QPT}, and an accompanying {\ESQPT} structure corresponding to the {\DoF} number ${f\I{=}1}$ is observed in the spectrum of excited states (${i\I{>}1}$).
This {\ESQPT} was first demonstrated in Ref.\,\cite{Puebl16}.
For ${\delta<1}$, the system even exhibits {\ESQPTs} of several types, in analogy to the richer {\ESQPT} structure of the extended Dicke model. 
As the size parameter that determines the sharpness of these critical effects is associated with $R$ rather than $N$, well developed {\ESQPT} structures can be studied even in a system with $N\I{=}1$ \cite{Hwang15,Puebl16}.
An example of such a spectrum is shown in panel (b) of Fig.\,\ref{bunch}. 

The above treatment of the Rabi regime is a specific application of the Born-Oppenheimer approximation based on an adiabatic separation of slow and fast modes of motions.
Many other examples exist in models overviewed in Sec.\,\ref{Finite}.
A general discussion of {\ESQPT} effects in such imbalanced systems will continue in Sec.\,\ref{Fisie}.

\section{Other collective many-body models}
\label{Finite}

Proceeding from the atom-field system discussed above, we overview in this section a variety of other many-body systems whose quantal spectra were proven to show similar {\ESQPT} singularities.
A common property of these systems is a moderate number $f$ of collective {\DoFs}, which implies that the limit of infinite number of constituents can be associated with the classical limit of collective dynamics. 

\subsection{Quasispin systems}
\label{Quasi}

We start with a class of models describing a collection of $N$ interacting two-state systems (qubits). 
Individual qubits, represented for instance by spin-$\frac{1}{2}$ particles or two-level atoms, are connected with identical two-dimensional Hilbert spaces ${\Hi^{(i)}={\mathbb C}^2}$ enumerated by ${i=1,2,...,N}$.
So the Hilbert space of the $N$-qubit system is ${\Hi=\bigotimes_{i=1}^{N}\Hi^{(i)}}$ with overall dimension ${d=2^N}$.  
Observables related to each qubit can be written in terms of Pauli matrices $\hat{\sigma}^{(i)}_{\alpha}$, with ${\alpha=(0,+,-)}$ or $(x,y,z)$, and the unit operator $\hat{I}^{(i)}$ acting in the respective space.
As the operators $\frac{1}{2}\hat{\sigma}^{(i)}_{\alpha}$ have commutation relations of spin, we unify the present kind of models under the term \uvo{quasispin}.
In this sense, the atoms in the Dicke model form a quasispin subsystem.
In the same way as described in Sec.\,\ref{EDM1}, each qubit can be endowed with an elementary Bloch sphere of radius $\frac{\sqrt{3}}{2}$ and can therefore be associated with a single {\DoF}.
The full $N$-qubit system has ${f=N}$ quantum {\DoFs} which do not become classical as ${N\to\infty}$.

Each qubit is supposed to have the same self-Hamiltonian such as ${\hat{H}_0^{(i)}=\frac{1}{2}\omega\hat{\sigma}^{(i)}_{z}}$, where $\omega$ is the difference between energy levels.
Hence the non-interacting part of an $N$-qubit Hamiltonian reads ${\hat{H}_0=\sum_{i=1}^{N}\hat{H}_0^{(i)}=\omega\hat{J}_{z}}$, cf. Eq.\,\eqref{quasis}. 
The interaction between qubits is usually assumed to be of the two-body type, so the coupling between $i$th and $i'$th qubits is written in terms of expressions like $[\hat{\sigma}^{(i)}_{\alpha}\hat{\sigma}^{(i')}_{\alpha'}+{\rm H.c.}]$.
For example, the celebrated {\em Ising model\/} associates the qubits with spins, assuming the Hamiltonian
\begin{equation}
\hat{H}=B\hat{J}_z-A\sum_{\langle ii'\rangle}\hat{\sigma}^{(i)}_{x}\hat{\sigma}^{(i')}_{x}
\label{Hising},
\end{equation}
where coefficient $B$ quantifies one-body interactions of individual spins with a magnetic field in direction $z$, and $A$ is the strength of mutual two-body interactions between spins.
We assume $A,B\I{>}0$.
The spins are located in vertices of a spatial lattice of dimension $D$ and their interactions involve only the spin pairs with the smallest distance (the closest neighbors); this is expressed by the symbol $\langle ii'\rangle$ in the sum.

The Ising model has the well known phase structure. 
For ${D\geq 2}$ lattices described by the Hamiltonian \eqref{Hising} with the ratio $\lambda\I{=}B/A$ below a certain critical value $\lambda_{\rm c}$, there exists a critical temperature $T_{\rm c}(\lambda)$ at which the system shows a {\TPT}  between the ferromagnetic ($T\I{<}T_{\rm c}$) and paramagnetic ($T\I{>}T_{\rm c}$) thermal phases. 
These phases are characterized, respectively, by non-zero and zero expectation values of the order parameter $\hat{J}_x$.
Quantitative description of this transition requires the use of the renormalization group theory \cite{Ma76}.
At $\lambda\I{=}\lambda_{\rm c}$, where $T_{\rm c}\I{=}0$, the system exhibits (for arbitrary $D$) a {\QPT} between the ferromagnetic ($\lambda\I{<}\lambda_{\rm c}$) and paramagnetic ($\lambda\I{>}\lambda_{\rm c}$) ground-state phases \cite{Sachd99}.
States in the ferromagnetic phase (in both thermal and quantum regimes) spontaneously break the symmetry of the system under the discrete rotation around the $z$ axis by angle $\pi$.
However, the equality $f\I{=}N$ disables observations of {\ESQPTs} as the condition $N\I{\to}\infty$ implies $f\I{\to}\infty$ and does not lead to the classical limit.
The {\ESQPTs} appear in quasispin systems only if a kind of collectivity reduces the number of effective {\DoFs} to a moderate value not increasing with $N$.

The commonly discussed example of this collective behavior is the case of infinite-range interactions,  i.e., interactions connecting with the same strength any pair of spins in the ensemble.
Such models are often called {\em fully connected}.
The Hamiltonian \eqref{Hising} modified in this way reads
\begin{equation}
\hat{H}=B\hat{J}_z-\frac{A}{N}\sum_{i<i'}\hat{\sigma}^{(i)}_{x}\hat{\sigma}^{(i')}_{x}=B\hat{J}_z-\frac{A}{N}\,\hat{J}_{x}^2+\frac{A}{4}
\label{HLip0},
\end{equation}
where we used the collective quasispin operators from Eq.\,\eqref{quasis}.
The interaction strength is attenuated with increasing $N$ to prevent complete dominance of the interaction term in the total energy for $N\I{\to}\infty$.  
Hamiltonian \eqref{HLip0} is a special case of what is called the {\em Lipkin-Meshkov-Glick model}, or simply the Lipkin model \cite{Lipki65}.
It replaces the multiple SU(2) algebras of quasispin operators $\frac{1}{2}\hat{\sigma}^{(i)}_{\alpha}$ associated with  individual qubits by a single SU(2) algebra of the collective quasispin operators $\hat{J}_{\alpha}$. 
Hence a unique collective Bloch sphere can be attributed to the whole system, reducing the number of {\DoFs} to $f=1$.

Hamiltonian \eqref{HLip0} and all its variations written in terms of collective operators $\hat{J}_{\alpha}$, 
\begin{equation}
\hat{H}=\omega\hat{J}_{z}+
\sum_{\alpha,\alpha'}\frac{{\gamma}_{\alpha\alpha'}}{2N}\left(\hat{J}_{\alpha}\hat{J}_{\alpha'}+\hat{J}^{\dag}_{\alpha'}\hat{J}^{\dag}_{\alpha}\right)
+\dots
\label{Hsu2},
\end{equation}
where $\omega$ and $\gamma_{\alpha\alpha'}$ (and so on) are coefficients at linear, quadratic (and eventually higher) terms, conserve the total squared quasispin $\hat{\vecb{J}}^2$.
Subspaces of the full Hilbert space $\Hi$ with a fixed value of $j$ are dynamically invariant, i.e., if the initial state is in this subspace, the evolved state never goes out.
To investigate the dynamics of the model,  one can therefore choose any of these subspaces.

Properties of subspaces with different $j$ were discussed for instance in Ref.\,\cite{Cejna16}.
As already mentioned in Sec.\,\ref{EDM1}, the values of $j$ are integers (for $N$ even) or half-integers (for $N$ odd) between 0 and $\frac{1}{2}N$.
For each $j$, the total quasispin projection takes $2j\I{+}1$ values from $-j$ to $+j$.
The match with the dimension $d\I{=}2^N$ of the full space $\Hi$ is achieved because the subspaces of states with a given $j$ appear in multiple replicas $\Hi_{j}^{(s)}$ enumerated by number $s$, so we can write: $\Hi=\bigoplus_{j=j_{\rm min}}^{j_{\rm max}}\bigoplus_{s=1}^{\nu_{j}}\Hi_{j}^{(s)}$.
The multiplicity $\nu_{j}$ of the space with given $j$ is a very large number for $0\I{\ll}j\I{\ll}\frac{1}{2}N$.
The subspaces differ in the symmetry with respect to exchange of qubits.  
Each value $j$ is associated with a particular form of the Young tableaux and various replicas $\Hi_{j}^{(s)}$ correspond to various permutations of qubits within this tableaux (note that due to the symmetry of operators $\hat{J}_{\alpha}$ under the qubit exchange, the above-mentioned dynamical invariance concerns each of the replica subspaces separately).
For a given $j$, the number $N\I{-}2j$ counts qubits that appear in exchange antisymmetric states, mutually compensating their contributions to the total quasispin projection $m$, while the number $2j$ represents unpaired active qubits which contribute to the value of $m$.
The subspace $\Hi_{j_{\rm max}}$ with the maximal number $2j\I{=}N$  is totally symmetric under the exchange of any pair of qubits and therefore unique in $\Hi$.
This subspace represents the common choice for a majority of dynamical studies.

Models based on the SU(2) algebra, which all belong to the family of Lipkin-like models, are probably the most frequently used playgrounds in various {\ESQPT}-related studies.
Besides an elegant mathematical formulation, relatively undemanding numerical solubility (moderate-dimensional Hilbert space  $\Hi_{j_{\rm max}}$) and large versatility of such models, the main advantage of this choice is the lowest possible number of collective {\DoFs}, $f\I{=}1$, for which the {\ESQPT} signatures are most pronounced.
They can be detected even for moderate values of the naturally defined size parameter $N$.
Many {\ESQPTs} disclosed in various Lipkin and related models can be found in Refs.\,\cite{Leyvr05,Reis05,Riber07,Riber09,Lopez11,Puebl13,Engel15,Sinde17,Romer17,Opatr18,Monda20} (identification of {\ESQPTs} and description of underlying mechanisms), \cite{Webst18,Relan18,Kloc19} (thermodynamic consequences) and \cite{Relan18,Peres09,Puebl15,Kopyl15b,Santo16,Kopyl17,Wang17,Zimme18,Wang19,Wang19b,Puebl20} (dynamic effects).

\begin{figure}[t!]
\begin{flushright}
\includegraphics[width=\textwidth]{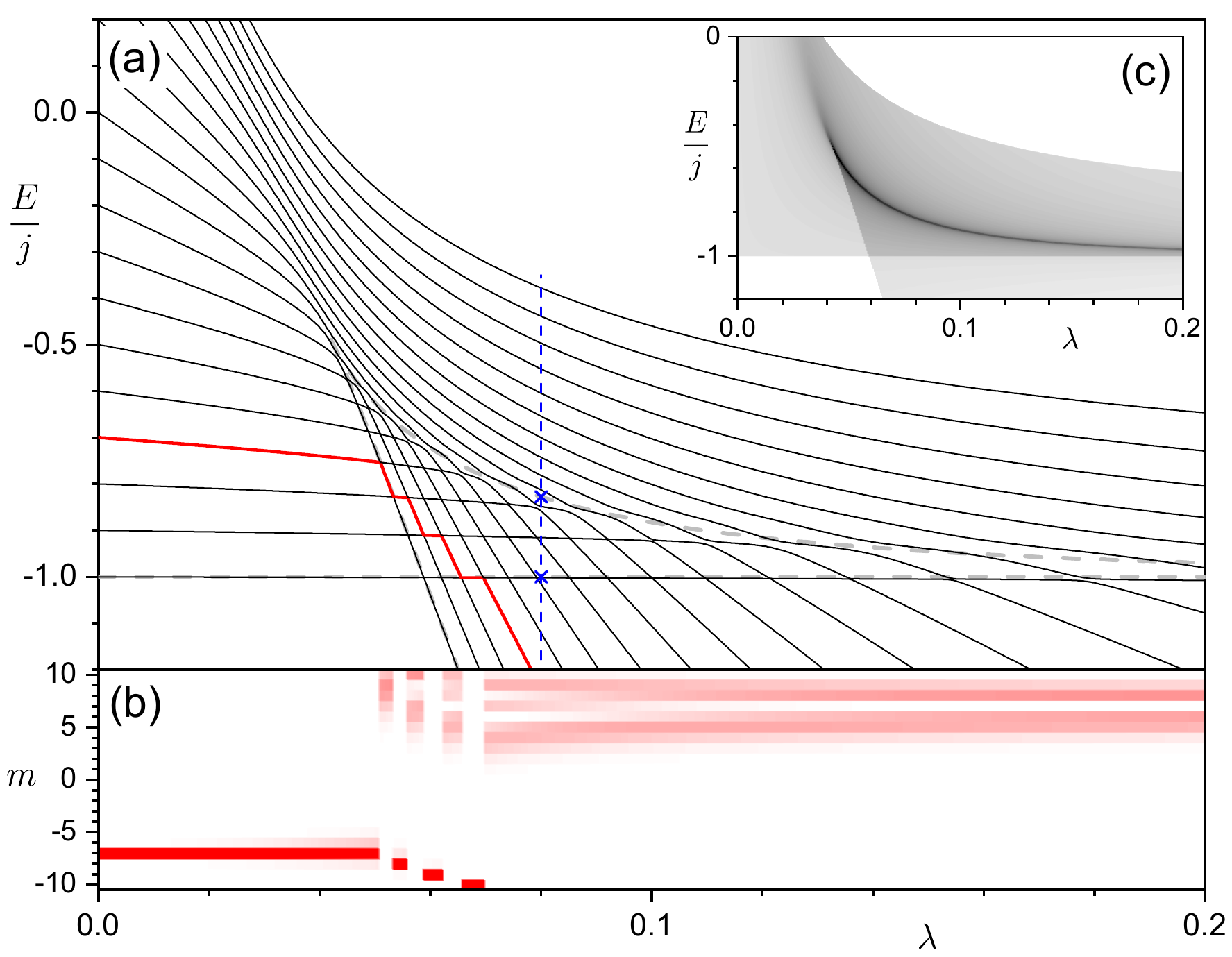}
\end{flushright}
\caption{
The {\ESQPT} structures in the spectrum of the Lipkin Hamiltonian \eqref{HLipkin} with $\chi\I{=}4$ and running $\lambda$.
Panel~(a): Energy levels (full curves) for $N\I{=}2j\I{=}20$ and energies of stationary points of the classical-limit Hamiltonian (gray dashed lines). Straight dashed lines correspond to two minima and the bent dashed line to the saddle point between them. The crossing of minima at $\lambda_{\rm c}\I{=}\frac{1}{17}$  indicates the first-order {\QPT}. The higher of the two minima on both sides of the {\QPT} and the saddle point generate two types of {\ESQPT} that show up in the smoothed level density in panel (c). The vertical line (with crosses at {\ESQPTs}) marks $\lambda$ used in Figs.\,\ref{liporder} and \ref{neher}.
Panel~(b): Evolving structure of the 4th eigenstate from panel (a) expressed by the overlap $|\scal{m}{\psi_4(\lambda)}|^2$ of the instantaneous eigenstate with the basis of $\hat{J}_z$ eigenstates $\ket{m}$ 
(darker shades mean larger overlaps). We note a sequence of sharp changes at each avoided crossing of levels.
Panel~(c): The smoothed ($N\I{\to}\infty$) level density $\overline{\rho}(E)$ (darker shades mean larger values) showing the {\ESQPT} non-analyticities. We observe a step-like singularity at the higher of the two Hamiltonian minima and a logarithmic divergence at the saddle point.
}\label{lipkac}
\end{figure}

Below we will often consider the Lipkin Hamiltonian 
\begin{equation}
\hat{H}=\hat{J}_z-\frac{\lambda}{N}\left[\hat{J}_x+\chi\left(\hat{J}_z+\frac{N}{2}\right)\right]^2
\label{HLipkin}.
\end{equation}
This model will double the Dicke and related models in the role of the principal illustrative examples of this review (it will be used in Figs.\,\ref{lipkac},\,\ref{liporder},\,\ref{epka} and \ref{Quo1}--\ref{neher} below).
While the interaction strength $\lambda\I{>}0$ is considered as the running parameter, the additional parameter $\chi$ is kept constant.
The spectrum of Hamiltonian \eqref{HLipkin} is mirror-symmetric with respect to transformation $\chi\I{\to}-\chi$, so it is sufficient to consider only the case $\chi\I{\geq}0$.
For $\chi\I{=}0$, the model has the second-order ground-state {\QPT} with the subsequent {\ESQPT} [cf.\,Eq.\,\eqref{HLip0}]. 
In this case the system conserves the parity $\hat{\Pi}\I{=}(-1)^{\hat{J}_z+j}$.
For $\chi\I{\neq}0$, the parity symmetry is broken and the model exhibits the first-order {\QPT} with the associated {\ESQPTs}.
The first- or second-order ground-state {\QPT} from the non-interacting to the interacting phase happens at the critical interaction strength ${\lambda_{\rm c}(\chi)=(1\I{+}\chi^2)^{-1}}$ \cite{Cejna16}. 
An example of the {\ESQPT} spectral structures accompanying a first-order {\QPT} for this Hamiltonian is depicted in Fig.\,\ref{lipkac}.
Note that similar {\ESQPT} patterns (but for $f\I{>}1$ shifted to higher derivatives of the level density) are typical near generic first-order {\QPTs} \cite{Cejna08,Stran14,Stran15,Macek19}.
The spectrum showing the {\ESQPT} behind the second-order {\QPT} point can be seen in Fig.\,\ref{epka}(a) below.

A straightforward extension of the quasispin models results from {\em coupling\/} of two or generally $n\geq 2$ quasispin subsystems.
Such models work with sets of quasispin operators $\{\hat{\vecb{J}}_k\}_{k=1}^{n}$ associated with each subsystem, using Hamiltonians analogous to \eqref{Hsu2} but with quadratic and eventually higher-order terms combining quasispin operators of various subsystems.
The number of {\DoFs} is $f\I{=}n$, but in some cases additional integrals of motion reduce the number of effective {\DoFs} in the relevant subspaces. 
A system of $n\I{=}2$ coupled Lipkin models and its {\ESQPTs} were studied in Ref.\,\cite{Garci17}.
Both subsystems were sent to the classical limit simultaneously by applying the limit $j_1,j_2\I{\to}\infty$ to both quasispin size quantum numbers.
In contrast, in the study of Ref.\,\cite{Relan16}, an integrable two-spin elliptic Gaudin model was analyzed with a strong imbalance between the sizes of both subsystems. 
This is analogous to the Rabi model discussed in Sec.\,\ref{EDM3}.
The $f\I{=}1$ {\ESQPT} structures were reported in absence of the ground-state {\QPT} within the selected parametric form of the Hamiltonian.

Another interesting modification of the above approaches was discussed in Ref.\,\cite{Gessn16}. 
The model considered there is a spin model with {\em variable-range interaction}, which realizes an intermediate case between the Ising and Lipkin regimes of two-level lattice systems.
Assuming a ${D=1}$ spin chain with spatially regularly arranged sites enumerated consecutively by integer ${i=1,2...,N}$, one can consider the Hamiltonian of the following form
\begin{equation}
\hat{H}=\hat{J}_z-\lambda\sum_{i<i'}\frac{1}{|i\I{-}i'|^K}\,\hat{\sigma}^{(i)}_{x}\hat{\sigma}^{(i')}_{x}
\label{Hvaran},
\end{equation}
where $|i\I{-}i'|$ measures the distance between the sites $i$ and $i'$, and the power $K\I{\geq}0$ is an adjustable parameter which determines the range of the spin-spin interaction. 
For $K\I{=}0$ we have the familiar fully connected Lipkin model, c.f.\,Hamiltonian~\eqref{HLip0}, which in its high-$j$ subspaces shows precursors of the $f\I{=}1$ {\ESQPT}.
However, as $K$ increases from zero, the range of interactions decreases and $j$ ceases to be a good quantum number.
The Hamiltonian has to be diagonalized in the whole ${d=2^N}$ dimensional Hilbert space and its spectrum breaks the huge degeneracy caused by states in various replica subspaces $\Hi_{j}^{(s)}$. 
This means that all $f\I{=}N$ {\DoFs} become relevant and the {\ESQPT} precursors get washed out.
For $K\I{\gg}1$ the system approaches the limit of the nearest-neighbor spin-spin interactions of the Ising type.
The same model in the dynamical context, also related to {\ESQPTs}, was recently considered, e.g., in Ref.\,\cite{Leros20}.
Note that for studying the effects of local interactions in collective systems, instead of the Hamiltonian \eqref{Hvaran} one may use the Dicke Hamiltonian from Eqs.\,\eqref{quasis} and \eqref{HDic} with the single atom-field interaction strength $\lambda$ replaced by atom-dependent strengths $\lambda^{(i)}$.

\subsection{Interacting boson systems}
\label{Ibms}

The SU(2) algebra of quasispin operators can be represented in a bosonic language, either via the Holstein-Primakoff mapping
\begin{equation}
\hat{J}_0=\hat{b}^{\dag}\hat{b}\!-\!\frac{1}{2}N,\quad\hat{J}_+=\hat{b}^{\dag}(N\!-\!\hat{b}^{\dag}\hat{b})^{\frac{1}{2}},\quad\hat{J}_-=(N\!-\!\hat{b}^{\dag}\hat{b})^{\frac{1}{2}}\hat{b},
\label{HoP}
\end{equation}
which uses creation and annihilation operators $\hat{b}^{\dag}$ and $\hat{b}$ of a single-type scalar bosons, or via the Schwinger mapping
\begin{equation}
\hat{J}_0=\frac{1}{2}\left(\hat{t}^{\dag}\hat{t}-\hat{s}^{\dag}\hat{s}\right)\,,\quad\hat{J}_+=\hat{t}^{\dag}\hat{s}\,,\quad\hat{J}_-=\hat{s}^{\dag}\hat{t}\,,
\label{Schw}
\end{equation}
written in terms of creation and annihilation opeators $\hat{s}^{\dag},\hat{s}$ and $\hat{t}^{\dag},\hat{t}$ of two (scalar and pseudoscalar) bosons, see Ref.\,\cite{Klein91}.
In both cases, the representation realizes a subspace of $\Hi$ with a selected quasispin size value $j=\frac{1}{2}N$, where integer $N$ in mapping \eqref{HoP} is associated with a maximal number of $b$ bosons, while $N$ in mapping \eqref{Schw} represents the total number of $s$ and $t$ bosons.
Using number operators $\hat{N}_{\bullet}$  of the respective boson types, we link the states $\ket{j,m}$ with $m=-j,-j\!+\!1,...,+j$ to the $\hat{N}_b$ eigenstates with $N_b=0,1,...,N$, or to the $\hat{N}_s$ and $\hat{N}_t$ eigenstates with $(N_s,N_t)=(N,0),(N\!-\!1,1),...,(0,N)$.

Switching the size parameter $N\I{=}1,2,3,...$ into a dynamical quantity is equivalent to moving from the special unitary algebra SU(2) with three generators to the full unitary algebra U(2) with four generators.
For the Schwinger mapping \eqref{Schw}, the fourth generator is obviously the total boson number operator $\hat{N}=\hat{s}^{\dag}\hat{s}+\hat{t}^{\dag}\hat{t}$, which commutes with all the other generators and its eigenvalue can be fixed at any $N$.
For the Holstein-Primakoff mapping \eqref{HoP}, the additional commuting generator that carries a chosen value $N$ can be associated with the number operator $\hat{N}_a\I{=}\hat{a}^{\dag}\hat{a}$ of a new boson~$a$.
The full unitary algebra forms the {\em dynamical algebra\/} of the bosonic system in the sense that any observable, including the Hamiltonian, can be written exclusively in terms of the generators of this algebra.

With the aid of the above mappings, any quasispin Hamiltonian \eqref{Hsu2} can be cast in the form of a Hamiltonian describing an interacting system of bosons.
The Hamiltonians resulting from the Holstein-Primakoff mapping with substitution $N\mapsto\hat{a}^{\dag}\hat{a}$ in general change the total number of $a$ and $b$ bosons (conserving $\hat{N}_a$ but changing $\hat{N}_b$) and contain unlimited $k$-body terms of boson operators resulting from an expansion of the square-root factors.  
On the other hand, the Hamiltonians obtained from the Schwinger mapping \eqref{Schw} conserve the total number of $s$ and $t$ bosons and, if we restrict to linear plus quadratic quasispin Hamiltonians, they contain only one- plus two-body terms.
The latter approach opens a path to direct generalizations to systems with more than two boson types, or more precisely to systems with the single-boson Hilbert space of a general dimension ${n=2,3,4,...}$, which are associated with unitary dynamical algebras U($n$).
The general one- plus two-body bosonic Hamiltonian reads
\begin{equation}
\hat{H}=\sum_{k=1}^{n}\varepsilon_{k}\ \hat{b}^{\dag}_k\hat{b}_k+
\!\!\!\sum_{k,l,k',l'=0}^{n}\!\!\!\!
\frac{{\nu}_{klk'l'}}{N}\ \hat{b}^{\dag}_k\hat{b}^{\dag}_{k'}\hat{b}_l\hat{b}_{l'}
\label{Hibm},
\end{equation}
where operators $\hat{b}_k^{\dag}$ and $\hat{b}_k$ create and annihilate a boson in the $k$th state ($k\I{=}n$, e.g., denoting the scalar boson $s$) and $\varepsilon_{k}$ and $\nu_{klk'l'}$, respectively, represent single-particle energies and two-body interaction parameters.
The $N$ denominator in the interaction term, in analogy to Eqs.\,\eqref{HLip0} and \eqref{Hsu2}, prevents dominance of this term in large-$N$ cases. 
Three- and more-body interactions, if needed, can be added in an obvious way.

The nomenclature of these models is as follows:
The cases with ${n=3}$ and $4$ correspond to two kinds of {\em vibron models\/} used in molecular physics.
The ${n=3}$ vibron model \cite{Iache96} works with a one-component boson $s$ and a two-component boson $(\tau_+,\tau_-)$ and is applied to describe the bending vibrational mode of polyatomic molecules (including the three-atomic ones, such as H$_2$O).
The ${n=4}$ vibron model \cite{Iache95} makes use of the scalar boson $s$ and a vector boson $(p_{-1},p_0,p_{+1})$ and simulates vibrations and rotations of two-atomic molecules.
The case with ${n=5}$ would have a four-component boson $(\pi_{--},\pi_-,\pi_+,\pi_{++})$ in addition to the boson $s$, but this combination is not used in any specific model at present.
The case with ${n=6}$ is the {\em interacting boson model\/} of nuclear physics \cite{Iache87}, which is based on the scalar boson $s$ and a quadrupole tensor boson $(d_{-2},d_{-1},d_0,d_{+1},d_{+2})$.
The $s$ boson represents a pair of nucleons with angular momentum 0, while the $d$ boson is interpreted as a pair with angular momentum 2 and simultaneously related to the quadrupole collective {\DoFs} of nuclei.
More sophisticated nuclear models are obtained by adding to the $s+d$ system some other types of bosons, namely $p$, $d$ and/or $f$ with angular momenta 1, 3 and 4, respectively, or by separating boson species associated with the proton and neutron {\DoFs} \cite{Iache87}.
The latter extension to interacting boson models with two or more \uvo{fluids} exploits the standard coupling formalism mentioned above in the context of atom-field and quasispin models.
It is also used in the framework of molecular vibron models \cite{Iache08}.

A considerable advantage of the the finite-$n$ interacting boson models is the fact that their Hilbert space $\Hi$ has a finite dimension which grows only polynomially (${\propto N^{n-1}}$) with the natural size parameter $N$. 
This greatly facilitates the exact numerical solution.
Another positive feature is a straightforward construction of the classical limit $N\I{\to}\infty$.
The coordinate-momentum representation of the interacting boson Hamiltonians \eqref{Hibm} is obtained with the aid of Eq.\,\eqref{QP} with ${\eta=N}$, which yields a system with $f\I{=}n$\, {\DoFs}.
Moreover, fixing the conserved total number of bosons on a selected value $N$ allows to perform a canonical transformation which reduces the number of active {\DoFs} to ${f=n-1}$.
The transformation represents a generalization of the link between Schwinger and Holstein-Primakoff mappings described above for the ${n=2}$ case.
It is formally equivalent to the transformation \eqref{QP} performed for all components of all bosons except the boson $s$ (so we obtain operators $\hat{q}_i$ and $\hat{p}_i$ with ${i=1,...,n\I{-}1}$), and to replacing both $\hat{s}^{\dag}$ and $\hat{s}$ operators by ${[N-\frac{1}{2}\sum_{i}(\hat{q}_i^2+\hat{p}_i^2)]^{1/2}}$.  
For details see Refs.\,\cite{Blaiz78} and \cite{Macek19}. 

An important role in the formulation of the inteacting boson models is played by {\em dynamical symmetries} \cite{Iache14}.
These generalizations of invariant symmetries are associated with specific decompositions of the dynamical algebra $\Gr_{\rm D}\equiv{\rm U}(n)$ into chains of subalgebras ${\Gr_{\rm D}\supset\Gr^{(k)}_{1}\supset\Gr^{(k)}_{2}\supset\dots\supset\Gr_{\rm S}}$.
The highest algebra $\Gr_{\rm D}$ as well as the lowest algebra $\Gr_{\rm S}$, which generates the invariant symmetry of the system, are fixed and can be usually connected by several chains, here enumerated by superscript~$k$.
The Hamiltonian written solely in terms of Casimir invariants of the algebras involved in one of such chains possesses the corresponding dynamical symmetry and can be proven to be integrable \cite{Zhang95}.
We note that all invariants within a single chain represent mutually commuting integrals of motion.

It is customary to study transitional Hamiltonians of the general form
\begin{equation}
\hat{H}(\lambda)=(1\I{-}\lambda)\,\hat{H}(0)+\lambda\,\hat{H}(1),\quad \lambda\in[0,1],
\label{Hds}
\end{equation}
where the single variable $\lambda$ drives the system along a path in the multiparameter space of the model between two distinct dynamical symmetries associated with Hamiltonians $\hat{H}(0)$ and $\hat{H}(1)$.
The {\QPT} that eventually appears along the selected path is then linked to phases defined through these limiting dynamical symmetries.
For example, the U(6) interacting boson model has three dynamical symmetries denoted according to the highest subgroup in the respective chain as U(5), SO(6) and SU(3) \cite{Iache87}.
In most studies, the space of all model control parameters is reduced by considering special paths that realize transitions U(5)--SU(3) (with first-order {\QPT}),  U(5)--SO(6) (with second-order {\QPT}), and SO(6)--SU(3) (non-critical crossover). 
The algebraic formulation of collective quantum systems will be further discussed in Sec.\,\ref{Algeb}.

The nuclear ${n=6}$ interacting boson model in a certain subset of its parameter space, namely along the path between the U(5) and SO(6) dynamical symmetries, was instrumental in the early studies of {\ESQPTs} \cite{Cejna06,Cejna07,Heinz06,Macek06,Cejna09}.
The model in this regime [i.e., for Hamiltonians with vanishing contribution of the SU(3) Casimir invariant] is a special case of a more general class of many-body systems with {\em pairing interaction}, which are integrable and solvable by algebraic methods \cite{Dukel04a}.
This is related to imperfect breaking of the limiting U(5) and SO(6) dynamical symmetries since the Casimir invariant of the SO(5) subalgebra (belonging to both limiting dynamical symmetry chains) remains an integral of motion along the whole path.
For the ground state, which has zero angular momentum associated with the SO(3) invariant symmetry subalgebra, the SO(5) symmetry is spontaneously broken at the second-order {\QPT} from the U(5) to SO(6) phase.
For excited states with zero SO(3) angular momentum, the SO(5) symmetry is broken along an {\ESQPT} critical borderline, which starts at the ground-state {\QPT} critical point and propagates towards the SO(6) limit, where it reaches the high-energy end of the spectrum.
The quantum phases separated by the critical borderline have been characterized by the U(5) and SO(6) quasi dynamical symmetries,  the word \uvo{quasi} emphasizing an approximate character of these symmetries between the limiting points of the path \cite{Rowe04,Macek10,Macek14}.
While the full ${n=6}$ model with fixed $N$ has $f\I{=}5$ {\DoFs} (three of them associated with rotations and two with quadrupole vibrations of nuclei), the subset of states with zero values of the SO(3) and SO(5) invariants corresponds to ${f_{\rm eff}=1}$ effective {\DoF}.
Hence, in the spectra of the corresponding states we observe generic {\ESQPT} singularities in the zeroth derivative of the smoothed level density.

\begin{figure}[t!]
\begin{flushright}
\includegraphics[width=\textwidth]{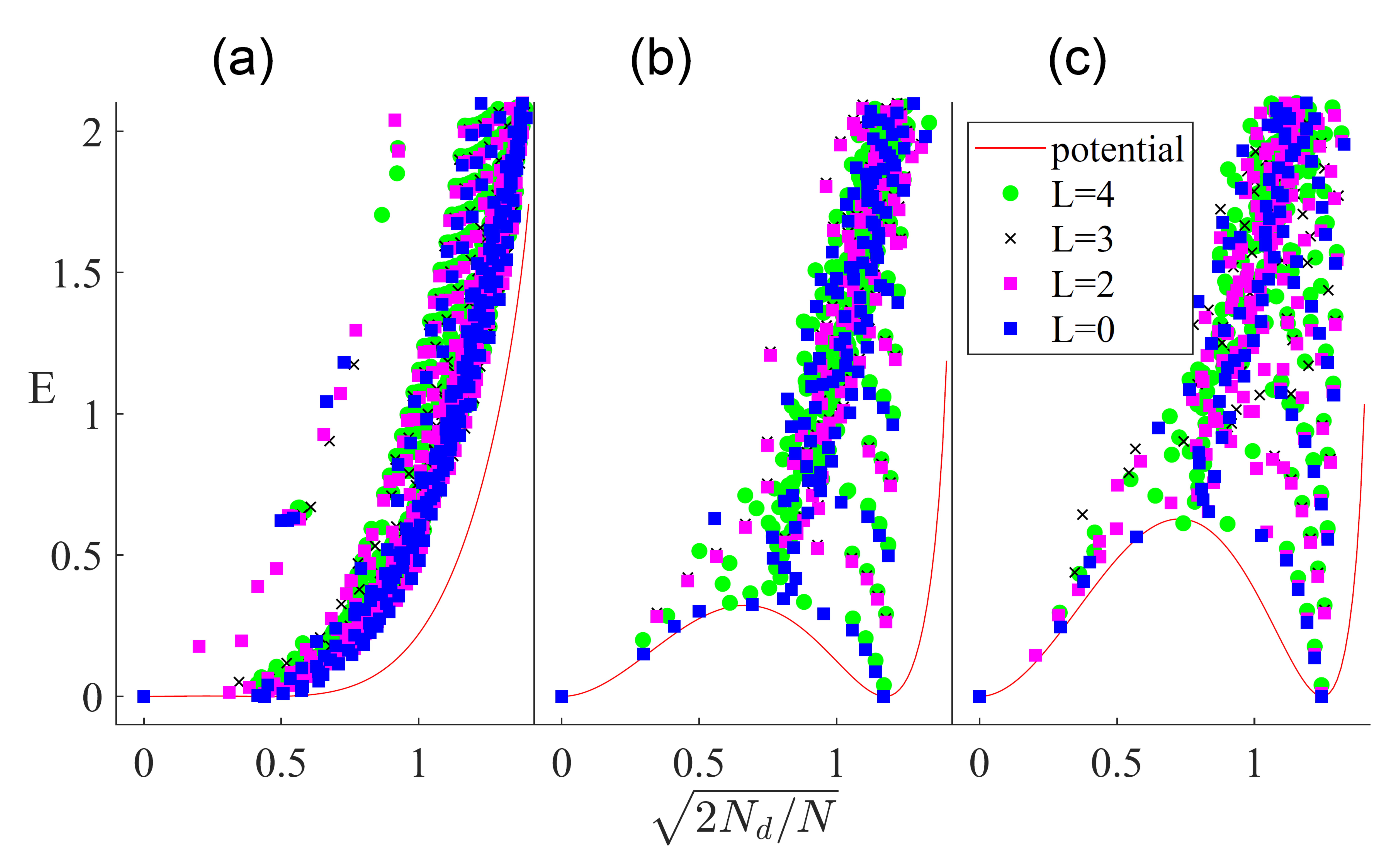}
\end{flushright}
\caption{Peres lattices of $\sqrt{2\ave{\hat{N}_d}_i/N}$ (which is related to a generalized coordinate of the system) in the ${n=6}$ interacting boson model at the critical point of the first-order {\QPT} between SU(3) and U(5) ground-state phases. States with angular momenta from ${L=0}$ to 4 are included. The total number of bosons ${N=50}$. The Hamiltonian and further details are described in Ref.\,\cite{Macek14}. Three panels correspond to different parameter choices leading to different heights of the phase-separating potential barrier  (the respective \lq\lq potential energy\rq\rq\ curves are drawn in each panel). The top of the barrier [invisibly small in panel (a)] is connected with an {\ESQPT}, which affects the expectation values. The patterns in panels (b) and (c) can be compared to analogous but sharper case of the $n\I{=}2$ Lipkin model in Fig.\,\ref{liporder}(c) below. The repetition of $\cap$-like structures observed here is due to partial separability of the system (cf.\,Sec.\,\ref{Fisie} and Ref.\,\cite{Macek19}).}
\label{ibm1st}
\end{figure}

The {\ESQPTs} for Hamiltonians of the $n\I{=}6$ interacting boson model, which are not of the pairing form and contain the SU(3) Casimir invariant, have been analyzed in Refs.\,\cite{Zhang16,Zhang17,Macek19}.
For subsets of states with zero SO(3) angular momentum the system has ${f_{\rm eff}=2}$ effective {\DoFs} (connected with quadrupole vibrations) and the {\ESQPT} signatures affect the first derivative of the smoothed level density.
Peres lattices at the first-order ground-state {\QPT} critical point of the U(5)--SU(3) path are illustrated in Fig.\,\ref{ibm1st} for various parameter settings.
Note that the quantity $\propto\ave{\hat{N}_d}^{1/2}_i$ displayed in the lattice is related to the expectation value of a generalized coordinate of the system.
Both minima of the \uvo{potential energy} curve (which, as seen in the figure, are degenerate at the critical point) are associated with one of the competing ground-state phases \cite{Macek14}.
The three panels show lattices for different heights of the phase-separating barrier. 
The {\ESQPT} at energy coinciding with the top of the barrier is seen as the merge of two branches of eigenstates localized in both minima into a single branch with $\ave{\hat{N}_d}^{1/2}_i$ corresponding to the barrier maximum. 

We point out that although energy spectra of the $n\I{=}6$ interacting boson model are commonly used to describe experimental data on rotational and vibrational states in even-even nuclei, including spectral effects of the transition from the spherical to deformed nuclear shape \cite{Cejna10}, the verification of the model's {\ESQPTs} in real spectra is hindered by the notorious complexity of the nuclear many-body system, whose higher excitations resist any simplified description in terms of a reduced set of {\DoFs}. 

\begin{figure}[t!]
\begin{flushright}
\includegraphics[width=0.85\textwidth]{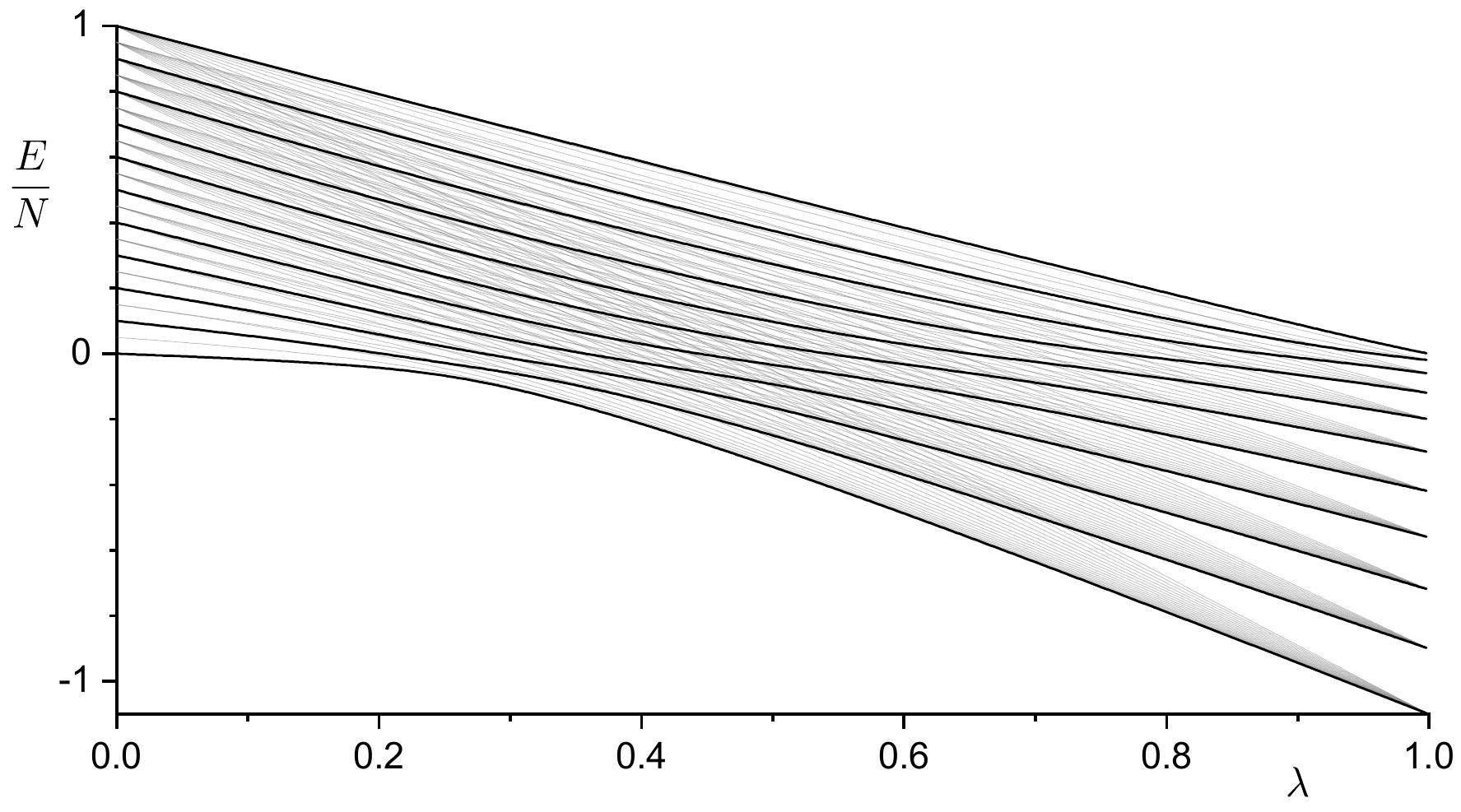}
\end{flushright}
\caption{The {\ESQPT} in the spectrum of the ${n=4}$ vibron model for ${N=20}$. 
The Hamiltonian has the form \eqref{Hds} with $\hat{H}(0)=\hat{C}_1[{\rm U(3)}]$ and $\hat{H}(1)=N^{-1}\hat{C}_2[{\rm SO(4)}]$, where $\hat{C}_1$ and $\hat{C}_2$ are the first- and second-order Casimir invariants of the respective algebras, see Ref.\,\cite{Iache95}. 
The spectrum contains states with all values of the SO(3) quantum number $l=0,1,...,N$ (ordinary 3D angular momentum).
Each level is ${(2l+1)}$ times degenerate in the angular-momentum projection $m$.
The levels with $l=0$ are drawn by thicker curves. 
This subspace corresponds to ${f_{\rm eff}=1}$, while the unconstrained model with all states has ${f=3}$. 
We observe the second-order ground-state {\QPT} at ${\lambda_{\rm c}=\frac{1}{5}}$ continued by an {\ESQPT} at $E_{\rm c}=0$ (a peak in the level density of ${l=0}$ states and an effect in the second derivative of the level density of all states). 
Spontaneous breaking of the SO(3) symmetry appears in the quantum phase with lower energy. 
Practically the same spectrum is obtained for an analogous Hamiltonian of the $n=3$ vibron model with ${f=2}$, where the SO(2) angular momentum corresponds to 2D rotations.}
\label{3Dvibron}
\end{figure}

The phase structure of $n\I{<}6$ interacting boson models is similar to that of the $n\I{=}6$ model in the U(5)--SO(6) pairing regime \cite{Cejna07b}.
These models are also analyzed along transitional paths between dynamical symmetries U(${n\I{-}1}$) and SO($n$).
The symmetry preserved along the whole path, which is spontaneously broken for the ground state at the second-order {\QPT}, is connected with the SO(${n\I{-}1}$) invariant.
For the $n\I{=}3$ and $n\I{=}4$ vibron models, the invariant expresses conserved 2D or 3D angular momentum, for the $n\I{=}2$ boson-based Lipkin model the conserved quantity is the parity $\hat{\Pi}=(-1)^{\hat{N}_t}$ associated with the number of $t$ bosons (cf.\,Sec.\,\ref{Quasi}). 
The spectra of all these models show {\ESQPTs} between quantum phases characterized by U(${n\I{-}1}$) and SO($n$) quasi dynamical symmetries.
As in the previous case,  restriction to states with vanishing SO(${n\I{-}1}$) invariant leads to $f_{\rm eff}\I{=}1$, so the {\ESQPTs} again affect the zeroth derivative of smoothed level density in these subsets of states.  
Besides the $n\I{=}2$ Lipkin model, which was already discussed in Sec.\,\ref{Quasi}, the theoretical studies of {\ESQPTs} were focused mostly on the $n\I{=}3$ model \cite{Capri08,Berna08,Berna10,Lares11,Lares13}. 
The results for $n\I{=}4$ are similar, an example of the spectrum being shown in Fig.\,\ref{3Dvibron}.
We stress that static signatures of {\ESQPT} in the $n\I{=}3$ model have been verified experimentally in the spectra of various molecules, see Refs.\,\cite{Winne05,Zobov05,Lares11,Lares13,Khalo20}.

Similar analyses apply to a more general class of bosonic pairing models with the dynamical algebra U($n_1\I{+}n_2$) encompassing two kinds of bosons with $n_1$ and $n_2$ components \cite{Capri08,Capri11}.
The Hamiltonians connect dynamical symmetries SO($n_1\I{+}n_2$) and ${\rm U}(n_1)_1\oplus{\rm U}(n_2)_2$,  called after the largest subalgebras of the corresponding chain, and conserve the Casimir invariant of the ${\rm SO}(n_1)_1\oplus{\rm SO}(n_2)_2$ subalgebra. 
This yields practically the same {\QPT} and {\ESQPT} effects as the above-discussed special cases with $n_1\I{=}1$, see Refs.\,\cite{Capri08,Capri11} and paper 27 in \cite{Carr10}.
The subscripts 1 and 2 distinguish algebras associated with both bosonic species.
Note that here and below we use a direct sum for algebras, implying a direct product for the corresponding groups. 

Another use of the interacting boson models with a limited dimension of the single-particle space is in the physics of Bose-Einstein condensation \cite{Pitae03,Angli02,Gardi14}, namely in the framework of the well known {\em Bose-Hubbard model\/} \cite{Fishe89}. 
The model in its general form describes a system of ultracold interacting Bose atoms trapped in a lattice of $N$ sites (see also Sec.\,\ref{Spati}).
Here we mention explicitly a two-site system of an atomic condensate in a double-well potential, which is applied as a model of the Josephson junction \cite{Milbu97}.
This and analogous systems are discussed in the literature particularly in connection with their capability to create macroscopic quantum superpositions and their potential use in quantum computation \cite{Cirac98}.
Experimental realization of such regimes in a two-site system was reported in Ref.\,\cite{Albie05}.
The bosons associated with both sites marked by signs $+$ and $-$ are created and annihilated by operators $\hat{b}^{\dag}_{\pm}$ and $\hat{b}_{\pm}$, and the Hamiltonian is assumed to have the following form:
\begin{equation}
\hspace{-18mm}
\hat{H}=\varepsilon_+\hat{b}^{\dag}_+\hat{b}_+\!+\!\varepsilon_-\hat{b}^{\dag}_-\hat{b}_-
-\tau \left(\hat{b}^{\dag}_+\hat{b}_-\!+\!\hat{b}^{\dag}_-\hat{b}_+\right)
+\frac{U}{N}\left(\hat{b}^{\dag}_+\hat{b}^{\dag}_+\hat{b}_+\hat{b}_+\!+\!\hat{b}^{\dag}_-\hat{b}^{\dag}_-\hat{b}_-\hat{b}_-\right).
\label{HBH}
\end{equation}
Here, $\varepsilon_{\pm}$ are the site energies, $\tau\I{\geq}0$ is a strength of inter-site hopping, and $U$ quantifies the intra-site interaction (repulsive or attractive for $U\I{>}0$ or $<\!0$, respectively).

The system with attractive intra-site interactions exhibits a {\QPT} between two distinct ground-state phases: the conductive (superfluid) phase with delocalized atoms (observed for $|U|\I{\ll}\tau$) and the trapped (Mott insulator) phase with atoms localized at individual sites (for $|U|\I{\gg}\tau$).
In the symmetric case with $\varepsilon_+=\varepsilon_-$, the transition is of the second order and corresponds to a spontaneous breaking of the ground-state symmetry expressed by a conserved parity operator exchanging populations $N_+$ and $N_-$ of both sites.
Similar localization and symmetry breaking effects can be observed for excited states of the two-site system with repulsive intra-site interactions \cite{Julia10}.
The spectrum of the model \eqref{HBH} with both signs of $U$ contains {\ESQPTs}, which can be described by the same semiclassical techniques as those for the more general Hamiltonians \eqref{Hibm}.
Since the two-site system can be reduced to $f\I{=}1$, the {\ESQPT} singularity appears directly in the zeroth derivative of the smoothed level density \cite{Bychek18}.

A formally similar model with the hopping term involving pairs of atoms instead of single ones has been applied to intraband tunneling processes of the Bose-Einstein condensate in periodic 2D optical lattices \cite{Shche09}.
And yet another two-site boson model with Hamiltonian of the Lieb-Lingier type was recently analyzed in Ref.\,\cite{Nitsc20}.
Both these models also exhibit clear $f\I{=}1$ {\ESQPT} signatures.

Finally, a model with three interacting boson species $b_{\pm}$ and $b_0$ is applied in a~simplified description of the spin-1 (\uvo{spinor}) Bose-Einstein condensate \cite{Kawag12}.
Instead of the ${b_+\leftrightarrow b_-}$ hopping, the Hamiltonian of this model contains interactions that induce the ${b_0b_0\leftrightarrow b_+b_-}$ changes, so the difference of boson numbers ${\hat{D}=\hat{N}_+\I{-}\hat{N}_-}$ is conserved in addition to the total number of bosons ${\hat{N}=\hat{N}_+\I{+}\hat{N}_-\I{+}\hat{N}_0}$. 
The {\ESQPTs} in this model, which because of the conservation of $\hat{D}$ can be characterized by the effective {\DoF} number  ${f_{\rm eff}=1}$, were recently studied in Ref.\,\cite{Feldm20}. 
The spinor condensates represent another promising tool for experimental studies of {\ESQPTs}.

\subsection{General algebraic systems}
\label{Algeb}

The above considerations can be extended to any algebraically formulated model.
The algebraic method \cite{Iache14} anchors the description of any specific system of interest to the definition of a suitable dynamical algebra $\Gr\equiv\{\hat{G}_l\}_{l=1}^{L}$, which is associated with an $L$-component Lie group of unitary transformations that embrace arbitrary dynamics of the system.
This algebra consists of a set of Hermitian generators $\hat{G}_l$ satisfying the closure relation $[\hat{G}_k,\hat{G}_l]=\ii\sum_m s_{klm}\hat{G}_m$ with real structure constants $s_{klm}$.
Various combinations of these generators must allow one to express any relevant physical observable of the system.
In particular the Hamiltonian can be written as $\hat{H}=H(\{\hat{G}_l\})$, where $H$ on the right-hand size denotes an unspecified function of arguments $\hat{G}_l$.
We assume that the number of generators $L$ is finite.
The number $R\leq L$ of mutually commuting generators, called the rank of algebra $\Gr$, is related to the number of quantum numbers needed to unambiguously label basis states in the associated Hilbert space. 
So in typical situations, this number in the classical limit coincides with the number $f$ of {\DoFs} of the system.

We have already seen examples of dynamical algebras for the systems discussed above.
It was the SU(2) algebra with generators $(\hat{J}_+,\hat{J}_-,\hat{J}_0)$ for the quasispin systems, and U($n$) algebras for the interacting boson systems with $n$ bosonic species. 
The latter algebras are generated by products of creation and annihilation operators $\hat{b}^{\dag}_k\hat{b}_l$ and hence conserve the total boson number $\hat{N}\I{=}\sum_{k=1}^{n}\hat{N}_k$.
Coupled systems are described by direct sums of the corresponding algebras.
In particular, we mention dynamical groups of the type ${\rm SU}(2)_1\oplus{\rm SU}(2)_2$ describing coupled quasispin systems, or ${\rm U}(n_1)_1\oplus{\rm U}(n_2)_2$ applied to \uvo{two-fluid} interacting boson systems with $n_1$- and $n_2$-component bosons conserving separately both boson numbers $\hat{N}_1$ and $\hat{N}_2$.
For coupled atom-field systems from Sec.\,\ref{Open}, we use the dynamical algebra ${\rm SU}(2)_{\rm A}\oplus{\rm HW}(1)_{\rm B}$, where ${\rm SU}(2)_{\rm A}$ stands for the standard quasispin algebra of the atomic subsystem and ${\rm HW}(1)_{\rm B}\equiv\{\hat{b}^{\dag},\hat{b},\hat{I}\}$ (where $\hat{I}$ is the unit operator) for the Heisenberg-Weyl algebra of the field subsystem.

With the aid of the algebraic approach, the class of the interacting boson models can be elegantly generalized to describe combined {\em atom-molecule systems} which enable mutual conversions of bosonic atoms and bosonic molecules built from two or more atoms, or interchanges of bosonic molecules composed of different numbers of atoms. 
This kind of boson models can be used in the description of Bose-Einstein condensation involving molecules \cite{Zolle02} and similar models are also applied in quantum optics \cite{Karas94}.
Such systems do not conserve the number $\hat{N}$ of bosons (atoms and molecules), but rather the number of elementary constituents $\hat{\mathcal{N}}=\sum_{k=1}^{n}m_k\hat{N}_k$, where $\hat{N}_k$ is the number of bosons of the $k$th type comprising of $m_k$ atoms. 
Therefore they can no longer be associated with the $\hat{N}$-conserving dynamical algebras U($n$).

As shown in Refs.\,\cite{Peres11a,Dukel04}, the simplest case with $n\I{=}2$ bosonic species characterized by atomic numbers $(m_1,m_2)\I{=}(1,2)$ can be associated with the dynamical algebra ${\rm SU}(1,1)_1\oplus{\rm HW}(1)_2$, where the non-compact algebra ${\rm SU}(1,1)_1$, generated by operators
\begin{equation}
\hat{I}_0=\frac{1}{2}\left(\hat{b}_1^{\dag}\hat{b}_1\I{+}\frac{1}{2}\right),\quad
\hat{I}_+=\frac{1}{2}\hat{b}_1^{\dag}\hat{b}_1^{\dag},\quad 
\hat{I}_-=\frac{1}{2}\hat{b}_1\hat{b}_1
\end{equation}
[obeying the same commutation relations as the SU(2) generators except the opposite sign of the commutator between the rising and lowering generators], describes the first boson (atom) and its pairwise creation and annihilation, while the Heisenberg-Weyl algebra ${\rm HW}(1)_2$ is built of the creation and annihilation operators of the second boson (molecule).
An alternative \cite{Graef15,Lee10,Karas94} is to use for the whole system a single polynomially deformed algebra SU(2)$_{\rm pd}$ expressed by generators
\begin{equation}
\hspace{-10mm}
\hat{K}_0=\frac{1}{4}\left(\hat{b}_1^{\dag}\hat{b}_1\!-\!2\hat{b}_2^{\dag}\hat{b}_2\right),\quad
\hat{K}_+=\mathcal{N}^{-\frac{1}{2}}\hat{b}_1^{\dag}\hat{b}_1^{\dag}\hat{b}_2,\quad
\hat{K}_-=\mathcal{N}^{-\frac{1}{2}}\hat{b}_2^{\dag}\hat{b}_1\hat{b}_1
\end{equation}
(with $\mathcal{N}$ standing for the selected value of the conserved number operator $\hat{\mathcal{N}}$), whose commutators involving $\hat{K}_0$ coincide with standard SU(2) ones, but $[\hat{K}_+,\hat{K}_-]$ results in a polynomial expression in $\hat{K}_0$ and $\hat{\mathcal{N}}$.
In the models based on both these approaches, $f\I{=}1$ {\ESQPT} structures were clearly identified in quantum spectra \cite{Peres11a,Graef15}.
The latter approach can be generalized to systems with arbitrary values of $n$ and $\{m_i\}$, again with {\ESQPT} effects potentially present \cite{Graef16}.
In particular, {\ESQPTs} in the $n\I{=}3$, $(m_1,m_2,m_3)=(1,1,2)$ extension of the two-site Bose-Hubbard Hamiltonian of the type \eqref{HBH} were analyzed in Ref.\,\cite{Relan14}.

The algebraic formulation can be naturally extended also to {\em interacting fermion systems\/} formulated in terms of algebras of fermionic operators.
Let us note that fermionic systems, due to the anticommutation relations of fermion creation and annihilation operators, are generally described by superalgebras rather than algebras, but if relevant dynamical operators involve only the even sector of the fermionic superalgebra, the standard algebraic description can be applied.
Giving just the most trivial examples, we point out that the original formulation of the Lipkin model \cite{Lipki65} was through the fermionic realization of the SU(2) algebra 
\begin{equation}
\hspace{-15mm}
\hat{J}_0=\frac{1}{2}\sum_{i=1}^N\left(\hat{a}_{i+}^{\dag}\hat{a}_{i+}\!-\!\hat{a}_{i-}^{\dag}\hat{a}_{i-}\right),\quad
\hat{J}_+=\sum_{i=1}^N\hat{a}_{i+}^{\dag}\hat{a}_{i-},\quad
\hat{J}_-=\sum_{i=1}^N\hat{a}_{i-}^{\dag}\hat{a}_{i+},
\label{Lipfer} 
\end{equation}
where operators $\hat{a}_{i\pm}^{\dag}$ and $\hat{a}_{i\pm}$ create and annihilate fermions in states $i=1,2,...,N$ on two $N$-fold degenerate levels distinguished by symbols $+$ and $-$.
To obtain the familiar $2^N$ dimensional Hilbert space known from the qubit realization of the quasispin algebra \eqref{quasis}, one needs to set the total conserved number of fermions $\hat{N}_a=\sum_{s=\pm}\sum_{i=1}^N\hat{a}_{is}^{\dag}\hat{a}_{is}$ to the value that coincides with the capacity $N$ of a single $\pm$ level, and allow just a single particle to be present in each couple of states $i+$ and $i-$.

Another fermionic incarnation of the SU(2) algebra appears in a schematic pairing model. 
Here the operators $\hat{a}_{i\pm}^{\dag}$ and $\hat{a}_{i\pm}$ describe fermions on a single level, with states $i+$ and $i-$, being conjugate with respect to time reversal (e.g., single-particle states with positive and negative projections of the total angular momentum).
Introducing a fermion-pair creation operator $\hat{A}^{\dag}=\sum_{i=1}^N(-1)^{N-i}\hat{a}_{i+}^{\dag}\hat{a}_{i-}^{\dag}$, and the corresponding Hermitian conjugate annihilation operator $\hat{A}$, we have
\begin{equation}
\hat{J}_0=\frac{1}{2}\left(\hat{N}_a\I{-}N\right),\quad
\hat{J}_+=\hat{A}^{\dag},\quad
\hat{J}_-=\hat{A}.
\label{bcsfer} 
\end{equation}
These operators do not conserve the particle number and to obtain the Hilbert space $\Hi$ of dimension $2^N$, we assume a basis in which each pair of states $i+$ and $i-$ is either empty or occupied.
The quasispin size quantum number is parametrized as $j=j_{\rm max}-v$, where the quantum number $v$ is called seniority.

The {\ESQPTs} were explicitly studied in a certain generalization of the above SU(2) fermionic pairing system, namely in a two-level  pairing model, see Refs.\,\cite{Capri08,Capri11} and paper 27 in \cite{Carr10}.
The two levels are supposed to have total (orbital plus spin) half-integer angular momenta $\ell_1$ and $\ell_2$, and the respective fermion pair creation/annihilation operators are $\hat{A}_1^{\dag},\hat{A}_1$ and $\hat{A}_2^{\dag},\hat{A}_2$.
The Hamiltonian is written as
\begin{equation}
\hat{H}=\sum_{k=1,2}\varepsilon_k\hat{N}_k+\sum_{k,l=1,2}\nu_{kl}\hat{A}_k^{\dag}\hat{A}_l,
\label{Hpair}
\end{equation}
where $\varepsilon_k$ is the single-particle energy of the $k$th level, $\hat{N}_k$ the respective occupation number operator, and $\nu_{kl}$ expresses pairing interaction strengths.
The dynamical algebra of the system is identified with ${\rm U}(n_1\!+\!n_2)$, where $n_k=2\ell_k\!+\!1$.
It is composed of all combinations of fermion creation and annihilation operators conserving the total particle number.
Hamiltonians of the form \eqref{Hpair} interpolate between dynamical symmetries corresponding to the dynamical algebra decompositions starting with the algebra ${\rm U}(n_1)\oplus{\rm U}(n_2)$, on one side of the transition, and with the symplectic algebra ${\rm Sp}(n_1\!+\!n_2)$, on the other side.
The situation is similar to the case of bosonic pairing systems discussed in Sec.\,\ref{Ibms}.
The system exhibits a {\QPT} between both dynamical symmetries (independent-particle and paired forms of the ground state) and an {\ESQPT} on the paired side of the transition.
Due to the residual dynamical symmetry, which remains unbroken across the whole transitional path, and the consequent integrability of the system, the effective {\DoF} number for suitable subsets of states is reduced so that the {\ESQPT} spectral singularity is seen in the zeroth derivative of the associated level density.
In contrast to the bosonic case, the Pauli exclusion principle enables one to reach the infinite size limit $N=N_1\!+\!N_2\to\infty$ only if the angular momentum quantum number $\ell_1$ or $\ell_2$ becomes infinite.
The numerically observed finite-size precursors of both {\QPT} and {\ESQPT} effects are nevertheless very clear \cite{Capri11}.

In this overview of {\ESQPT}-related models and systems, we have to mention mixed interacting many-body systems composed of bosons and fermions. 
In nuclear physics, such {\em Bose-Fermi systems\/} naturally appear in the framework of the interacting boson-fermion model applied to nuclei with odd numbers of protons or neutrons \cite{Iache91}.
This model separately conserves the numbers of bosons and fermions, and its dynamical algebra can be associated with ${\rm U}(6)_{\pi}\oplus{\rm U}(6)_{\nu}\oplus{\rm U}(2k)_{\rm F}$, where the indices distinguish unitary algebras of proton ($\pi$) and neutron ($\nu$) bosons (paired fermions of the respective type), and that of the odd-fermion (F), proton or neutron, on the valence shell with total capacity $2k$.
In the analyses of the phase structure,  the bosonic subsystem is sent to the asymptotic size regime with the number of bosons $N_{\pi}\!+\!N_{\nu}\to\infty$, and the coupling with the single-fermion subsystem (whose quantum character is preserved) results in multiple overlapping semiclassical spectra with modified quantum critical properties \cite{Petre11}.

Similar Bose-Fermi systems exist also in condensed matter physics. 
The most challenging are the systems that allow for mutual conversions between boson-like particles and pairs of fermions.
This happens for instance in the so-called bilayer model \cite{Morei08,Figue10}, which describes a {\QPT} from a fermionic phase of unbound electron-hole pairs to a bosonic phase of electron-hole bound states. 
Experimental signatures of such transitions were reported in Ref.\,\cite{Eisen04}.
Introducing the fermion (particle or hole) creation and annihilation operators for states enumerated by $i=1,2,...,\frac{1}{2}M$ on two fermionic \uvo{layers} denoted as $+$ and $-$, one can write down a fermion pair creation operator $\hat{A}^{\dag}=\sum_{i=1}^{N/2}\hat{a}_{i-}^{\dag}\hat{a}_{i+}^{\dag}$.
This pair couples to a bosonic excitation quantum created by $\hat{b}^{\dag}$.
Employing the constraint on the conserved quantity $\hat{\mathcal{N}}=\hat{N}_a+2\hat{N}_b$ (where $\hat{N}_a$ and $\hat{N}_b$ are fermion and boson total number operators, respectively) fixed at the overall capacity of the fermion state space $M$, one can identify the operators
\begin{equation}
\hspace{-10mm}
\hat{J}_0=\frac{1}{2}\hat{N}_b-\frac{1}{4}\hat{N}_a,\quad 
\hat{J}_+=\hat{N}_b^{-\frac{1}{2}}\,\hat{b}^{\dag}\hat{A},\quad
\hat{J}_-=\hat{A}^{\dag}\,(\hat{N}_b\!+\!1)^{-\frac{1}{2}}\hat{b}
\end{equation}
with generators of the familiar SU(2) algebra.
Spectral signatures of the $f=1$ {\ESQPT} for a Hamiltonian written in terms of these generators were described in Refs.\,\cite{Morei08,Figue10}. 

\subsection{Size parameter and classical limit}
\label{Size}

Having viewed (from bird's perspective, of course) in the preceding paragraphs a large variety of interacting many-body systems based on the algebraic description, let us finally address the general relation between the infinite-size and classical limits of these systems.
We have already highlighted the equivalence of these limits for the specific systems above.
The same conclusion holds, in principle, for any quantum system described through a finite-dimensional dynamical Lie algebra (and can be extended to systems with deformed dynamical algebras).

The definition of the {\em size parameter} of the system is rather obvious in many cases.
For instance, it coincides with the number $N$ of qubit sites (or the quasispin size $j$) in qubit systems based on the quasispin formalism (Sec.\,\ref{Quasi}). 
It represents the conserved total number $N$ of bosons or fermions for systems based on the particle-number conserving unitary algebras (Secs.\,\ref{Ibms} and \ref{Algeb}).
Or it can be a combination of the type $\mathcal{N}=\sum_km_kN_k$ in atom-molecule or boson-bifermion systems (Sec.\,\ref{Algeb}).
On the other hand, some systems require a more careful analysis, like e.g. the asymmetric Rabi regime of the atom-field system (Sec.\,\ref{EDM3}).

The choice of a general size parameter $\aleph$ in an arbitrary quantum system with a finite {\DoF} number $f$ follows some common rules.
The size parameter depends on a characteristic energy $\mathcal{E}$ that sets the energy scale relevant in the given context.
This is clearly not an exactly determined quantity, but its rough knowledge is essential.
A trivial example is a harmonic oscillator in which the size parameter, a typical number of phonons, increases with the selected energy scale $\mathcal{E}$.
As argued in Ref.\,\cite{Stran14}, the general size parameter reflects the overall volume $\Omega(\mathcal{E})$ of the domain activated at the selected scale $\mathcal{E}$ in the $2f$-dimensional phase space.
The number $\aleph$ is obtained by expressing this volume in units of elementary quantum cells, namely:
\begin{equation}
\aleph^f\sim\frac{\Omega(\mathcal{E})}{\hbar^f}
\label{alef}.
\end{equation}
This formula is applicable in all the above examples.
For instance, in a system of $n$ bosonic species with comparable single-particle energies $\varepsilon_k\approx\bar{\varepsilon}_k$ and interactions $\nu_{klk'l'}\ll\bar{\varepsilon}_k$, see Eq.\,\eqref{Hibm}, the volume in the $f=n$ phase space reads $\Omega(\mathcal{E})\approx(\mathcal{E}/\bar{\varepsilon}_k)^n$, which for $\mathcal{E}=N\bar{\varepsilon}_k$ yields $\aleph\approx N$.
In Sec.\,\ref{Fisie}, we will return to these considerations in connection with general coupled systems.

We now follow the general algebraic treatment sketched in Sec.\,\ref{Algeb}.
Once the size parameter $\aleph$ is chosen, the model must be checked (or modified) to satisfy some {\em scaling properties}.
In particular, one can introduce the scaled Hamiltonian $\hat{h}=\hat{H}/\aleph$ and scaled generators $\hat{g}_l=\hat{G}_l/\aleph^{\kappa_l}$, where $\kappa_l>0$ are suitable powers, and enforce validity of the formula $\hat{h}=H(\{\hat{g}_l\})$.
If the original Hamiltonian does not satisfy this relation, its parameters must be redefined so that the transformed Hamiltonian does.
Consider as an example the extended Dicke model with $\aleph=N$, where the scaled generators read $\hat{J}_{\alpha}/N$ and $\hat{b}^{\dag}/N^{1/2}$, $\hat{b}/N^{1/2}$, so to satisfy the proper behavior of the scaled Hamiltonian $\hat{h}$ we write the atom-field interaction parameter in Eq.\,\eqref{HDic} as $\lambda/N^{1/2}$.
Similarly, scaled bosonic generators of unitary groups are $\hat{b}_k^{\dag}\hat{b}_l/N$, where $N=\aleph$ is the total number of bosons, hence the parameters that measure the overall strength of two-body interactions must have the form $\sim\lambda/N$, see Eq.\,\eqref{Hibm}.
These requirements ensure that the one-, two- and $n$-body terms of the unscaled Hamiltonian keep roughly the same energy proportions as $N$ is varied.

If one considers the scaled quantities instead of the bare ones, the role of the dynamical algebra generators $\{\hat{G}_l\}$ is transmitted to the scaled generators $\{\hat{g}_l\}$.
Their commutation relations
\begin{equation}
\bigl[\hat{g}_k,\hat{g}_l\bigr]=\ii\sum_m\aleph^{\kappa_m-\kappa_k-\kappa_l}s_{klm}\hat{g}_m
\label{scalecomu}
\end{equation}
usually lead to trivial asymptotic behavior ${\lim_{\aleph\to\infty}[\hat{g}_k,\hat{g}_l]=0}$. 
This holds for all the models studied above, and in general if ${\kappa_k+\kappa_l>\kappa_m}$ for all index combinations for which the structure constants ${s_{klm}\neq 0}$. 
Also the case of polynomially deformed dynamical algebras can be treated in this way. 
This means that for large $\aleph$ the scaled quantities become quasiclassical and the infinite-size limit ${\aleph\to\infty}$ of the system coincides with the {\em classical limit}.
The identification of {\QPT} and {\ESQPT} phenomena in such systems therefore depends on the analysis of the classical dynamics.
Among the vast literature dealing with classical limits of quantum systems we mention Ref.\,\cite{Bulga90}, where a general method is described assigning classical canonical variables and the associated phase space to algebraically based systems, and Ref.\,\cite{Klein91}, which shows that the coordinate-momentum resolution of a set of generators can be done via a boson realization of the dynamical algebra.

We have therefore arrived to a somewhat paradoxical result:
For algebraic systems with finite {\DoF} numbers, all quantum critical effects---which belong to the ${\aleph\to\infty}$ emergent phenomena---must  be rooted in purely classical properties.
This is true for the determination of the phase structure of the system and for the classification of its phase transitions. 
However, we emphasize that the classical-like behavior for large values of $\aleph$ applies only to scaled quantities, while the original unscaled quantities keep their quantum properties, in particular near the {\QPT} and {\ESQPT} critical points.
Here the quantum finite-size effects are important for any value of $\aleph$.

\section{Classical origins of excited-state singularities}
\label{Singul}

In this section we develop a general formalism for identification and classification of an {\ESQPT} in the classical limit of the quantum system. 
At first we overview the effects in the smoothed level density.
Then we show that, due to non-analytic evolutions of Hamiltonian eigenstates across the {\ESQPT}, similar effects appear in the smoothed level flow and in energy densities of various observables.
Finally we discuss some {\ESQPT} precursors in the oscillatory component of level density.

\subsection{Smoothed level density and classification of excited-state singularities}
\label{Leden}

We assume a quantum system which in the classical limit has $f$ {\DoFs} and is described by a Hamiltonian function $H(\vecb{q},\vecb{p})$, where $\vecb{q}\equiv(q_1,q_2,\dots,q_f)$ and $\vecb{p}\equiv(p_1,p_2,\dots,p_f)$ are mutually conjugate sets of canonical coordinates and momenta.
For points $(\vecb{q},\vecb{p})$ in the classical phase space $\Ph$ we introduce a simplified notation with a single symbol $\vecb{x}$, that in individual components means $(q_1,...,q_f,p_1,...,p_f)=(x_1,x_2,...,x_{2f})$.

As we saw above, the most significant signatures of {\ESQPTs} appear in the smoothed level density.
It was introduced in Eq.\,\eqref{leds}, which however contained a somewhat arbitrary smoothening function $\overline{\delta}(E)$.
A unique determination of the smoothed level density follows from a semiclassical analysis. 
The exact level density \eqref{led} of a bound quantum system can be expressed via the Fourier transform of the trace of the evolution operator $\hat{U}(t)$, namely as $\rho(E)=(2\pi\hbar)^{-1}\int_{-\infty}^{+\infty}dt\ {\rm Tr}[\hat{U}(t)]e^{\ii Et/\hbar}$.
Using a semiclassical approximation based on the path integral in the evaluation of $\hat{U}(t)$, one gets a unique resolution of $\rho(E)$ into the smooth and oscillatory components \cite{Haake10,Gutzw71,Berry76}.
While the smooth component $\overline{\rho}(E)$ is obtained from the contribution of so-called zero-length orbits, the oscillatory component $\widetilde{\rho}(E)$ is approximated by a~sum of contributions from classical periodic orbits (see Sec.\,\ref{Fisie}).
The formula for the {\em smooth component}, also called {\em Weyl's law\/} \cite{Haake10}, reads
\begin{equation}
\hspace{-20mm}
\overline{\rho}(E)=\frac{1}{(2\pi\hbar)^f}\!\int d^{2f}\!\vecb{x}\ \delta\bigl(E\!-\!H(\vecb{x})\bigr)
=\frac{\partial}{\partial E}\biggl[\frac{1}{(2\pi\hbar)^f}\!\underbrace{\int d^{2f}\!\vecb{x}\ \theta\bigl(E\!-\!H(\vecb{x})\bigr)}_{\Omega(E)}\biggr]
\label{smole},
\end{equation}
where $\theta$ is the step function.
The quantity in square brackets is a smoothed cumulative number of eigenstates for all energies below the given value $E$, and coincides with the volume $\Omega(E)$ of the classical phase space accessible for systems with energy not exceeding the value $E$ divided by the elementary quantum volume $(2\pi\hbar)^f$. 

The dimension of the first integration in Eq.\,\eqref{smole} can be reduced by one if we perform a transformation from the original phase-space coordinates $\vecb{x}$ to new ones $\tilde{\vecb{x}}\equiv(\tilde{x}_1,...,\tilde{x}_{2f-1},\tilde{x}_{2f})\equiv(\vecb{y},\tilde{x}_{2f})$, for which the component $\tilde{x}_{2f}$ points along the $2f$-dimensional gradient $\vecb{\nabla}H(\vecb{\vecb{x}})$ of the Hamiltonian on the hypersurface $H(\vecb{\vecb{x}})=E$.
This hypersurface coincides with the $(2f\!-\!1)$-dimensional manifold obtained by setting $\tilde{x}_{2f}=0$ and varying the remaining components of $\tilde{\vecb{x}}$ denoted by $\vecb{y}$.
Thus we have
\begin{equation}
\overline{\rho}(E)=\frac{1}{(2\pi\hbar)^f}\!\int d^{2f-1}\vecb{y}\ \left|{\rm Det}\,\frac{\partial\vecb{x}}{\partial\tilde{\vecb{x}}}\right|\ 
|\vecb{\nabla}H(\vecb{\vecb{y}},0)|^{-1}
\label{smole1},
\end{equation}
where $\partial\vecb{x}/\partial\tilde{\vecb{x}}$ is the Jacobian matrix of the $\vecb{x}\mapsto\tilde{\vecb{x}}$ transformation.
The integral in Eq.\,\eqref{smole1} is not well defined at the classical stationary points, where the gradient vanishes.
At these points we can expect non-analyticities of the smoothed level density.

Any classical stationary point $\vecb{x}^{(i)}$, where the superscript $i=1,2,...,M$ distinguishes all stationary points of the system, obeys the defining condition $\frac{\partial}{\partial x_k}H(\vecb{x})|_{\vecb{x}=\vecb{x}^{(i)}}\equiv\partial_k H(\vecb{x})|_{\vecb{x}=\vecb{x}^{(i)}}=0$ for $k=1,2,...,2f$, that is $\vecb{\nabla}H(\vecb{x})|_{\vecb{x}=\vecb{x}^{(i)}}=0$.
An important role in the classification of any stationary point $\vecb{x}^{(i)}$ is played by the eigenvalues of the Hessian matrix $\partial_k\partial_l H(\vecb{x})|_{\vecb{x}=\vecb{x}^{(i)}}$.
If all these eigenvalues are non-zero numbers, i.e., if the Hessian matrix has a non-vanishing determinant, we have a non-degenerate stationary point.
This means that the behavior of the Hamiltonian function $H(\vecb{x})$ around $\vecb{x}^{(i)}$ is locally quadratic.
On the other hand, if at least one of the eigenvalues is zero, i.e., if the determinant of the Hessian matrix vanishes, the stationary point is called degenerate.
The local dependence of the Hamiltonian function near such a point is flat in the directions determined by eigenvectors associated with the vanishing eigenvalues.

\begin{figure}[t!]
\begin{flushright}
\includegraphics[width=0.85\textwidth]{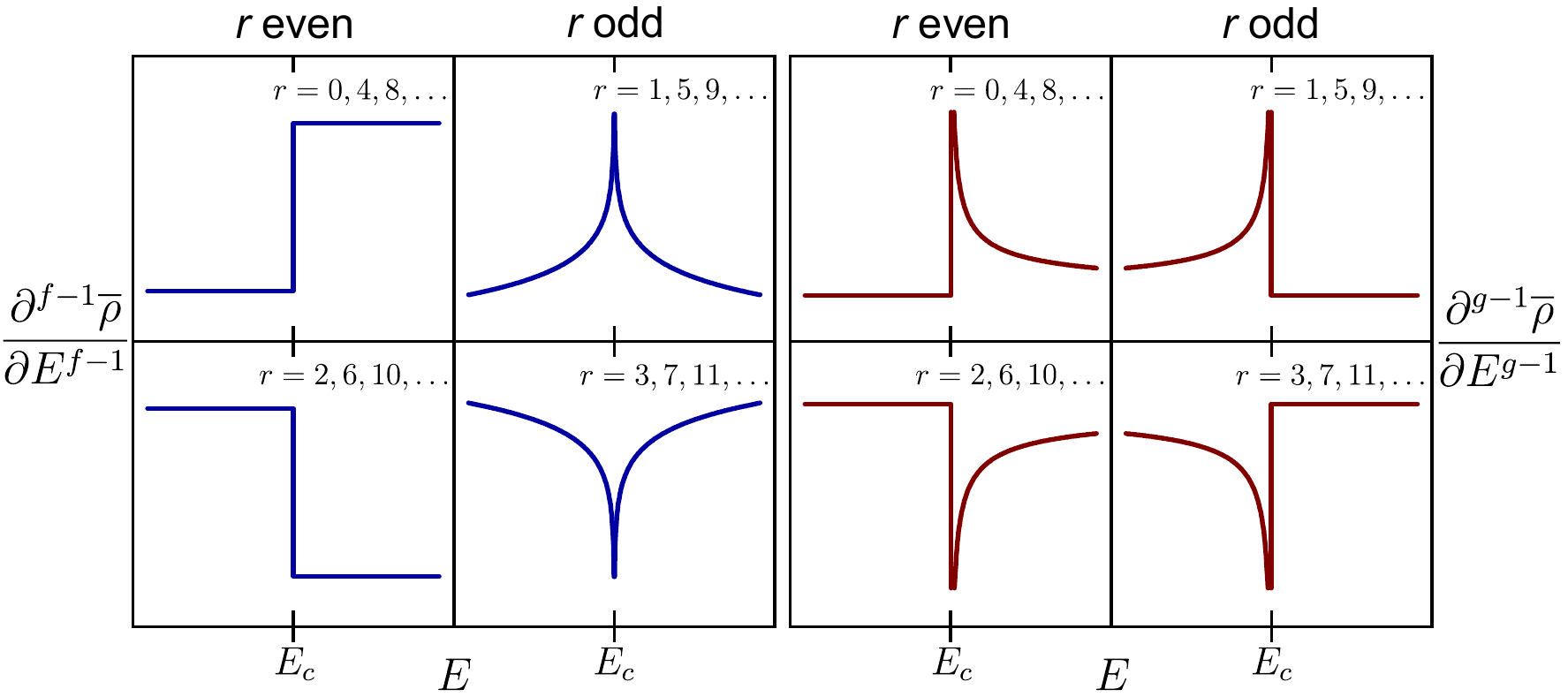}
\end{flushright}
\caption{Typology of {\ESQPT} singularities related to non-degenerate stationary points of the classical Hamiltonian in the $(f\I{-}1)$st or $(g\I{-}1)$st derivative of the smoothed level density. The panel on the left corresponds to stationary points in the phase-space of even dimension $2f$, the panel on the right shows effects of stationary points in a space with odd dimension $2g\I{-}1$. Adapted from Ref.\,\cite{Stran17}.}
\label{Typ}
\end{figure}

For any {\em non-degenerate stationary point\/} $\vecb{x}^{(i)}$, the singular contribution to integrals \eqref{smole} and \eqref{smole1} from a close vicinity of $\vecb{x}^{(i)}$ can be calculated explicitly.
Each of such points is characterized by an integer $r\in[0,2f]$ called {\em index}, which is the number of negative eigenvalues of the Hessian matrix.
Hence a non-degenerate stationary point with index $r$ in the phase space of dimension $2f$ represents a local minimum for $r=0$, local maximum for $r=2f$, while for intermediate values of $r$ it is a multidimensional saddle point with $r$ and $(2f\!-\!r)$ independent directions for which it behaves as a local maximum and minimum, respectively.
The effect of index-$r$ non-degenerate stationary point at energy $E_{\rm c}^{(i)}$ on the smoothed level density was determined in Ref.\,\cite{Stran16} using the Morse lemma.
It is expressed as
\begin{equation}
\hspace{-10mm}
\frac{\partial^{f-1}}{\partial E^{f-1}}\ \overline{\rho}(E)\propto h(E)+
\left\{\begin{array}{ll}
(-1)^{r/2}\ \theta(E\!-\!E_{\rm c}^{(i)}) & {\rm for\ } r {\rm\ even},\\
(-1)^{(r+1)/2}\ \ln|E\!-\!E_{\rm c}^{(i)}| & {\rm for\ } r {\rm\ odd},
\end{array}\right.
\label{clasfull}
\end{equation}
where $h(E)$ is an unspecified smooth function resulting from non-singular contributions to the integrals in Eqs.\,\eqref{smole} and \eqref{smole1} from the phase space beyond a vicinity of $\vecb{x}^{(i)}$.
Formula \eqref{clasfull} is a more explicit form of Eq.\,\eqref{clas}, showing how the signs of the step and logarithmic singularities in the $(f\!-\!1)$st derivative depend on the index $r$.
Such singularities are schematically sketched in the left panel of Fig.\,\ref{Typ}.
This enables us to formulate a general classification of the {\ESQPT} effects connected to non-degenerate stationary points in terms of a pair of integers $(f,r)$.
Returning to the $f=2$ extended Dicke model and its phase diagram in Fig.\,\ref{dede}, we can show that {\ESQPTs} between the {\D}--{\TC} and {\TC}--{\N} quantum phases correspond to non-degenerate stationary points of the classical-limit Hamiltonian with indexes $r=1$ and 2, respectively.
This is consistent with the observed singularities in $\overline{\rho}(E)$.
The {\ESQPT} between the {\N}--{\S} phases cannot be classified in this way as the corresponding stationary point lies at the maximal value of $j_z$ and its Hessian matrix is undefined, but as shown in Ref.\,\cite{Kloc17a}, the singularity of $\overline{\rho}(E)$ is of the same type as for an $r=2$ stationary point. 

The dimension $2f$ of the phase space is always an even number.
However, considering integrals \eqref{smole} formally also in spaces with an odd dimensions $(2g\I{-}1)$, where $g=1,2,3...$, we find \cite{Stran16} that index-$r$ non-degenerate stationary points lead to singular dependencies obeying the formula 
\begin{equation}
\hspace{-20mm}
\frac{\partial^{g-1}}{\partial E^{g-1}}\ \overline{\rho}(E)\propto h(E)+
\left\{\begin{array}{ll}
(-1)^{r/2}\ \theta(E\!-\!E_{\rm c}^{(i)})\ |E\!-\!E_{\rm c}^{(i)}|^{-1/2} & {\rm for\ } r {\rm\ even},\\
(-1)^{(r-1)/2}\ \theta(E_{\rm c}^{(i)}\!-\!E)\ |E\!-\!E_{\rm c}^{(i)}|^{-1/2} & {\rm for\ } r {\rm\ odd}.
\end{array}\right.
\label{clasfullodd}
\end{equation}
How can these dependencies be relevant for the {\ESQPT} classification?
We will see in Sec.\,\ref{Spati} that the same technique for evaluating the smoothed level density as presented above is used in spatially extended $D$-dimensional periodic systems, but with the integration of the form \eqref{smole} performed in the quasi-momentum space instead of phase space.
If the quasi-momentum space has an odd dimension, the singularities of the type \eqref{clasfullodd} occur. 
They are schematically depicted in the right panel of Fig.\,\ref{Typ}. 

The above classification of {\ESQPT} singularities in the level density concerns solely non-degenerate stationary points.
We stress that such {\ESQPTs} are the most common ones as the non-degenerate stationary point represents the generic type among all stationary points.
Indeed, the vanishing of one or more eigenvalues of the Hessian matrix requires additional constraints which for a typical stationary point are not satisfied or get violated by an infinitesimal perturbation of the Hamiltonian.
Nevertheless, in some special cases the {\em degenerate stationary points}, i.e., those showing a non-quandratic behavior of $H(\vecb{x})$ near $\vecb{x}^{(i)}$, appear in the phase space.
Effects of such points are remarkable since, in contrast to Eq.\,\eqref{clasfull}, they occur in lower that $(f\I{-}1)$st derivatives of $\overline{\rho}(E)$.
Because there is no unique classification of degenerate stationary points in arbitrary dimensions, the singularities they generate in the level density must be described case by case. 

Consider as an example a degenerate separable local minimum at a point $\vecb{x}^{(i)}$.
The Hamiltonian function in a vicinity of this point can be written as $H(\vecb{x})\approx E_{\rm c}^{(i)}\!+\!\sum_{k=1}^{2f} c_k(x_k\!-\!x_k^{(i)})^{K_k}$, where $c_k$ are some positive constants and $K_k\geq 2$ powers that quantify the flatness of the minimum. 
This stationary point produces an extra cumulative phase-space volume $\Omega(E)$ for energies above $E_{\rm c}^{(i)}$, see Eq.\,\eqref{smole1}.
Because the linear size of the additional phase space along the $k$th axis grows with energy as $(E-E_{\rm c}^{(i)})^{1/K_k}$, the smoothed level density develops a singularity of the following form,
\begin{equation}
\hspace{-5mm}
\overline{\rho}(E)\propto g(E)+\theta(E\I{-}E_{\rm c}^{(i)})\ (E\I{-}E_{\rm c}^{(i)})^{L},
\quad
L=\sum_{k=1}^{2f}\frac{1}{K_k}-1
\label{degen},
\end{equation}
where $g(E)$ is a smooth function.
It is clear that if $K_k\I{=}2$ $\forall k$, we obtain $L=f-1$, consistently with the case of a non-degenerate minimum, see Eq.\,\eqref{clasfull} with $r\I{=}0$. 
However, if some of the powers $K_k$ are larger than 2, the {\ESQPT} singularity becomes stronger than in the non-degenerate case.
For instance, consider a system with $f\I{=}2$.
If $(K_1,K_2,K_3,K_4)\I{=}(4,4,4,4)$ (pure quartic minimum of the full Hamiltonian), the step-like dependence at the critical energy appears already in the zeroth derivative of $\overline{\rho}(E)$.
The same conclusion holds for $(K_1,K_2,K_3,K_4)\I{=}(\infty,\infty,2,2)$ (square well in coordinates and quadratic minimum in momenta). 
On the other hand, for $(K_1,K_2,K_3,K_4)\I{=}(4,4,2,2)$ (quartic minimum in coordinates and quadratic in momenta) we obtain a continuous square-root increase of $\overline{\rho}(E)$ with a power-law divergence of $\frac{\partial}{\partial E}\overline{\rho}(E)$ at $E\I{=}E_{\rm c}^{(i)}$.

The most familiar classical Hamiltonian function consists of a quadratic kinetic term with a mass parameter $M$ and a potential energy $V$:
\begin{equation}
H(\vecb{q},\vecb{p})=\frac{|\vecb{p}|^2}{2M}+V(\vecb{q})
\label{Hnorm}.
\end{equation}
It turns out that a convenient size parameter for systems with this type of classical limit is $\aleph=\sqrt{M}/\hbar$ \cite{Stran14}.
Stationary points of Hamiltonians \eqref{Hnorm} can be disclosed and classified by analyzing solely the potential function $V(\vecb{q})$.
This reduces the problem from the full $2f$-dimensional phase space to just an $f$-dimensional configuration space, implying that the index $r$ of stationary points can take only values between 0 and $f$.
A general analysis of {\ESQPTs} in such systems was presented in Ref.\,\cite{Stran14}.

However, quantum many-body systems usually yield in their classical-limit Hamiltonians $H(\vecb{q},\vecb{p})$ more complicated momentum dependencies than in Eq.\,\eqref{Hnorm} (cf.\,Sec.\,\ref{Ibms}).
Search for stationary points of such Hamiltonians is more difficult.
It is often performed with the aid of a function $\mathcal{V}(\vecb{q})\equiv H(\vecb{q},\vecb{p}=0)$, which looks analogous to the above potential $V(\vecb{q})$ (cf.\,Fig.\,\ref{ibm1st}).
However, the function $\mathcal{V}(\vecb{q})$ does not, in general, coincide with the potential energy of the system since vanishing of the canonical momentum $\vecb{p}$ does not generally imply vanishing of the mechanical momentum $\propto\dot{\vecb{q}}$.
Even if the function $H(\vecb{q},\vecb{p})$ depends (perhaps due to its time-reversal symmetry) only on even powers of $\vecb{p}$, so that $\vecb{p}=0$ implies $\dot{\vecb{q}}=0$, one cannot exclude additional $\vecb{p}\neq 0$ solutions of the $\dot{\vecb{q}}=0$ constraint.
In this situation, some of the stationary points of $H(\vecb{q},\vecb{p})$ may fail to satisfy the condition $\partial_k\mathcal{V}(\vecb{q})=0$.
This leads to a possibility of {\ESQPT} singularities rooted in somewhat counter-intuitive $\vecb{p}\neq 0$ stationary points.
Such {\ESQPTs} were indeed detected in the U(6) interacting boson model on the first-order {\QPT} path between U(5) and SU(3) dynamical symmetries \cite{Macek19}. 

Stationary points in the classical phase space of the system are not the only classical source of {\ESQPTs}.
Some of these singularities are connected with effects of the {\em phase-space boundary}.
For many-body systems with a finite-dimensional Hilbert space $\Hi$ the overall volume of the corresponding classical phase space $\Ph$ must also be finite.
For instance, the phase space associated with an interacting boson system at a fixed value $N$ of the total number of bosons (see Sec.\,\ref{Ibms}) is a ball (interior of a $2f$-dimensional sphere with ${f=n-1}$) with the radius equal to $\sqrt{2}$ for the usual choice of ${\eta=N}$ in Eq.\,\eqref{QP}.
The finiteness of the phase space $\Ph$ implies the existence of a boundary $\partial\Ph$, which can cause additional spectral singularities.
In particular, as shown in Ref.\,\cite{Macek19}, stationary points of the Hamiltonian function on the $(2f\I{-}1)$-dimensional boundary manifold lead to singularities in the smoothed level density calculated from Eq.\,\eqref{smole}. 
We stress that stationary points of $H(\vecb{x})|_{\vecb{x}\in\partial\Ph}$ are not true stationary points of the system (excluding exceptional cases)  as the derivative of $H(\vecb{x})$ along the normal direction is not specified.
Classification of {\ESQPTs} connected with the phase-space boundary has not yet been given.
The studied special cases show that non-degenerate stationary points on $\partial\Ph$ lead to weaker singularities than non-degenerate stationary points inside $\Ph$.
However, degenerate stationary points on $\partial\Ph$ are likely because the extremal values of coordinates and momenta on the boundary usually yield somehow restricted behavior of $H(\vecb{x})|_{\vecb{x}\in\partial\Ph}$.
Effects of degenerate stationary points on $\partial\Ph$ may compete with those of non-degenerate stationary points inside $\Ph$.
Examples of boundary-related {\ESQPTs} were demonstrated in the U(6) interacting boson model along the U(5)--SU(3) transition \cite{Macek19}.

\subsection{Level flow, expectation values  and structure of eigenvectors}
\label{Leflo}

So far the level density has been studied mostly as a function of energy $E$.
Now we fully activate also the variation of the system's control parameter $\lambda$.
We assume that with running $\lambda$ individual energy levels $E_i(\lambda)$ just move up or down along the energy axis, i.e., that they do not appear or disappear at any value of $\lambda$.
This can be expressed by a continuity equation in which the roles of coordinate $x$ and time $t$ are played by $E$ and $\lambda$, respectively.
For the exact density defined  in Eq.\,\eqref{led} the corresponding flow reads $\jmath(\lambda,E)=\sum_i\frac{dE_i}{d\lambda}(\lambda)\,\delta(E\!-\!E_i(\lambda))$.
Considering the smoothed density \eqref{leds} written in terms of the smoothed (differentiable) $\delta$-functions $\overline{\delta}(E\!-\!E_i(\lambda))$, we can also define the {\em smoothed flow}:
\begin{equation}
\overline{\jmath}(\lambda,E)=\sum_i\frac{dE_i(\lambda)}{d\lambda}
\ \overline{\delta}\bigl(E-E_i(\lambda)\bigr)
\label{flos}.
\end{equation}
The continuity equation
\begin{equation}
\frac{\partial}{\partial\lambda}\,\overline{\rho}(\lambda,E)+\frac{\partial}{\partial E}\,\overline{\jmath}(\lambda,E)=0
\label{conti}
\end{equation}
can then be proven by elementary evaluation of the derivatives involved.
Note that the smoothed flow can be also expressed as $\overline{\jmath}(\lambda,E)=\overline{\rho}(\lambda,E)\overline{\phi}(\lambda,E)$ via a smoothed \uvo{velocity field} $\overline{\phi}(\lambda,E)$ of levels in the spectrum, called the {\em flow rate} \cite{Stran14}.

\begin{figure}[t!]
\begin{flushright}
\includegraphics[width=0.85\textwidth]{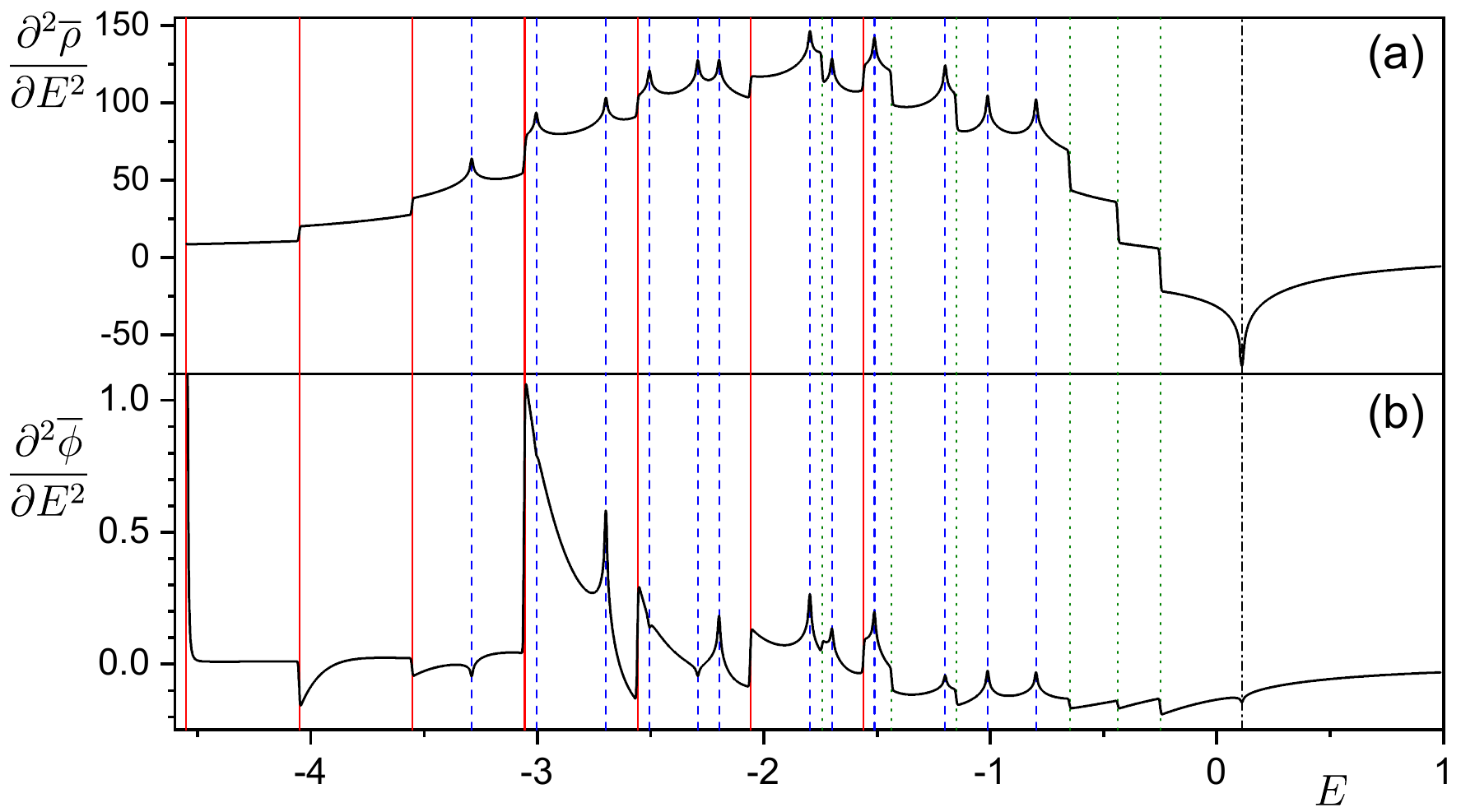}
\end{flushright}
\caption{Singularities in the second energy derivatives of (a) the smoothed level density and (b) the smoothed flow rate for a toy $f\I{=}3$ Hamiltonian $\hat{H}=\frac{1}{2}|\hat{\vecb{p}}|^2\I{-}2|\hat{\vecb{q}}|^2\I{+}\frac{1}{4}\hat{q}_1\I{+}\frac{1}{2}\hat{q}_2\I{+}(\frac{3}{4}\I{+}\lambda)\hat{q}_3\I{+}\hat{q}_1^4\I{+}\hat{q}_2^4\I{+}\hat{q}_3^4$ at $\lambda\I{=}0$. Energies of stationary points are marked by vertical lines. Adapted from Ref.\,\cite{Stran16}.}
\label{Badhair}
\end{figure}

The mutual relation of $\overline{\rho}(\lambda,E)$ and $\overline{\jmath}(\lambda,E)$ or $\overline{\phi}(\lambda,E)$ following from the continuity equation \eqref{conti} leads to the conclusion that in generic situations both quantities share the same type of non-analyticity \cite{Stran16}.
Indeed, assuming that both smoothed level density and flow are $k$ times differentiable functions, the continuity equation implies that $\frac{\partial}{\partial\lambda}\frac{\partial^k}{\partial E^k}\overline{\rho}+\frac{\partial^{k+1}}{\partial E^{k+1}}\overline{\jmath}=0$.
Consider now a discontinuity or divergence in the $(k\!+\!1)$st energy derivative of the smoothed level density occurring at an {\ESQPT} critical borderline given by a smooth curve $E\I{=}E_{\rm c}(\lambda)$ in the $\lambda\times E$ plane.
If we choose a point $(\lambda,E)$ at the critical borderline so that $\frac{d}{d\lambda}E_{\rm c}(\lambda)\I{\neq}0$, the derivative $\frac{\partial}{\partial\lambda}\frac{\partial^{k}}{\partial E^{k}}\overline{\rho}(\lambda,E)$ must show the same type of non-analyticity as $\frac{\partial^{k+1}}{\partial E^{k+1}}\overline{\rho}(\lambda,E)$.
The differentiated continuity equation implies that this singularity must be compensated by the same type of singularity in the derivative $\frac{\partial^{k+1}}{\partial E^{k+1}}\overline{\jmath}(\lambda,E)$ of the smoothed flow.
The same conclusion holds also for the smoothed flow rate $\overline{\phi}(\lambda,E)$ \cite{Stran16}.
We stress that the signs and sizes of singular terms in the dependencies $\overline{\rho}(\lambda,E)$, $\overline{\jmath}(\lambda,E)$ and $\overline{\phi}(\lambda,E)$ differ but the type of non-analyticity (discontinuity or divergence) is the same.
The systems with $f\I{=}1$ must be treated separately as their stationary points affect typically the zeroth derivative of the level density.
It turns out that in this case a divergence of $\overline{\rho}(\lambda,E)$ causes an indefinite singularity of $\overline{\phi}(\lambda,E)$ \cite{Stran15}.

These conclusions are illustrated in Fig.\,\ref{Badhair}.
Here we see singularities in $\frac{\partial^{2}}{\partial E^{2}}\overline{\rho}(\lambda,E)$ and $\frac{\partial^{2}}{\partial E^{2}}\overline{\phi}(\lambda,E)$ at a number of non-degenerate stationary points of a toy $f\I{=}3$ Hamiltonian of the form \eqref{Hnorm}.
The singularities in the smoothed level density agree with Eq.\,\eqref{clasfull}, taking into account the indexes $r=0,...,3$ assigned to individual stationary points. 
The corresponding singularities in the smoothed flow rate are of the same types, but are often turned upside down.
For details see Ref.\,\cite{Stran16}. 

The slopes of individual energy levels $E_i(\lambda)$ can be expressed via the Hellmann-Feynman formula
\begin{equation}
\frac{dE_i(\lambda)}{d\lambda}=\biggl\langle\psi_i(\lambda)\biggl|\frac{\partial\hat{H}}{\partial\lambda}\biggr|\psi_i(\lambda)\biggr\rangle\equiv\bigl\langle\hat{H}'\bigr\rangle_i
\label{HeFe},
\end{equation} 
where $\ket{\psi_i(\lambda)}$ is the Hamiltonian eigenvector assigned to the $i$th eigenvalue at the parameter value $\lambda$.
The smoothed flow in Eq.\,\eqref{flos} is then rewritten in the form of Eq.\,\eqref{Asmodep} with the general quantity $A$ substituted by the Hamiltonian derivative with respect to $\lambda$. 
Hence we write $\overline{\jmath}(\lambda,E)=\overline{H'}(\lambda,E)$, where $H'$ stands for the quantity associated with operator $\frac{\partial}{\partial\lambda}\hat{H}$.
Therefore, returning to Fig.\,\ref{nipere} above, we see that the curve depicting the smoothed energy dependence of quantity  $\hat{H}'(\delta)$ is also the smoothed flow of the spectrum.
This explains why its non-analyticites at the {\ESQPT} critical energies are of the same type as those in the dependence of $\overline{\rho}(\lambda,E)$, see Fig.\,\ref{dede}.

We stress that singular dependencies of $\overline{H'}(\lambda,E)$ at the {\ESQPT} critical borderlines must be attributed to singular changes in the structure of eigenstates.
For linear Hamiltonians the operator $\frac{\partial}{\partial\lambda}\hat{H}$ is independent of $\lambda$ and all variations of its expectation values with $\lambda$ and $E$ are solely due to the variations of $\ket{\psi_i(\lambda)}$.
Therefore, the {\ESQPTs} caused by non-degenerate stationary points are likely to affect the $(f\I{-}1)$st derivative of the smoothed energy density from Eq.\,\eqref{Asmodep} for almost any generic observable.

\begin{figure}[t!]
\begin{flushright}
\includegraphics[width=0.85\textwidth]{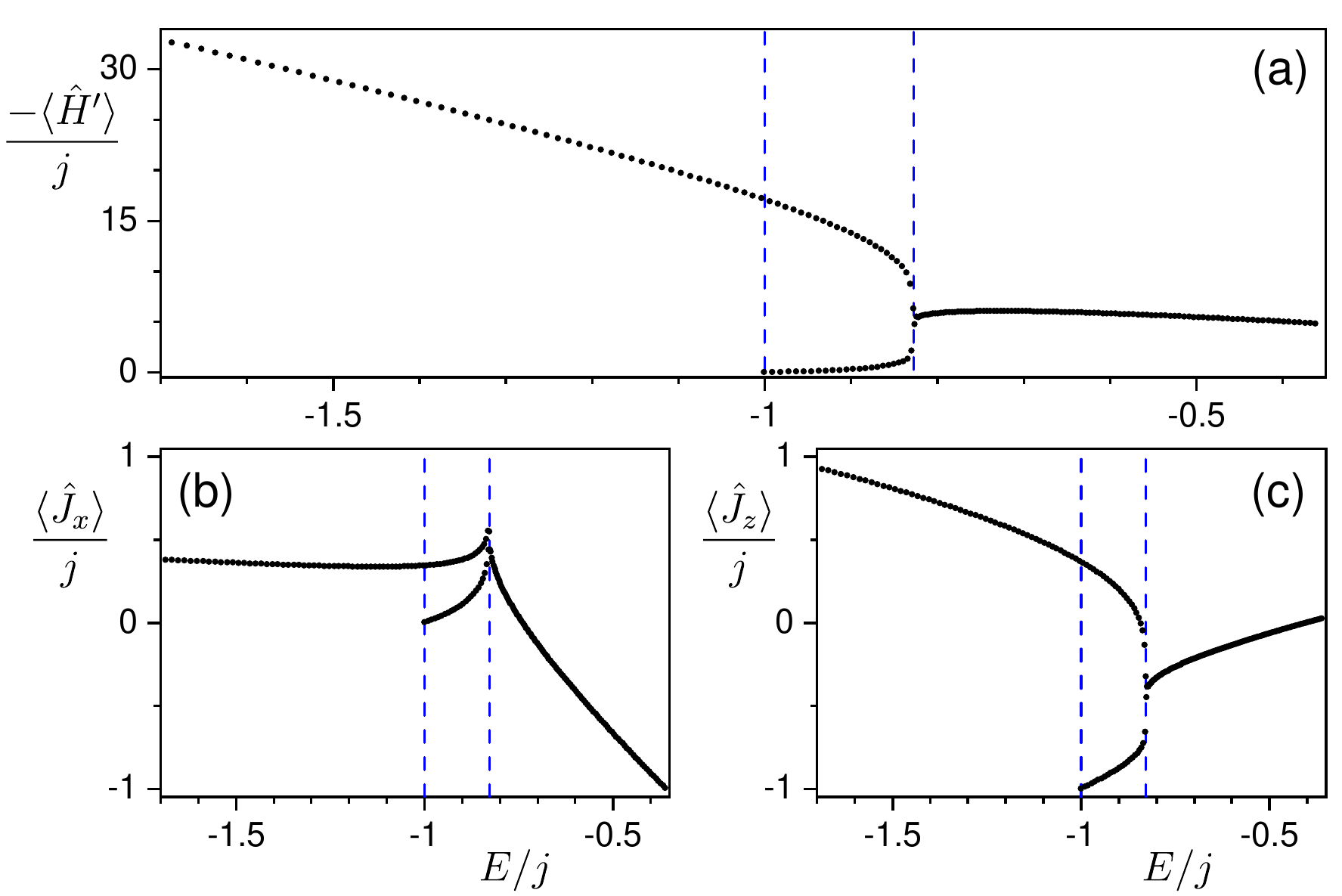}
\end{flushright}
\caption{Expectation values of various observables in eigenstates of the Lipkin Hamiltonian from Fig.\,\ref{lipkac} with $\lambda\I{=}0.8$ (vertical line in Fig.\,\ref{lipkac}) and $N\I{=}2j\I{=}200$.
Since $f\I{=}1$, Peres lattices localize along curves. 
Panel~(a) corresponds to observable $-\hat{H}'=N^{-1}(\hat{J}_x\I{+}4\hat{J}_z\I{+}2N)^2$, panel (b) to $\hat{J}_x$ and panel~(c) to $\hat{J}_z$ (note that $\ave{\hat{J}_y}_i\I{=}0$ $\forall i$).
The {\ESQPT} with index $r\I{=}0$ at $E_{{\rm c}1}=-j$ adds a second branch of points to each lattice, so it produces a jump of the smoothed value of the respective observable. 
The {\ESQPT} with $r\I{=}1$ at $E_{{\rm c}2}>-j$ is seen as a merge of both branches, yielding a singular energy derivative of the smoothed dependence (cf.\,Fig.\,\ref{ibm1st}).
Note that the latter singularity differs from that of $\overline{\rho}(\lambda,E)$ as this is the exceptional $f=1$ case (see the text).
}
\label{liporder}
\end{figure}

An estimate of expectation values revealing the classical origins of their {\ESQPT} singularities can be obtained via the {\it phase-space integration}.
Indeed, the smoothed energy density $\overline{A}(\lambda,E)$ of a quantity $\hat{A}$ in the energy spectrum for large values of the size parameter $\aleph$ is given by the phase-space average of the corresponding classical observable $A(\vecb{q},\vecb{p})\I{\equiv}A(\vecb{x})$ over the $H(\lambda,\vecb{x})\I{=}E$ manifold:
\begin{equation}
\overline{A}(\lambda,E)=\frac{1}{(2\pi\hbar)^f}\int d^{2f}\!\vecb{x}\ A(\vecb{x})\,\delta\bigl(E\!-\!H(\lambda,\vecb{x})\bigr)
\label{clAve}.
\end{equation}
This quantity represents the overall density of quantity $A$ at given energy, i.e., its average in a single state times the density of states. 
The single-state average reads
\begin{equation}
\overline{a}(\lambda,E)=\frac{\overline{A}(\lambda,E)}{\overline{\rho}(\lambda,E)}=\frac{\int d^{2f}\!\vecb{x}\ A(\vecb{x})\,\delta\bigl(E\!-\!H(\lambda,\vecb{x})\bigr)}{\int d^{2f}\!\vecb{x}\ \delta\bigl(E\!-\!H(\lambda,\vecb{x})\bigr)}
\label{clave}.
\end{equation}

In particular, using the last formula in $f\I{=}1$ systems, we obtain 
\begin{equation}
\overline{a}(\lambda,E)=\sum_{\vecb{o}}
\underbrace{\frac{T_{\vecb{o}}}{\sum_{\vecb{o}'}T_{\vecb{o}'}}}_{w_{\vecb{o}}}\ \underbrace{\frac{1}{T_{\vecb{o}}}\int_0^{T_{\vecb{o}}}dt\,A\bigl(\vecb{x}_{\vecb{o}}(t)\bigr)}_{\ave{A}_{\vecb{o}}}
\label{clave1},
\end{equation}
where $\ave{A}_{\vecb{o}}$ is the time average of $A$ over a classical orbit $\vecb{x}_{\vecb{o}}(t)$ with period $T_{\vecb{o}}$ (all these quantities depend implicitly on $\lambda$ and $E$).
We assume that, in general, there is more primitive periodic orbits at given energy $E$, and we sum over them (index $o$) with weight factors $w_{\vecb{o}}$.
Note that in some calculations equal weights for all orbits are used \cite{Engel15}.
The existence of different orbits at the same energy (like the orbits located in different potential wells in the case of a double-well potential) generates distinct branches of the expectation values $\ave{\hat{A}}_i$, which are given by individual time averages $\ave{A}_{\vecb{o}}$ along the corresponding orbits.
Energies of the stationary points, at which the orbits undergo non-analytic change (e.g., bifurcate to both wells in the example of a double-well potential), locate non-analyticities of the dependence $\overline{a}(\lambda,E)$ and splittings of the $\ave{\hat{A}}_i$ branches.
Such effects are illustrated in Fig.\,\ref{liporder} for several choices of the observable $\hat{A}$ in the spectrum of the Lipkin model from Fig.\,\ref{lipkac}.
Instead of showing the dependencies of various averages $\ave{\hat{A}}_i$ and $\ave{\hat{B}}_i$ on energy, the work in Ref.\,\cite{Engel15} presents mutual dependencies of these observables on each other, which for $f\I{=}1$ systems (there the second-order {\QPT} Lipkin model) form curves with non-analyticities and bifurcations on the {\ESQPT} points.

A quantum perspective on the {\ESQPT}-induced non-analyticities of the flow of levels and expectation values of observables follows from singularities of the Hamiltonian eigensolutions known as {\em exceptional points} \cite{Kato66,Zirnba83,Moise11}.
In our case these singularities appear if the Hamiltonian parameter $\lambda$ is extended to the complex plane, i.e., is replaced by $\Lambda\in\mathbb{C}$.
Then one can identify a discrete set of special values of $\Lambda$ where some of the complex eigenvalues of the non-Hermitian Hamiltonian $\hat{H}(\Lambda)$ coalesce.
This typically affects arbitrary pairs of eigenvalues, but in less generic situations a single degeneracy can involve even a larger number of eigenvalues.  
Degeneracies in the non-Hermitian context differ from ordinary degeneracies of Hermitian Hamiltonians since, in typical cases, the eigenvectors corresponding to the degenerate eigenvalues coalesce, too, and the single eigenvector at the degeneracy point becomes self-orthogonal (in the sense of the scalar product including left and right eigenvectors) \cite{Kato66}.
The eigenvalues and eigenvectors near a pairwise exceptional point $\Lambda_{\rm EP}$ connecting levels $i$ and $i'$ can be expanded in powers $(\Lambda\I{-}\Lambda_{\rm EP})^{k/2}$, where $k$ is an integer. 
The lowest-order behavior is expressed by
\begin{eqnarray}
\hspace{-18mm}
E_{i}(\Lambda)\approx E_{0ii'}+\sqrt{\Lambda\I{-}\Lambda_{\rm EP}}\,E_{1ii'},&&
\quad
E_{i'}(\Lambda)\approx E_{0ii'}-\sqrt{\Lambda\I{-}\Lambda_{\rm EP}}\,E_{1ii'},
\label{exc1}\\
\hspace{-18mm}
\ket{\psi_{i}(\Lambda)}\approx\ket{\psi_{0ii'}}+\sqrt{\Lambda\I{-}\Lambda_{\rm EP}}\,\ket{\psi_{1ii'}},&&
\quad
\ket{\psi_{i'}(\Lambda)}\approx\ket{\psi_{0ii'}}-\sqrt{\Lambda\I{-}\Lambda_{\rm EP}}\,\ket{\psi_{1ii'}},
\label{exc2}
\end{eqnarray}
where $E_{0ii'}$ and $\ket{\psi_{0ii'}}$ stand for the common eigenvalue and the right eigenvector of both levels at $\Lambda\I{=}\Lambda_{\rm EP}$, while $E_{1ii'}$ and $\ket{\psi_{1ii'}}$ are the $k\I{=}1$ corrections.
Both expressions \eqref{exc1} and \eqref{exc2} consider only a single branch of $\sqrt{\Lambda\I{-}\Lambda_{\rm EP}}$, the two-valuedness of the square root function being taken into account by the $\pm$ signs of components $i$ and $i'$.
We see that the exceptional point $\Lambda_{\rm EP}$ corresponding to the pairwise degeneracy represents a square-root branching point for both eigenvalues and eigenvectors connected with the levels $i$ and $i'$ involved.
Following a path near $\Lambda_{\rm EP}$, one observes very fast variations of energies $E_{i}(\Lambda),E_{i'}(\Lambda)$ and wavefunctions $\ket{\psi_{i}(\Lambda)},\ket{\psi_{i'}(\Lambda)}$.

\begin{figure}[t!]
\begin{flushright}
\includegraphics[width=\textwidth]{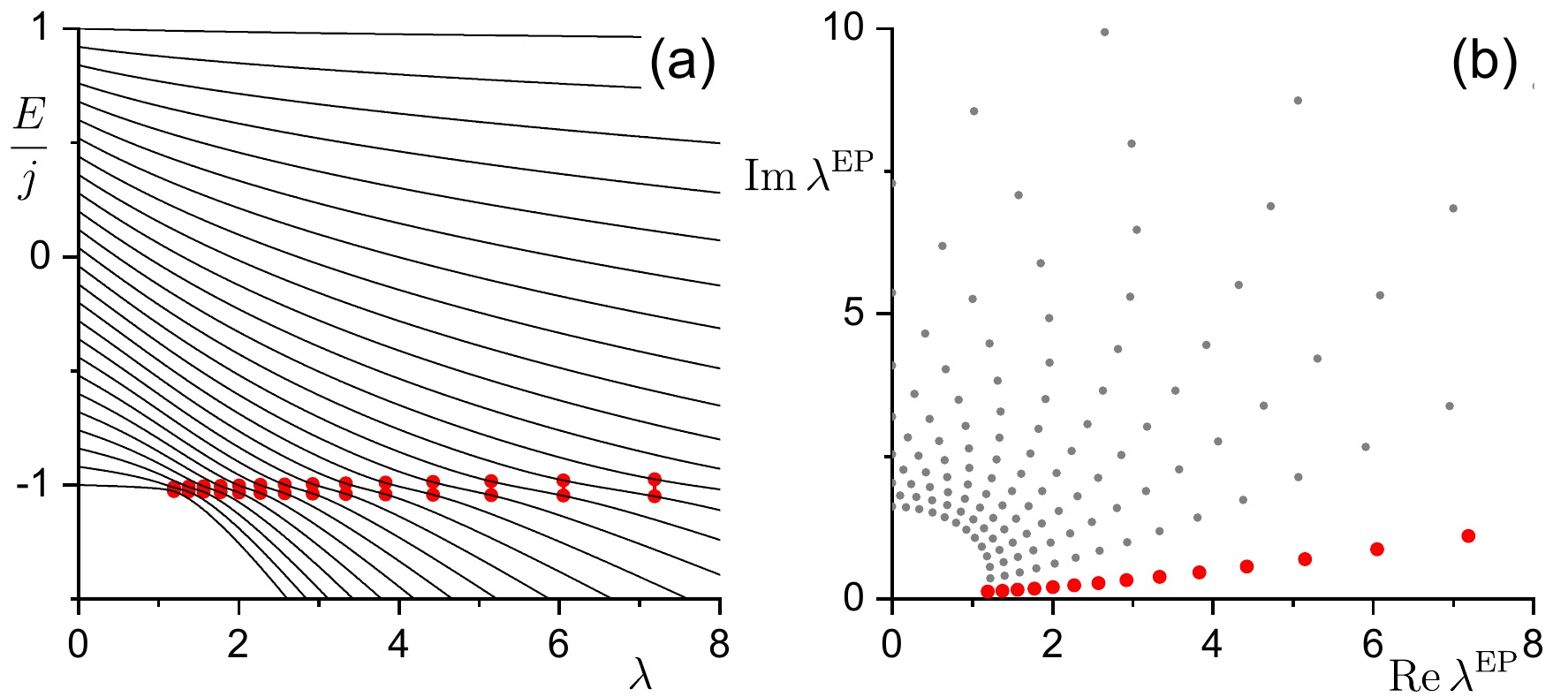}
\end{flushright}
\caption{The spectrum (a) and exceptional points (b) for positive-parity eigenstates of the Lipkin Hamiltonian \eqref{HLipkin} with $\chi\I{=}0$ and $N\I{=}2j\I{=}50$. The {\ESQPT} signaled by a sequence of avoided crossings of levels in panel (a) (highlighted by pairs of linked red dots) is connected with the lowest row of exceptional points in panel (b) (marked by thicker red dots). In this model, exceptional points come in complex conjugate pairs, so the ${\rm Im}\lambda\I{>}0$ pattern in panel (b) has a mirror symmetric ${\rm Im}\lambda\I{<}0$ image. In Ref.\,\cite{Sinde17}, the exceptional points are shown to algebraically converge to the real axis with increasing $N$, which correspondingly reduces the minimal energy gap of the avoided crossings in panel~(a).
}
\label{epka}
\end{figure}

These findings are important also for Hermitian Hamiltonians $\hat{H}(\lambda)$ with $\lambda\in\mathbb{R}$.
If an exceptional point $\Lambda_{\rm EP}$ of the complex-extended Hamiltonian $\hat{H}(\Lambda)$ appears close to the real axis, an avoided crossing of the corresponding levels is observed around $\lambda\I{=}{\rm Re}\Lambda_{\rm EP}$ \cite{Heiss90}.
The energy gap between both levels at their closest approach is proportional to ${\rm Im}\Lambda_{\rm EP}$.
Avoided crossings of levels, in general, are the places where eigenvectors undergo the most rapid evolution with $\lambda$.
This is seen from the formula
\begin{equation}
\hspace{-15mm}
\abs{\scal{\psi_i(\lambda\I{+}d\lambda)}{\psi_i(\lambda)}}^2=1-d\lambda^2\sum_{i'(\neq i)}\frac{\bigl|\matr{\psi_i(\lambda)}{\frac{\partial}{\partial\lambda}\hat{H}(\lambda)}{\psi_{i'}(\lambda)}\bigr|^2}{\bigl(E_i(\lambda)\I{-}E_{i'}(\lambda)\bigr)^2}+O(d\lambda^3)
\label{charate},
\end{equation}
which describes a local variation of an arbitrary eigenstate $\ket{\psi_i(\lambda)}$ of $\hat{H}(\lambda)$ under a small parameter change $d\lambda$  [cf.\,Eq.\,\eqref{metric} below].
This formula clearly indicates that the overlap between $\ket{\psi_i(\lambda)}$ and $\ket{\psi_i(\lambda+d\lambda)}$ quickly decays with increasing $d\lambda$ if some of the energy differences in the denominator become small (provided that the corresponding local matrix elements of $\frac{\partial}{\partial\lambda}\hat{H}$ are not negligible).
Indeed, it has been demonstrated that if an increasing number of exceptional points accumulates for an increasing size parameter $\aleph$ near a point $\Lambda=\lambda_{\rm c}\I{+}\ii 0$, the system in the $\aleph\I{\to}\infty$ limit develops a {\QPT} whose type (order) depends on the asymptotic density of exceptional points near the critical point \cite{Heiss88,Cejna05,Cejna07,Stran18}.

A similar conclusion probably holds also for {\ESQPTs}, which however require an accumulation of exceptional points close to the real axis within a broader interval of $\lambda$, where the {\ESQPT} critical borderline exists.
This statement is rather straightforward for the {\ESQPTs} connected with asymptotically diverging level density, where a very close mutual approach of energy levels is obvious.
However, even if the {\ESQPT} non-analyticity is of another type, for instance appears in a higher energy derivative of $\overline{\rho}(\lambda,E)$, sharp avoided crossings of a certain subset of levels seem to be a universal {\ESQPT} accompaniment. 
The distribution of exceptional points in {\ESQPT} systems has so far been analyzed only in the case of diverging level density, i.e., in models with a single effective {\DoF}.
One of them was the U(6) interacting boson model, where exceptional points were studied along the transition between U(5) and SO(6) dynamical symmetries for the subset of levels with zero values of SO(3) and SO(5) angular momenta \cite{Cejna07,Heinz06}, the other one was the parity-conserving Lipkin model \cite{Sinde17}.
The latter case is exemplified in Fig.\,\ref{epka}.
In both these systems the associated {\QPT} is of the second order and the subsequent {\ESQPT} is of the $(f,r)=(1,1)$ type. 
A more general analysis of the distribution of exceptional points for {\ESQPTs} of other types still needs to be performed.

The above considerations related to eigenstate variations across the {\ESQPT} critical energy bring us to the question concerning the possibility to unambiguously identify the quantum phases on both sides of the {\ESQPT}.
It is difficult and probably unfeasible in general cases to distinguish these phases by some physically meaningful \uvo{order parameters} following from some universal principles. 
These would be associated with observables whose expectation values $\ave{\hat{A}}_i$, or densities $\overline{A}(\lambda,E)$, are zero on one side of the {\ESQPT} and non-zero on the other.
Nevertheless, it is true that in numerous cases such order parameters exist.
This is so in the models in which the {\ESQPT} generalizes the {\QPT} of a continuous type by extending the phenomenon of spontaneous symmetry breaking from the ground state to individual excited states \cite{Puebl13,Puebl15}.
Consider as an example the {\ESQPT} linked to the local maximum of a symmetric $f\I{=}1$ double-well potential in coordinate~$\hat{q}$.
It is clear that the quantum phase with lower energy spontaneously breaks the parity with respect to the inversion of $\hat{q}$ (below the barrier the parity doublets become nearly degenerate in the semiclassical limit), so the order parameter can obviously be associated with $\hat{q}$ (we get expectation values ${\ave{\hat{q}}_i=0}$ or $\neq 0$, respectively, for states above or below the barrier).
This picture can be generalized to $f\I{>}1$ systems with a second-order ground-state {\QPT} whose critical point is associated with a degenerate minimum of the classical-limit Hamiltonian.
Such order parameters can indeed be introduced in the Dicke-like models, as well as in some of the quasispin and boson models discussed above.

However, the spontaneous symmetry breaking mechanism does not characterize an arbitrary {\ESQPT} in any system.
It fails even in some systems with a continuous ground-state {\QPT}, like in the complex quantum phase diagram of the extended (${0<\delta<1}$) Dicke model with multiple {\ESQPTs} (see Fig.\,\ref{dede}), or in the ${M=2j}$ subspace of the ${\delta=0}$ model [see Fig.\,\ref{bunch}(a)].
The systems that most commonly rule out the description in terms of the spontaneous symmetry breaking are those with the first-order ground-state {\QPT}.  
In such systems, a universal order parameter describing the associated {\ESQPTs} cannot be determined, but the distinction of quantum phases can still be based on differences in smoothed energy dependencies of diverse observables. 
The {\ESQPTs} are unambiguously identified by non-analyticities of these dependencies, as exemplified in Fig.\,\ref{liporder}.
Other examples of {\ESQPT} singularities in various expectation values can be found in a number of references, see e.g. Refs.\,\cite{Peres11b,Brand13,Puebl13,Basta14b,Engel15,Puebl16,Lobez16,Basta17,Kloc17b,Peres17,Zhu19} for the Dicke and related models, \cite{Basti14b,Kopyl15b,Engel15,Santo16} for the Lipkin model, \cite{Lares11,Lares13} for the vibron model, \cite{Macek14,Zhang16} for the nuclear interacting boson model, and many others.

\subsection{Oscillatory level density and finite-size effects in coupled systems}
\label{Fisie}

So far we have focused on the smoothed level densities, flows and energy densities of observables.
However, all these quantities have also the oscillatory components, which sometimes play important roles in the description of {\ESQPTs}.
The semiclassical approximation of the oscillatory level density $\widetilde{\rho}(E)=\rho(E)\!-\!\overline{\rho}(E)$ is expressed via so-called {\em trace formulas}, taking into account periodic orbits generated by the classical-limit Hamiltonian of the system \cite{Haake10}.
Returning to the notation in which the dependence on $\lambda$ is implicit, these formulas have the following general form:
\begin{equation}
\widetilde{\rho}(E)\approx\frac{1}{\pi\hbar}\sum_{\vecb{o}}\sum_{\iota=1}^{\infty}\frac{T_{\vecb{o}}(E)}{C_{\vecb{o}\iota}(E)}\, \cos\left(\frac{\iota I_{\vecb{o}}(E)}{\hbar}-\varphi_{\vecb{o}\iota}\right)
\label{trace}.
\end{equation}
Here the sums run over all primitive periodic orbits ${\vecb{o}\equiv\{\vecb{x}(t)\}_{t=0}^{T_{\vecb{o}}}}$ at energy $E$ with period $T_{\vecb{o}}(E)$ and their multiple repetitions $\iota=1, 2,...$, the quantity $I_{\vecb{o}}(E)=\oint_{\vecb{o}}\vecb{p}\cdot d\vecb{q}$ represents an action along the primitive orbit, and $\varphi_{\vecb{o}\iota}$ is a phase shift.
The form of quantity $C_{\vecb{o}\iota}(E)$ depends on whether the system is chaotic or integrable, the resulting formulas being called after Gutzwiller \cite{Gutzw71} or Berry and Tabor \cite{Berry76}, respectively, for an overview see Ref.\,\cite{Haake10}.
It expresses the degree of \uvo{stability} of the given orbit, in the sense of a chaotic or integrable system, and ensures that more stable orbits give larger contributions to $\widetilde{\rho}(E)$ than less stable ones.

By its definition, the oscillatory level density $\widetilde{\rho}(E)$ has zero mean, expressing only positive and negative deviations of the real density from the average represented by $\overline{\rho}(E)$.   
The energy scale of these fluctuations for each periodic orbit $(\vecb{o},\iota)$ is determined by the immediate slope of the function inside the cosine term of Eq.\,\eqref{trace}.
The energy period $\Delta E_{\vecb{o}\iota}\approx 2\pi\hbar/(\iota\frac{d}{dE}I_{\vecb{o}}(E))$ decreases to zero in the classical limit $\hbar\to 0$, so the classical oscillatory level density is an infinitely fast fluctuating function which gets washed out by smoothening over an infinitesimally narrow energy interval.
Hence in the limit $\hbar\to 0$, equivalent to $\aleph\to\infty$, only the smooth component $\overline{\rho}(E)$ remains relevant.
However, the oscillatory component $\widetilde{\rho}(E)$ is important in finite-size cases, and as discussed below, can produce even stronger finite-size precursors of {\ESQPTs} than visible for the given value of $\aleph$ in the smooth density $\overline{\rho}(E)$. 
These effects arise in systems composed of two or more subsystems that are fully or partially separable.

Consider a fully {\em separable system\/} with the Hilbert space ${\Hi=\Hi_1\otimes\Hi_2}$ composed of subspaces $\Hi_k$ corresponding to the $k=1$ and 2 subsystems, and with a quantum Hamiltonian ${\hat{H}=\hat{H}_1+\hat{H}_2}$ (more precisely ${\hat{H}=\hat{H}_1\otimes\hat{I}_2+\hat{I}_1\otimes\hat{H}_2}$, where $\hat{H}_{k}$ and $\hat{I}_{k}$ are the Hamiltonian and the unit operator, respectively, in $\Hi_k$).
Also the classical Hamiltonian is a sum of the respective parts ${H(\vecb{x})=H_1(\vecb{x}_1)+H_2(\vecb{x}_2)}$ in a $2f$-dimensional phase space written as the Cartesian product $\Ph=\Ph_1\I{\times}\Ph_2$ of phase spaces of both subsystems, with the overall {\DoF}s being sum of the subsystem's {\DoF}s, $f=f_1\I{+}f_2$.
Due to additivity of energies, the total level density $\rho(E)$ of the whole system is a convolution of level densities $\rho_1(E_1)$ and $\rho_2(E_2)$ of the subsystems, namely
\begin{equation}
\rho(E)=\int\!\!\int dE_1 dE_2\,\rho_1(E_1)\rho_2(E_2)\delta(E_1\I{+}E_2\I{-}E)
\label{sepade0}.
\end{equation}
For the smoothed and oscillatory components of the total level density we obtain \cite{Stran15}:
\begin{eqnarray}
\hspace{-22mm}
\overline{\rho}(E)&=&\!\!\!\int\!\! dE_1\, \overline{\rho}_1(E_1)\overline{\rho}_2(E\I{-}E_1),
\label{sepade}\\
\hspace{-22mm}
\widetilde{\rho}(E)&=&\!\!\!
\int\!\! dE_1\, \overline{\rho}_1(E_1)\widetilde{\rho}_2(E\I{-}E_1)
\I{+}\!
\int\!\! dE_1\, \widetilde{\rho}_1(E_1)\overline{\rho}_2(E\I{-}E_1)
\I{+}\!
\int\!\! dE_1\,\widetilde{\rho}_1(E_1)\widetilde{\rho}_2(E\I{-}E_1),
\nonumber
\end{eqnarray}
where $\overline{\rho}_k(E)$ and $\widetilde{\rho}_k(E)$ are smooth and oscillatory level densities of the $k$th subsystem.
Note that all integrals in Eq.\,\eqref{sepade} are taken from the minimal energy $E_{1{\rm min}}$ of the subsystem 1 up to $(E\I{-}E_{2{\rm min}})$ given by the minimal energy $E_{2{\rm min}}$ of subsystem 2.

To define size parameters of separable systems, we refine the general treatment described in Sec.\,\ref{Size}. 
We chose a characteristic energy scale $\mathcal{E}$ and its \uvo{typical} partitioning ${\mathcal{E}=\mathcal{E}_1+\mathcal{E}_2}$ to the subsystems.
This is of course ambiguous, but we usually choose nearly equal splitting $\mathcal{E}_1\I{\sim}\frac{1}{2}\mathcal{E}\I{\sim}\mathcal{E}_2$.
Following the formula \eqref{alef} and employing miltiplicability of volumes $\Omega_k(E_k)$ in the $\Ph_1\I{\times}\Ph_2$ phase space, we get:
\begin{equation}
\aleph^f\equiv\frac{\Omega(\mathcal{E})}{\hbar^{f}}=\frac{\Omega_1(\mathcal{E}_1)}{\hbar^{f_1}}\,\frac{\Omega_2(\mathcal{E}_2)}{\hbar^{f_2}}\equiv\aleph_1^{f_1}\,\aleph_2^{f_2}
\label{sizesep}.
\end{equation}
Here $\aleph$ is the size parameter of the whole system at energy scale $\mathcal{E}$, and $\aleph_1$ and $\aleph_2$ are size parameters of the subsystems at scales $\mathcal{E}_1$ and $\mathcal{E}_2$.
To see how it works, let us look for example on the Dicke model \eqref{HDic} with subsystems 1 and 2 identified with the bosonic field and the ensemble of $N$ atoms, respectively. 
Consider first the balanced regime (Sec.\,\ref{EDM1}) with $\omega\I{\sim}\omega_0$ represented by a common average $\bar{\omega}$.
The characteristic energy can be chosen as twice the saturation energy of the atomic subsystem in absence of interaction, $\mathcal{E}=2N\omega_0$ (this energy range contains all {\ESQPTs} described in Sec.\,\ref{EDM2}), and the equal partitioning ${\mathcal{E}_1=\mathcal{E}_2=N\omega_0\approx N\bar{\omega}}$ yields equal size parameters $\aleph_1\I{=}\aleph_2\I{=}\aleph\I{=}N$.
On the other hand, the Rabi regime in the strongly imbalanced case with ${R=\omega_0/\omega\gg 1}$ (Sec.\,\ref{EDM3}) results in different values.
The same choice of energy scales $\mathcal{E}$, $\mathcal{E}_1$ and $\mathcal{E}_2$ as above leads now to ${\aleph_1\I{=}NR}$ (field) and ${\aleph_2\I{=}N}$ (atoms), while the total size parameter is ${\aleph=N\sqrt{R}}$. 

Such imbalanced settings of coupled systems are in our main focus here.
Suppose without loss of generality that $\aleph_1\gg\aleph_2$.
This means that subsystem 1 is much closer to classicality, so its energy levels $E_{1i}$ are much denser than energy levels $E_{2i}$ of the subsystem 2 and the oscillatory level density $\widetilde{\rho}_1(E_1)$ oscillates much \uvo{faster} than $\widetilde{\rho}_2(E_2)$.
Under these conditions, the second and third integrals in the second line of Eq.\,\eqref{sepade} can be neglected.
For the sake of concreteness we assume that: (i) a non-degenerate stationary point with index $r_1$ in the phase space of subsystem~1 induces the corresponding {\ESQPT} non-analyticity in $\frac{\partial^{f_1-1}}{\partial E_1^{f_1-1}}\overline{\rho}_1(E_1)$ at energy $E_{1{\rm c}}\I{>}E_{1{\rm min}}$, and (ii) the only stationary point of subsystem~2 is a non-degenerate global minimum (index $r_2\I{=}0$) at $E_{2{\rm min}}$, so $\overline{\rho}_2(E_2)$ has no {\ESQPT} singularity.
Then the whole system exhibits at the total energy ${E=E_{\rm c}=E_{1{\rm c}}+E_{2{\rm min}}}$ the same type of non-analyticity as subsystem~1 but in the derivative $\frac{\partial^{f-1}}{\partial E^{f-1}}\overline{\rho}(E)$.
This conclusion follows from the analysis of the first line in Eq.\,\eqref{sepade} as well as from the properties of the corresponding stationary point with index ${r=r_1+r_2=r_1}$ in the coupled phase space $\Ph_1\times\Ph_2$.
The singularity in $\overline{\rho}(E)$ is shifted up to the $(f_1\I{+}f_2\I{-}1)$st derivative, consistently with the larger {\DoF} number of the whole system, so it is weaker than that in $\overline{\rho}_1(E_1)$.
However, summing both the smooth and oscillatory components of the level density in Eq.\,\eqref{sepade} yields (neglecting the integrals involving quickly oscillating functions) 
\begin{equation}
\rho(E)\approx\sum_{i}\overline{\rho}_1(E-E_{2i})
\label{sumde},
\end{equation}
where we see a cumulative repetition of the level density of subsystem~1 above each discrete level $E_{i2}$ of subsystem~2.
The density in Eq.\,\eqref{sumde} shows a multiple occurrence of the singularity in the $(f_1\I{-}1)$st derivative at energies $E=E_{1{\rm c}}\I{+}E_{2i}$ for ${i=1,2,...}$.

\begin{figure}[t!]
\begin{flushright}
\includegraphics[width=0.88\textwidth]{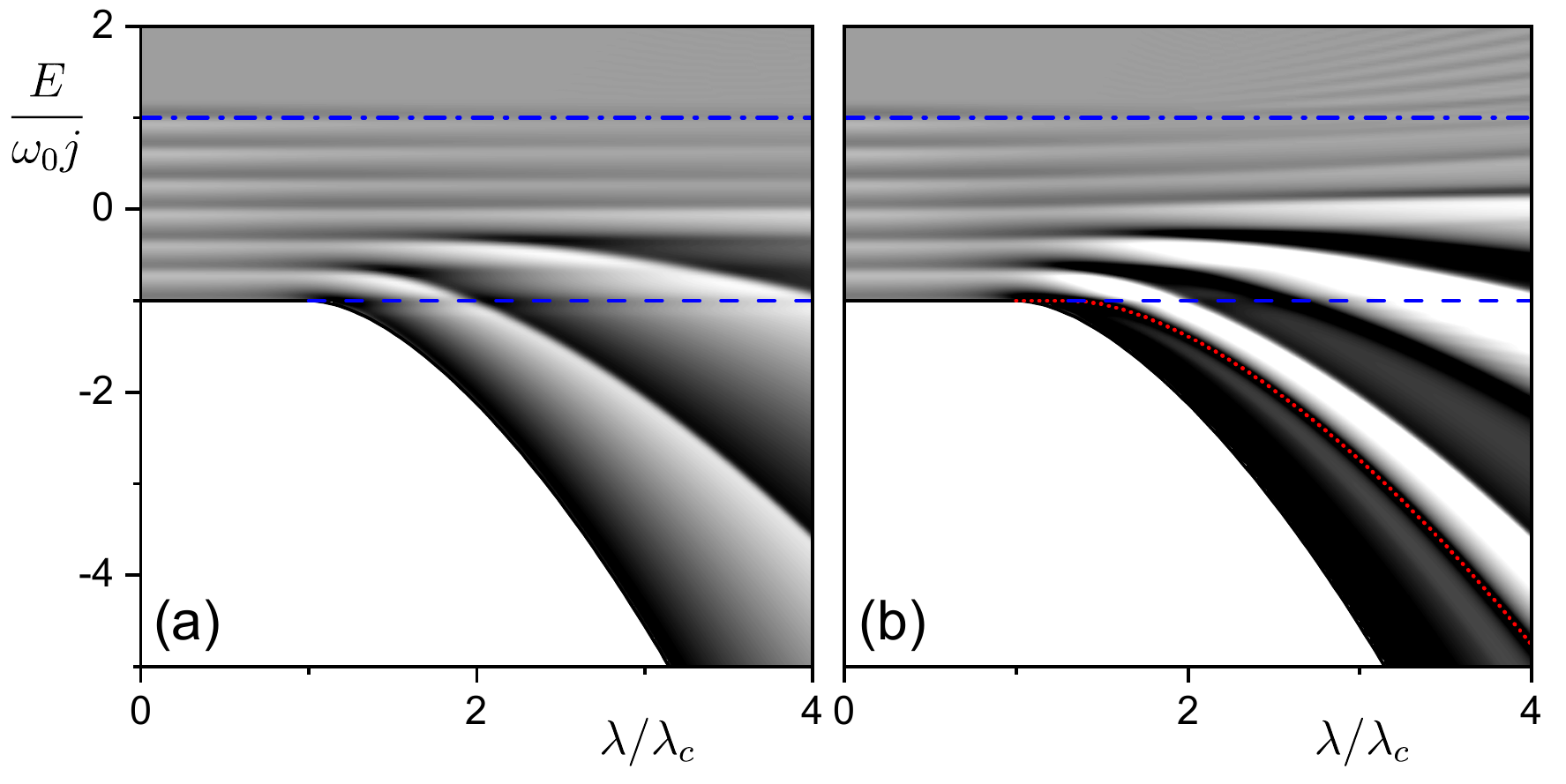}
\end{flushright}
\caption{Oscillatory component of the level density of the extended Dicke model \eqref{HDic} in a strongly imbalanced regime with $\omega\I{=}1$, $\omega_0\I{=}30$ for $N\I{=}2j\I{=}6$. 
Panel (a) corresponds to $\delta\I{=}1$, panel (b) to $\delta\I{=}0.132$ (cf.\,Fig.\,\ref{dede}). 
We know from Sec.\,\ref{EDM3} that the full spectrum is approximated by a pile-up of field spectra corresponding to individual quasispin projections.
This affects the oscillatory level density, which is calculated as ${\widetilde{\rho}(E)=\rho_{\sigma}(E)-\overline{\rho}(E)}$, where $\rho_{\sigma}(E)$ is the exact density folded with a Gaussian of width $\sigma\I{=}0.04$ and $\overline{\rho}(E)$ is the semiclassical density \eqref{smole}.
Dashed and dotted lines, respectively, mark {\ESQPT} borderlines associated with jumps and peaks of $\frac{\partial}{\partial E}\overline{\rho}(E)$.
Similar structures appear repeatedly in $\widetilde{\rho}(E)$, whose positive (negative) values correspond to dark (light) shades. 
}
\label{Rab}
\end{figure}

Similar analyses as presented above would apply also in other situations, e.g., in cases with multiple and/or degenerate stationary points in $\Ph_1$, or with additional stationary points in $\Ph_2$. 
The general result is an enhancement of {\ESQPT} signatures in strongly imbalanced separable systems.
These effects represent only finite-size precursors (subsystem~2 is small) and would disappear in the full semiclassical limit.  
It is worth noting that the effect is contained in the first term of the second-line formula in Eq.\,\eqref{sepade}, which is on the contrary neglected for balanced systems.
As shown in Ref.\,\cite{Stran15}, this term is not included in the quasiclassical formula \eqref{trace} as the corresponding periodic orbits are scarce.

Exact separability implies full integrability of the system.
However, it has been shown \cite{Stran15} that even an approximate separability can produce effects very close to those described above, although in less perfect forms.
Islands of separable classical orbits, i.e., orbits composed of independent components corresponding to different {\DoFs}, can coexist with the others in the phase space of partially (even strongly) chaotic systems.
If one of the {\DoFs} has an {\ESQPT}, the presence of such islands---including maybe infinitesimally small ones---leads to appearance of the above-described {\ESQPT} precursors.
Interestingly, if subsystem~1 shows more stationary points above the global minimum, the orbits located in the phase-space domains near these points generate selectively only the precursors of the corresponding {\ESQPT} \cite{Stran15}.

There exist numerous examples of separable or partially separable many-body systems with the above type of {\ESQPT} related finite-size effects.
We have already described the Rabi regime of the atom-field system \cite{Puebl16,Relan16b} (Sec.\,\ref{EDM3}).
Oscillatory structures in the spectrum of this system caused by imbalance of elementary excitation energies of the atomic and field subsystems are shown in Fig.\,\ref{Rab}.
We have also mentioned the two-spin Gaudin model \cite{Relan16} (Sec.\,\ref{Quasi}) and the interacting boson-fermion models (Sec.\,\ref{Ibms}).
Recently, a neat example of separability-induced {\ESQPT} effects was presented in the U(6) interacting boson model along the first-order {\QPT} path between U(5) and SU(3) dynamical symmetries \cite{Macek19}.
Indications of these effects can be observed in the Peres lattices shown in panels (b) and (c) of Fig.\,\ref{ibm1st}.

\section{Thermodynamic consequences}
\label{Thermo}

Critical effects in excited states must show up in thermodynamic properties of the system.
At the first sight, there seems to be a direct link between the {\ESQPTs} and the {\TPTs}, i.e., standard phase transitions induced by thermal excitations.
However, it turns out that although these two types of criticality are related, they represent distinct phenomena which can be applied only in mutually incompatible contexts. 
In this section, we overview the presently known thermodynamic aspects of {\ESQPTs}, noting that research in this field still needs to be continued.
In all formulas below we set the Boltzmann constant $k_{\rm B}$ to unity.

\subsection{Thermodynamics of finite quantum systems}

In spite of the early application of quantum mechanics in thermodynamics through Einsten's 1905 solution of the specific heat problem, fundamental reconsideration of the laws of thermodynamics in the quantum framework became a subject of systematic research only relatively recently, see e.g. Ref.\,\cite{Binde19}.
This is related to the fact that the present experimental quantum simulation techniques make quantum effects in thermodynamics accessible to experimentation and potentially applicable in quantum technologies \cite{Gardi14}.
Quantum effects, including {\QPTs} and {\ESQPTs}, can serve as resources for some particular functionalities of the corresponding nanostructures, but they can also generate obstacles.
Let us mention, for example, a recent analysis \cite{Kloc19} of a quantum heat engine based on the quasispin system, in which quantum criticality (both {\QPT} and {\ESQPT}) played a negative role for the engine function. 

We start our discussion with some general relations between thermodynamic quantities and properties of quantum energy spectra $E_i(\lambda)$.
The Hamiltonian control parameter $\lambda$ appears explicitly in all expressions.
In {\em canonical thermodynamics}, we assume that the quantum system exchanges its energy with an external heat bath, which leads to the familiar form of the occupation probabilities of individual levels, 
\begin{equation}
\hspace{-15mm}
p_i(\lambda,T)=\frac{e^{-E_i(\lambda)/T}}{Z(\lambda,T)},\quad Z(\lambda,T)=\sum_i e^{-E_i(\lambda)/T}=\int dE\,\rho(\lambda,E)e^{-E/T},
\label{canon}
\end{equation}
where $Z(\lambda,T)$ is the canonical partition function.
The thermal distribution of energy of the system $w_T(\lambda,E)=\sum_i\delta\bigl(E\I{-}E_i(\lambda)\bigr)p_i(\lambda,T)$ 
then in a smoothed form reads
\begin{equation}
\overline{w}_T(\lambda,E)=\overline{\rho}(\lambda,E)\frac{e^{-E/T}}{Z(\lambda,T)}
\label{canE}.
\end{equation}

Consider a quantity ${A(\lambda,E)=\sum_i A_i(\lambda)\,\delta(E\I{-}E_i(\lambda))}$ with values $A_i(\lambda)$ assigned to individual energy levels.
Here, in contrast to the previous considerations [cf.\,Eqs.\,\eqref{Asmodep},\,\eqref{clAve},\,\eqref{clave}], we do not aim at standard quantum mechanical observables $\hat{A}$, but rather associate $A_i(\lambda)$ with some definite characteristics of individual energy levels, e.g., with their slopes ${\frac{d}{d\lambda}E_i(\lambda)\equiv E'_i(\lambda)}$ or curvatures ${\frac{d^2}{d\lambda^2}E_i(\lambda)\equiv E''_i(\lambda)}$.
A~smoothed energy density reads 
\begin{equation}
\overline{A}(\lambda,E)=\sum_i A_i(\lambda)\,\overline{\delta}\bigl(E\I{-}E_i(\lambda)\bigr)
\equiv {\overline a}(\lambda,E)\,\overline{\rho}(\lambda,E),
\label{Adepka}
\end{equation}
where we introduce also a smoothed single-level contribution ${\overline a}(\lambda,E)$, which represents a local average of values $A_i(\lambda)$ assigned to individual levels near energy~$E$ [cf.\,Eq.\,\eqref{clave}].
For the slopes $E'_i(\lambda)$, e.g., the overall density $\overline{A}(\lambda,E)$ is the smoothed flow $\overline{\jmath}(\lambda,E)$ from Sec.\,\ref{Leflo}, while the single-level contribution ${\overline a}(\lambda,E)$ is what we called the smoothed flow rate $\overline{\phi}(\lambda,E)$.
We also introduce thermal characteristics
\begin{equation}
\hspace{-15mm}
\ave{A(\lambda)}_T=\sum_i p_i(\lambda,T)\,A_i(\lambda)\,,\quad
\dis{A(\lambda)^2}_T={\ave{A(\lambda)^2}_T-\ave{A(\lambda)}^2_T}\,,
\label{Tavevar}
\end{equation}
which denote the thermal average of quantity $A(\lambda,E)$ and the corresponding variance at temperature~$T$.
The thermal average can be approximated through the smoothed single-level contribution ${\overline a}(\lambda,E)$ and the thermal energy distribution~\eqref{canE}:
\begin{equation}
\ave{A(\lambda)}_T\approx\int dE\ \overline{a}(\lambda,E)\,\overline{w}_T(E).
\label{Tave}
\end{equation}

The description of {\TPTs} in standard thermodynamics is based on the behavior of the free energy $F=\ave{E}_T-TS$, where  $S=-\sum_ip_i\ln p_i$ is the entropy associated with an arbitrary occupation probabilities $p_i$ of energy levels.
Minimization of the free energy (i.e., maximization of the entropy with $\ave{E}_T$ fixed) over all sets $\{p_i\}$ leads to the canonical occupation probabilities \eqref{canon}.
The equilibrium value of the free energy reads ${F(\lambda,T)=-T\ln Z(\lambda,T)}$.
Its temperature derivatives are given by
\begin{equation}
\hspace{-17mm}
\frac{\partial}{\partial T}F(\lambda,T)=-S(\lambda,T),\qquad
\frac{\partial^2}{\partial T^2}F(\lambda,T)=-\frac{\bidis{E(\lambda)^2}_{T}}{T^3}=-\frac{C(\lambda,T)}{T},
\label{dFdT}
\end{equation}
where ${S(\lambda,T)=\ave{E(\lambda)}_{T}/T+\ln Z(\lambda,T)}$ is the canonical entropy and $C(\lambda,T)$ the specific heat.
For derivatives of the free energy with respect to $\lambda$ we obtain relations 
\begin{equation}
\hspace{-17mm}
\frac{\partial}{\partial\lambda}F(\lambda,T)=\ave{E'(\lambda)}_{T},\quad
\frac{\partial^2}{\partial\lambda^2}F(\lambda,T)=\ave{E''(\lambda)}_{T}-\frac{\bidis{E'(\lambda)^{2}}_{T}}{T},
\label{dFdla}
\end{equation}
and a similar one for $\frac{\partial^2}{\partial\lambda\partial T}F(\lambda,T)$ \cite{Cejna08}. 

Formulas in Eqs.\,\eqref{dFdT} and \eqref{dFdla} interconnect standard thermodynamic functions describing the thermally equilibrated system and the thermal averages of quantities of the type \eqref{Adepka} characterizing the spectrum of energy levels.
This may form a bridge between thermodynamic and quantum phases of the system.
The strongest link can be expected near the thermodynamic limit, when the canonical distribution of energy \eqref{canE} is supposed to form a peak centered at energy  ${E\I{\approx}\ave{E(\lambda)}_{T}}$ whose width ${\Delta E\I{\approx}\sqrt{ \dis{E(\lambda)^2}_{T}}}$ satisfies the condition ${\Delta E\I{\ll}E}$.
This activates the microcanonical approximation, in which 
the canonical entropy is given by the original Boltzmann definition ${S(\lambda,T)\approx\ln\bigl[\overline{\rho}(\lambda,E)\Delta E\bigr]}$ and the thermal averages by the formula ${\ave{A(\lambda)}_T\approx\overline{a}\bigl(\lambda,E\I{=}\ave{E(\lambda)}_T\bigr)}$.

The microcanonical approximation of Eqs.\,\eqref{dFdT} and \eqref{dFdla} can be applied to phase transitions.
Consider a {\TPT} at a critical temperature $T\I{=}T_{\rm c}(\lambda)$ forming a phase separating curve in the ${\lambda\times T}$ plane and suppose that this curve is not parallel with either of the $\lambda$ and $T$ directions.
If the {\TPT} is of the first order, it shows up as a discontinuity in the the first derivatives of the free energy. 
This is equivalent to a jump in $\ln\overline{\rho}(\lambda,E)$ if the phase separatrix is crossed in the $T$ direction, see the first formula in Eq.\,\eqref{dFdT}, and to a jump in the flow rate $\ave{E'(\lambda)}_T$ if the separatrix is crossed in the $\lambda$ direction, see the first formula in Eq.\,\eqref{dFdla}.
These look like signatures of a generic {\ESQPT} in a $f\I{=}1$ system (see Secs.\,\ref{Leden} and \ref{Leflo}).
On the other hand, a second-order {\TPT} generates a discontinuity in the second derivatives of the free energy, that is in the first derivative of $\ln\overline{\rho}(\lambda,E)$ and $\ave{E'(\lambda)}_T$, resembling a generic {\ESQPT} for $f\I{=}2$. 
The second formulas in Eqs.\,\eqref{dFdT} and \eqref{dFdla} moreover imply singular evolutions of quantities $\dis{E(\lambda)^2}_T$, $\ave{E''(\lambda)}_{T}$, and $\dis{E'(\lambda)^{2}}_{T}$.

We have to quickly declare that part of the suggestions in the previous paragraph is misleading.
It is true that Eqs.\,\eqref{dFdT} and \eqref{dFdla} connect thermodynamic and spectral properties, and this is indeed interesting for systems that show {\TPTs} rooted in canonical thermodynamics.
However, the opposite implication---that each {\ESQPT} in a moderate-$f$ system leads to a {\TPT} in its canonical description---would be incorrect.
This is because the canonical thermodynamics {\em does not\/} converge to the microcanonical thermodynamics in the infinite-size limit ${\aleph\to\infty}$ of systems with finite {\DoF} numbers, i.e., in the systems for which the {\ESQPTs} are relevant.
Therefore, the canonical averages in the above relations cannot be replaced by the microcanonical ones and the suggested equivalence of canonical {\TPTs} and {\ESQPTs} breaks down. 

Finite-$f$ systems violate standard thermodynamic rules in many ways.
We immediately see that if the Hamiltonian is scaled so that the total energy is extensive in the size parameter $\aleph$, the average thermal energy per {\DoF} diverges with ${\aleph\to\infty}$.
This is in conflict with the equipartition theorem.
The discrepancies between canonical and microcanonical pictures in the infinite-size limit of finite-$f$ systems can be linked to anomalous forms of the canonical thermal energy distribution \eqref{canE}.
Indeed, searching for points satisfying ${\frac{\partial}{\partial E}\overline{w}_T(\lambda,E)\I{=}0}$, we obtain the equation
\begin{equation}
\frac{\partial}{\partial E}\ln\overline{\rho}(\lambda,E)=\frac{1}{T}
\label{microT}. 
\end{equation}
This implies that for the distribution $\overline{w}_T(\lambda,E)$ having a single maximum at energy increasing with $T$ we need that the microcanonical entropy $\ln\overline{\rho}(\lambda,E)$ is a monotonously increasing concave function of energy.
For the energy distribution satisfying ${\Delta E/E\I{\ll}1}$ we moreover need that ${C(\lambda,T)\I{\ll}\ave{E(\lambda)}^2/T^2}$.
All these conditions can be broken in finite-$f$ systems, and the most flagrant violation is observed in systems with {\ESQPTs}.

Non-standard thermal properties of collective (fully connected) quantum systems, such as quasispin systems with infinite-range interactions constrained to a single-$j$ subspace of states, became a subject of recent study. 
For example, anomalously wide (\uvo{non-concentrating}) thermal distributions of various quantities and insufficiency of thermal averages to capture relevant thermal properties were discussed in Ref.\,\cite{Webst18}. 
Anomalous thermalization properties, involving some memory effects connected with apparently thermalized states prepared by different non-equilibrium procedures, were reported in Ref.\,\cite{Relan18}. 
Although the role of {\ESQPTs} in many of these phenomena is not decisive, they are present in the models used and participate in the results obtained.
Examples of anomalous relaxation of systems dynamically excited to the {\ESQPT} energy region will be further discussed in Sec.\,\ref{quench}.

\subsection{Microcanonical singularities}
\label{Micro}

Microcanonical thermodynamics, already mentioned above, represents a natural ground for studying thermal critical properties of finite quantum systems \cite{Franz04,Kastn07,Kastn08,Gross01,Dunke06,Brody07}.
It assumes that the system is isolated, having a fixed energy (uniformly distributed within a very narrow interval), and that its thermalization proceeds only via internal interactions.
The microcanonical description is based on the microcanonical entropy ${\mathcal{S}(\lambda,E)=\ln\overline{\rho}(\lambda,E)}$ (neglecting an arbitrary additive constant). 
The definition of the microcanonical temperature, ${\mathcal{T}^{-1}(\lambda,E)=\frac{\partial}{\partial E}\mathcal{S}(\lambda,E)}$, parallels the basic relation for canonical temperature, ${T^{-1}=\frac{\partial}{\partial U}S}$, where $S$ is the canonical entropy and $U$ the internal energy.
If the canonical thermal energy distribution $\overline{w}_T(\lambda,E)$ forms a single peak,  we see from Eq.\,\eqref{microT} that the microcanonical temperature $\mathcal{T}$ assigned to energy $E$ is the canonical temperature $T$ for which $E$ is the peak (most probable) energy.
Under such circumstances, the microcanonical and canonical pictures shall converge to each other with increasing size $\aleph$.

However, as indicated above, for systems with small {\DoF} numbers, and specifically for systems with {\ESQPTs}, the microcanonical and canonical pictures remain inconsistent even in the $\aleph\I{\to}\infty$ limit.
In such systems, the setting of the microcanonical thermodynamics via Eq.\,\eqref{microT} is even problematic as this equation may be completely undefined, or may yield for a fixed temperature $T$ either multiple energy solutions, or no solution at all.
The microcanonical picture is undefined if $\ln\overline{\rho}(\lambda,E)$ shows jumps or divergences, like in typical {\ESQPTs} systems with $f\I{=}1$.
Then the right-hand side of microcanonical equation \eqref{microT} does not exist.
The cases with multiple or no solutions of Eq.\,\eqref{microT} in wide energy intervals apply if $\frac{\partial}{\partial E}\ln\overline{\rho}(\lambda,E)$ is non-analytic, as typically happens in $f\I{=}2$ {\ESQPTs} systems.
This leads to non-analytic distributions $\overline{w}_T(\lambda,E)$ which sometimes takes bimodal, multimodal or other exotic forms \cite{Stran14}.
As discussed in Ref.\,\cite{Cejna17}, multiple branches of the microcanonical temperature in narrow energy intervals can arise even from non-analyticities in $\frac{\partial^2}{\partial E^2}\ln\overline{\rho}(\lambda,E)$, so in typical {\ESQPT} systems with $f\I{=}3$.

\begin{figure}[t!]
\begin{flushright}
\includegraphics[width=\textwidth]{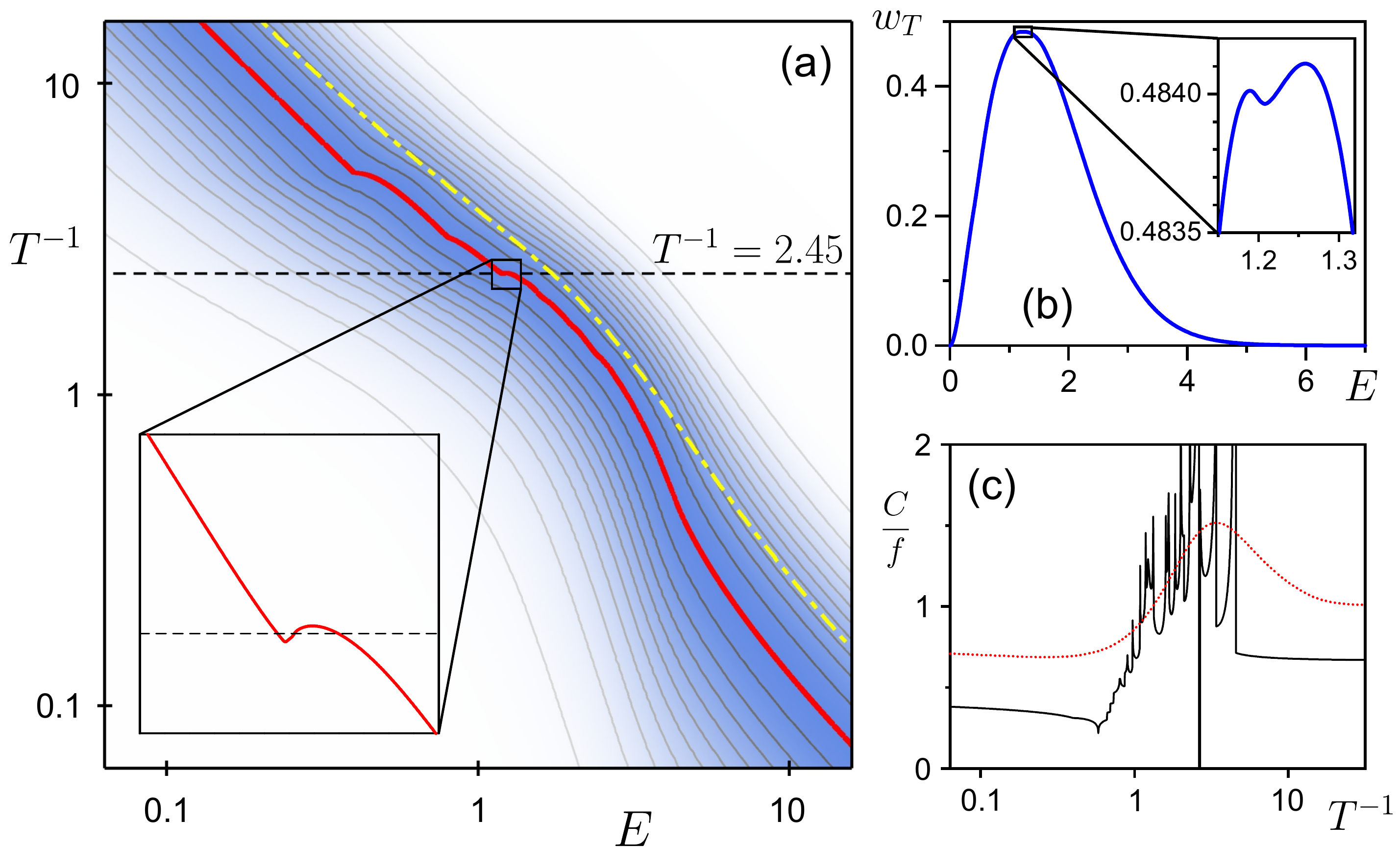}
\end{flushright}
\caption{Thermal properties of an $f=3$ system with a toy Hamiltonian $\hat{H}=\frac{1}{2}|\hat{\vecb{p}}|^2\I{-}2|\hat{\vecb{q}}|^2\I{+}\frac{1}{5}\hat{q}_1\I{+}\frac{2}{5}\hat{q}_2\I{+}\frac{3}{5}\hat{q}_3\I{+}\hat{q}_1^4\I{+}\hat{q}_2^4\I{+}\hat{q}_3^4$.
The system has {\ESQPTs} connected with 27 non-degenerate stationary points of the potential (cf.\,Fig.\,\ref{Badhair}). 
Panel~(a): The thermal energy distributions $\overline{w}_T(E)$ as a function of $E\times T^{-1}$ is visualized by the colored contour diagram; the darker band is the region with large values (the maximum normalized to unity).
The microcanonical and canonical caloric curves, respectively, are drawn by the thicker full line (red online) and the dotdashed line (yellow online).
Panel~(b): Thermal energy distribution at $T^{-1}\I{=}2.45$ with the region of bimodality (expanded in the inset) corresponding to the kink of the caloric curve expanded in the inset of panel (a).
Panel~(c): The microcanonical and canonical heat capacity (full and dotdashed line, respectively) as a function of $T^{-1}$.
Adapted from Ref.\,\cite{Cejna17}.
}
\label{Calor}
\end{figure}

The above-discussed effects can be visualized via so-called {\em caloric curves\/} in the ${E\times T}$ plane.
The canonical caloric curve displays the dependence of the canonical average energy $\ave{E(\lambda)}_T$ on  temperature $T$.
The microcanonical caloric curve is the dependence of $\most{E(\lambda)}_T$ on $T$, where $\most{E}_T$ stands for the most probable energy at temperature $T$ (maximum of the thermal energy distribution). 
The derivatives ${C(\lambda,T)\I{=}\frac{\partial}{\partial T}\ave{E(\lambda)}_T}$ and ${\mathcal{C}(\lambda,T)\I{=}\frac{\partial}{\partial T}\most{E(\lambda)}_T}$ represent the canonical and microcanonical {\em heat capacity}, respectively (in the microcanonical case, the dependence on $T$ shall be transformed to the corresponding dependence on energy~$[E]_T$).
Figure~\ref{Calor} shows examples of both canonical and microcanonical caloric curves and heat capacities, and the corresponding thermal energy distributions for an artificial $f\I{=}3$ system that has {\ESQPTs} associated with $3^f\I{=}27$ non-degenerate stationary points.
We observe in panel (c) that the microcanonical heat capacity has a chain of anomalous structures, associated with each {\ESQPT}, and is even negative and/or multivalued (undefined) in some narrow temperature intervals.
These structures result from the shape of the microcanonical caloric curve in panel (a) (transformed to the form with $E$ on the horizontal axis and $T^{-1}$ on the vertical axis) and reflect complex forms of the thermal energy distributions exemplified in panel (b).

The {\ESQPTs} in higher than second energy derivatives of $\ln\overline{\rho}(\lambda,E)$, which typically appear in systems with $f\I{>}3$, do not affect the existence and uniqueness of solutions of the microcanonical equation (this is however not guaranteed anyway).
The resulting non-analyticities of the microcanonical entropy and related thermodynamic functions at the microcanonical temperatures corresponding to the critical {\ESQPT} energies represent a special type of critical effects rooted exclusively in microcanonical thermodynamics.
They can have important consequences in the processes involving internal thermalization mechanisms in isolated systems with not too high values of $f$, but with increasing $f$ their significance fades away. 

Stationary points of the classical dynamics were nevertheless considered as a possible source of thermal critical behavior even for {\em standard thermodynamic systems\/} with ${f\sim\aleph\to\infty}$. 
This link was proposed in Ref.\,\cite{Franz04} and further investigated in Refs.\,\cite{Kastn07,Kastn08,Kastn08b,Kastn08c,Kastn09,Caset09}.
What makes these analyses relevant, in spite of expected weakening of effects of stationary points with increasing $f$, is the fact that the total number of stationary points of the system grows exponentially with~$f$.
This can be readily proven for a separable system with classical Hamiltonian $H(\vecb{q},\vecb{p})\I{=}\sum_{k=1}^f H_k(q_k,p_k)$.
If $m_k$ is the number of stationary points of each one-dimensional component $H_k$, the total number of stationary points is  $M\I{=}\prod_{k=1}^f m_k\I{=}\overline{m}^f$ (where $\overline{m}$ is the geometric average of $m_k$), and qualitatively the same conclusion holds also for non-separable systems.
Hence some infinitesimal intervals of scaled energy $E/\aleph$ for a large value of $f$ may exhibit very large accumulation of stationary points. 
Moreover, among the quickly diverging total number of stationary points there may be a sufficiently large subset of points that become  asymptotically flat (degenerate) in the $\aleph\I{\to}\infty$ limit.  
 
In these studies, the Hamiltonian was assumed to have the form \eqref{Hnorm} with a quadratic kinetic term and the potential energy function $V(\vecb{q})$ depending on an $f$-dimensional coordinate vector $\vecb{q}$.
The collection of values $\vecb{q}$ accessible for the system at a given energy, i.e., the coordinates for which ${V(\vecb{q})\leq E}$, forms the available configuration space.
In Ref.\,\cite{Franz04}, the occurrence of a {\TPT} at a certain temperature $T_{\rm c}$ is related to the change of topology of the available configuration space at the corresponding scaled microcanonical energy $E_{\rm c}/\aleph$.
These changes are indeed connected with stationary points of $V(\vecb{q})$.
However, because of the above-discussed exponential abundance of such non-analyticities and because of their typical effects appearing only in very high derivatives of thermodynamic functions, such \uvo{critical phenomena} do not have much significance in ${f\sim\aleph}$ systems.

The way how stationary points in the configuration space of an $f\I{\sim}\aleph$ system may give rise to a real {\TPT} affecting low derivatives of the thermodynamic functions in the infinite-size limit was explained in Refs.\,\cite{Kastn08b,Kastn08c,Kastn09}.
Roughly speaking, the proposed mechanism relies on a high enough density of degenerate stationary points in an infinitesimal interval of scaled energy near $E_{\rm c}/\aleph$, which is the microcanonical energy corresponding to the critical temperature $T_{\rm c}$.  
Necessary (but not sufficient) conditions for a {\TPT} to occur at this temperature read approximately as follows: (i) faster than exponential growth of the overall number of stationary points in the interval with increasing ${f\sim\aleph}$, and (ii) vanishing infinite-size limit of the quantity $|\overline{\mathcal{H}}(E_{\rm c})|^{1/\aleph}$, where $\overline{\mathcal{H}}(E_{\rm c})$ is a local average of ${{\rm Det}\,\partial_k\partial_l H(\vecb{q}^{(i)},\vecb{p}\I{=}0)}$ (determinant of the Hessian matrix of the Hamiltonian) over all stationary points $\vecb{q}^{(i)}$ in an infinitesimal vicinity of $E_{\rm c}$.
For the exact formulation of these conditions and for illustrative examples of systems where they can be applied see the quoted references.

\subsection{Effects of excited-state singularities in canonical thermodynamics}
\label{Esther}

Let us return to {\ESQPT} systems with moderate numbers of {\DoFs}.
In such systems, in contrast to the microcanonical case, the canonical thermodynamic functions must be fully analytic even in the ${\aleph\to\infty}$ limit. 
We have already seen that the canonical heat capacity of an {\ESQPT} system in Fig.\,\ref{Calor} shows no singularities, unlike its microcanonical counterpart.
As follows from numerical examples of Ref.\,\cite{Cejna17}, the canonical heat capacity of systems with {\ESQPTs} is often a non-monotonous function of temperature, with maxima and minima located in the temperature regions that yield the average thermal energy near the {\ESQPT} borderlines.
However, such structures may appear also for Hamiltonians which are close to but not really in the {\ESQPT} control parameter domains.
The link of {\ESQPTs} to these anomalies is therefore rather ambiguous.

Detailed {\em comparative studies\/} of canonical and microcanonical thermodynamic treatments, searching for mutual relations between {\TPTs} and {\ESQPTs}, were performed in the extended Dicke model from Sec.\,\ref{Open} \cite{Basta16,Peres17}.
Canonical thermodynamics of the superradiance models disclosing their {\TPT} was presented already in Refs.\,\cite{Wang73,Hepp73}, but at that time the {\ESQPTs} were not known.
A crucial assumption of all these analyses is that they consider the  full $2^N$ dimensional Hilbert space $\Hi$ including all subspaces $\Hi_{j}^{(s)}$ with the quasispin quantum number $j$ taking integer or half-integer values between $j_{\rm min}\I{=}0$ or $\frac{1}{2}$ and $j_{\rm max}\I{=}\frac{1}{2}N$ and $s=1,...,\nu_{j}$ counting all replicas of the given-$j$ subspace within $\Hi$  (see Sec.\,\ref{Quasi}).
The number of relevant {\DoFs} is then $f=N\I{+}1$ and the model shows standard behavior in the true thermodynamic limit $N\to\infty$ (with converging canonical and microcanonical results).

To describe thermodynamic properties of the extended Dicke model with Hamiltonian $\hat{H}(\lambda,\delta)$ from Eq.\,\eqref{HDic}, we use a method \cite{Keeli14} in which the partition function is evaluated by tracing the exponential of the Hamiltonian over the atomic subsystem only, while the field subsystem is kept in the Glauber coherent state $\ket{\alpha}\propto\exp(\alpha\hat{b}^{\dag})\ket{0}$ characterized by a parameter ${\alpha\in{\mathbb C}}$ such that $|\alpha|^2$ is an average number of photons in the cavity.
Thus we have 
\begin{equation}
Z(\lambda,\delta,T;\alpha)={\rm Tr}_{\rm A}\bmatr{\alpha}{e^{-\hat{H}(\lambda,\delta)/T}}{\alpha}=e^{-F(\lambda,\delta,T;\alpha)/T}.
\label{partialph}
\end{equation}
The true equilibrium is obtained by minimization of the \uvo{trial} value of the free energy  $F(\lambda,\delta,T;\alpha)$  in parameter $\alpha$, which therefore becomes an order parameter of the superradiant {\TPT} in the sense of the Landau theory.
If ${\omega\sim\omega_0}$, the value $|\alpha|^2$ is expected to be of order $N$ in the superradiant phase and zero in the normal phase.

Applying this procedure \cite{Kloc17a}, we identify two special values of temperature:
\begin{eqnarray}
\hspace{-15mm}
T_{\rm c}(\lambda,\delta)=\frac{\omega_0}{2\,{\rm arctanh}\left(\lambda_{\rm c}(\delta)^2/\lambda^{2}\right)}
\qquad
T_{0}(\lambda,\delta)=\frac{\omega_0}{2\,{\rm arctanh}\left(\lambda_{0}(\delta)^2/\lambda^{2}\right)}
\label{Tc0}\\
\hspace{8mm}{\rm for\ }\lambda\in\bigl(\lambda_{\rm c}(\delta),\infty\bigr),
\qquad
\hspace{20mm}{\rm for\ }\lambda\in\bigl(\lambda_{0}(\delta),\infty\bigr).
\nonumber
\end{eqnarray}
The value $T_{\rm c}(\lambda,\delta)$, with the critical coupling $\lambda_{\rm c}(\delta)$ from Eq.\,\eqref{lac}, represents the critical temperature of the superradiant {\TPT}, with the superradiant phase located below this temperature.
When the temperature increases across the critical value, the location $\alpha_0$ of the global minimum of the free energy $F(\lambda,\delta,T;\alpha)$ changes from ${|\alpha_0|>0}$ to ${\alpha_0=0}$. 
The temperature ${T_{0}(\lambda,\delta)\leq T_{\rm c}(\lambda,\delta)}$ from Eq.\,\eqref{Tc0}, with $\lambda_{0}(\delta)$ again defined in Eq.\,\eqref{lac}, is the temperature at which the free energy for ${\delta\in(0,1)}$ changes its form so that $F$ below $T_0$ has a complex conjugate pair $\alpha'_0$ and ${\alpha'_0}^*$ of degenerate saddle points on the imaginary axis of $\alpha$.
Remarkably, the coupling parameter value $\lambda_0(\delta)$ where the region with saddle points starts is also the point where the {\TC} phase appears in the quantum phase diagram associated with the $j\I{=}j_{\rm max}$ subspace of $\Hi$ (see Sec.\,\ref{EDM1}). 

The analysis in the full Hilbert space of the model takes into account all subspaces $\Hi_{j}^{(s)}$ with ${j\I{=}j_{\rm min},...,j_{\rm max}}$, including typically huge numbers of the subspace replicas.
Introducing a parameter ${\gamma\I{=}j/j_{\rm max}\I{=}2j/N}$,  which in the infinite-size limit becomes a~continuous variable ${\gamma\in[0,1]}$, we define a scaled coupling strength ${\lambda^{(\gamma)}\I{=}\lambda/\sqrt{\gamma}}$ and a~scaled energy ${E^{(\gamma)}\I{=}\gamma E}$.
It turns out \cite{Kloc17b} that from the perspective of the semiclassical analysis the point $(\lambda^{(1)},E^{(1)})$ in the $\lambda\times E$ plane for the $j\I{=}j_{\rm max}$ subspace is equivalent to the point $(\lambda^{(\gamma)},E^{(\gamma)})$ for a $j\I{<}j_{\rm max}$ subspace.
The set of points $(\lambda^{(\gamma)},E^{(\gamma)})$ with $\gamma\in[0,1]$ forms a smooth curve uniquely specified by the point $(\lambda^{(1)},E^{(1)})$ reached in the $\gamma\I{=}1$ limit.
Each of these curves connects points in $\lambda\times E$ with the same semiclassical properties in subspaces with various $j$.
Note that among the subspaces contributing to the given energy domain those with the the largest multiplicity $\nu_{j}$ (which for ${N\I{\to}\infty}$ means those with the lowest $j$) are most significant. 

\begin{figure}[t!]
\begin{flushright}
\includegraphics[width=0.8\textwidth]{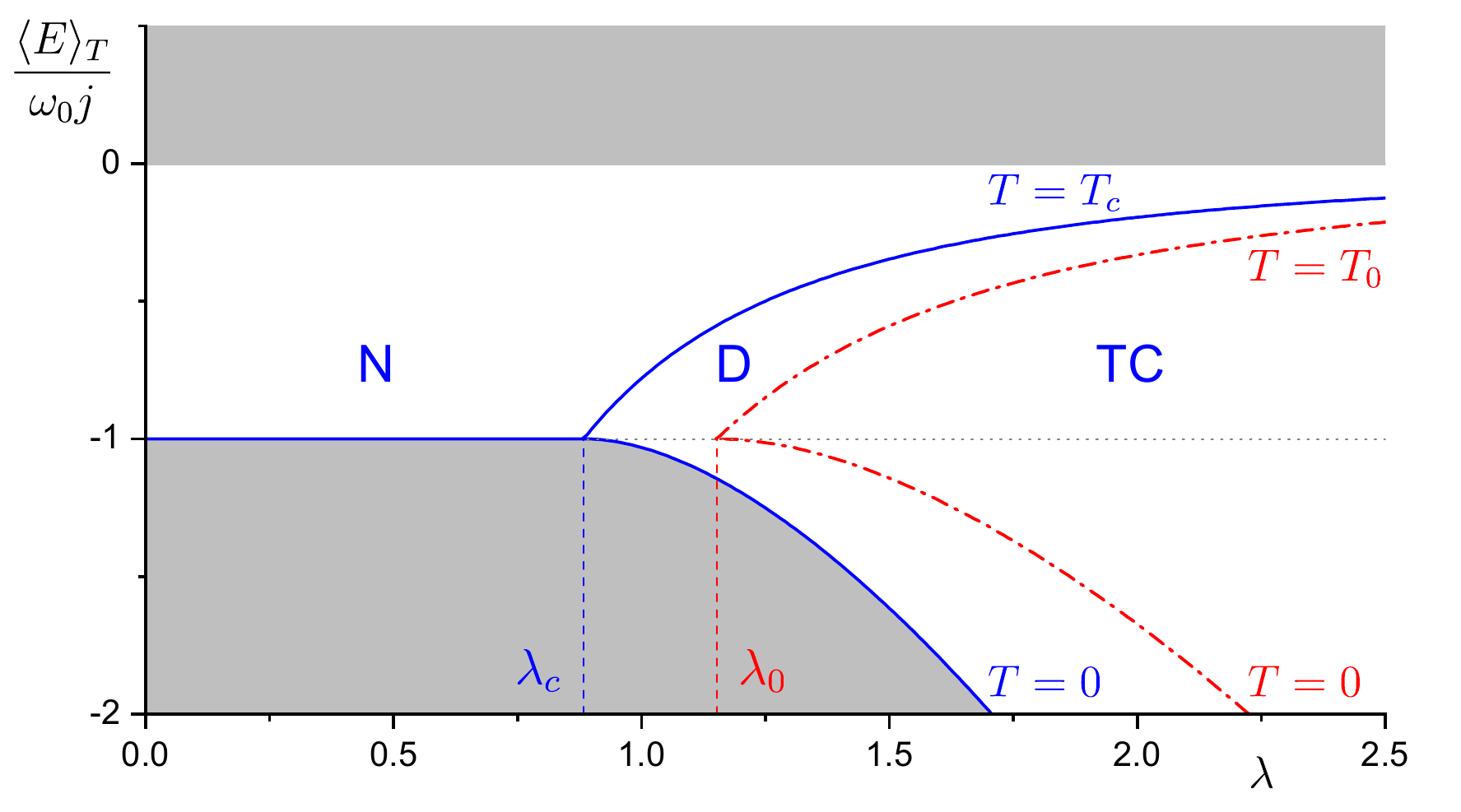}
\end{flushright}
\caption{
Canonical thermal phase diagram of the extended Dicke Hamiltonian \eqref{HDic} with ${\omega=\omega_0=1}$ and $\delta=0.132$ (cf.\,Fig.\,\ref{dede}). The vertical axis shows the thermal average of energy $\ave{E(\lambda)}_T$. Acronym {\N} denotes the normal phase, and acronyms {\D} and {\TC}, respectively, mark the superradiant phases without and with the \lq\lq false equilibria\rq\rq\ connected with saddle points of the free energy. 
The full (dashdotted) curves labeled by $T=0,T_{\rm c}$ ($T=0,T_0$) correspond to the global minima (saddle points) of the free energy at the respective temperatures.
The gray areas represent unreachable domains of $\ave{E(\lambda)}_T$. 
Based on Ref.\,\cite{Basta16}.
}
\label{Dictherm}
\end{figure}

As shown in Ref.\,\cite{Basta16}, the curve $(\lambda^{(\gamma)},E^{(\gamma)})$ passing for ${\gamma=1}$ the ground-state {\QPT} point $(\lambda,E)\I{=}(\lambda_{\rm c}(\delta),-j_{\rm max}\omega_0)$ is identical with the $\lambda\I{>}\lambda_{\rm c}(\delta)$ dependence of the thermal average of energy $\ave{E(\lambda,\delta)}_{T=T_{\rm c}(\lambda,\delta)}$ at the {\TPT} critical temperature.
Note that  the average energy at any temperature can be calculated from the formula
\begin{equation}
\ave{E(\lambda,\delta)}_{T}=T^2\frac{\partial}{\partial T}\ln Z(\lambda,\delta,T;{\alpha\I{=}\alpha_0})
\label{avEcan},
\end{equation} 
where the partition function \eqref{partialph} is evaluated at the global minimum $\alpha_0$ of the free energy $F(\lambda,\delta,T;\alpha)$.
This suggests that the {\TPT} in the Dicke model results from a~cumulative effect of ground-state {\QPTs} which occur in multiple subspaces $\Hi_{j}^{(s)}$.
An analogous interpretation was developed \cite{Basta16} also for the temperature $T_0(\lambda,\delta)$, but with the true equilibrium point $\alpha_0$ in Eq.\,\eqref{avEcan} replaced by the \uvo{false equilibria} associated with the saddle points $\alpha'_0$ and ${\alpha'_0}^*$ of the free energy landscape.
The average energy $\ave{E(\lambda,\delta)}'_{T=T_{0}(\lambda,\delta)}$, where the prime indicates that the formula \eqref{avEcan} is evaluated at the saddles instead of the global minimum, coincides with the curve ${(\lambda^{(\gamma)},E^{(\gamma)})}$ passing for ${\gamma=1}$ the point ${(\lambda,E)}={(\lambda_0(\delta),-j_{\rm max}\omega_0)}$.
Finally, the most important finding is that the curve $\ave{E(\lambda,\delta)}'_{T=0}$ (again evaluated at the saddles) is identical with the {\ESQPT} critical borderline $E_{\rm c}(\lambda,\delta)$ separating the {\D} and {\TC} quantum phases.
The canonical thermodynamic phase diagram in the plane ${\lambda\times\ave{E(\lambda,\delta)}_T}$ is shown in Fig.\,\ref{Dictherm}.

These results imply that only the {\D}-{\TC} {\ESQPT} borderline of the extended Dicke model has a clear counterpart in the canonical averages, while the other {\ESQPTs} seem invisible.
Though analogues of the {\D}, {\TC} and {\N} quantum phases exist in the thermodynamic phase diagram of Fig.\,\ref{Dictherm}, their arrangement and \uvo{critical} transitional energies $\ave{E(\lambda,\delta)}_T$ differ from those in Fig.\,\ref{dede}.
This is particularly surprising in the case of the  {\D}-{\N} {\ESQPT}, which is of the same type as the  {\D}-{\TC} one (diverging derivative of the level density).
It can be shown \cite{Basta16,Peres17} that the average energy of the canonical ensemble with an arbitrary temperature $T\in[0,\infty)$ satisfies the relation ${\ave{E(\lambda,\delta)}_T\leq0}$, which means that the {\N}-{\S} {\ESQPT} at ${E_{\rm c}=+j\omega_0}$ lies in an inaccessible energy region if the system is heated via interaction with a~thermal bath.  
Possible effects of {\ESQPTs} in thermal fluctuations were not studied. 

We can therefore conclude that although the above results undoubtedly disclose interesting relations between {\ESQPTs} and canonical thermodynamics, they do not set any direct {\ESQPT}-{\TPT} correspondence.
Thermodynamic consequences of {\ESQPTs} still require more investigations.

\section{Dynamic signatures}
\label{Dynam}

In this section, we discuss consequences of {\ESQPTs} in the dynamics induced by non-thermal excitations of the system, namely by excitations due to {\em external driving\/} of the control parameter in the Hamiltonian.
This potentially invokes a rather broad area of non-equilibrium effects \cite{Eiser15}, which we however outline only selectively.
First we describe the limit of very fast (instantaneous) parameter changes (so called quantum quenches). 
Then we investigate slow changes and conditions for excitation-free (adiabatic) driving.
At last, we look at driven dynamics in presence of dissipation.

\subsection{Quantum quench dynamics}
\label{quench}

Quantum quench is a diabatic limit of driving \cite{Sengu04,Calab06}.
It is a sudden change, \uvo{jump} of the control parameter $\lambda$ of a quantum system from an initial value $\lamin$ to a final value $\lamfi$.
The system is prepared in an initial state $\ket{\psiin}$, which is supposed to coincide with an eigenstate of the system's Hamiltonian $\hat{H}(\lambda)$ at ${\lambda=\lamin}$.
The usual (but not the only) choice of the initial state is the ground state of the initial Hamiltonian, ${\ket{\psiin}=\ket{\psi_{\rm gs}(\lamin)}}$.
Assuming a generic case with ${[\hat{H}(\lamin),\hat{H}(\lamfi)]\neq 0}$, we see that $\ket{\psiin}$ is not an eigenstate of the final Hamiltonian $\hat{H}(\lamfi)$ and hence evolves nontrivially with the elapsing time $t\in[0,\infty)$ (with ${t=0}$ setting the sharp time of the jump).
In the following, the Hamiltonian is supposed to be linear in~$\lambda$, having the form ${\hat{H}(\lambda)=\hat{H}(0)+\lambda\hat{H}'}$. 
Thus the final and initial Hamiltonians satisfy the homogeneous relation ${\hat{H}(\lamfi)=\hat{H}(\lamin)+\Delta\lambda\,\hat{H}'}$ depending on $\Delta\lambda=\lamfi\I{-}\lamin$.

The time evolution of the system after the quench depends on the overlaps of the initial state $\ket{\psiin}$ with individual eigenvectors $\ket{\psi_i(\lamfi)}$ of the final Hamiltonian.
To characterize properties of these overlaps, we introduce the strength function $W(E)$ (also called the local density of states) and the autocorrelation function $R(\varepsilon)$:
\begin{eqnarray}
W(E)=&&\sum_i\overbrace{\abs{\scal{\psi_i(\lamfi)}{\psiin}}^2}^{w_i}\delta\bigl(E\I{-}E_i(\lamfi)\bigr),
\label{Stre}
\\
R(\varepsilon)=&&\sum_i\sum_{i'}w_i\,w_{i'}\ \delta\bigl(\varepsilon\I{-}E_i(\lamfi)\I{+}E_{i'}(\lamfi)\bigr).
\label{Auto}
\end{eqnarray}
While the strength function $W(E)$ represents the energy distribution of the initial state in the final Hamiltonian spectrum, the autocorrelation function $R(\varepsilon)$ characterizes mutual correlation of overlaps of the initial state with the final eigenstates $\ket{\psi_i(\lamfi)}$ and $\ket{\psi_{i'}(\lamfi)}$ differing by energy $\varepsilon$.   
To simplify the notation, the specifications of $\ket{\psiin}$ and $\lamfi$ are skipped in $W(E)$ and $R(\varepsilon)$.

The survival probability of the initial state in the evolving final state can be expressed via Fourier images of the strength and autocorrelation functions:
\begin{equation}
\hspace{-20mm}
P(t)=\abs{\matr{\psiin}{e^{-\ii\hat{H}(\lamfi)t/\hbar}}{\psiin}}^2=\abs{\int dE\ W(E)\,e^{-\ii Et/\hbar}}^2=\int d\varepsilon\ R(\varepsilon)\,e^{\ii\varepsilon t/\hbar}.
\label{Psur}
\end{equation}
As shown in numerous studies (see, e.g., Refs.\,\cite{Sengu04,Calab06,Tavor17,Torre18,Kloc18,Lerma18}, paper 19 in Ref.\,\cite{Binde19}, and the references therein), the survival probability \eqref{Psur} in most situations undergoes a rather complex evolution.
After an initial decay and passage through a minimal overlap stage, it usually grows again and manifests an infinite aperiodic sequence of partial revivals and breakdowns of the initial state.
The evolution of $P(t)$ on a given time scale reflects the energy distribution \eqref{Stre} with resolution ${\Delta E\sim\hbar/t}$ determined by the energy-time uncertainty relation.
Thus in early stages of the evolution, at very short times, the dependence $P(t)$ reveals only a rough outline of the distribution $W(E)$, while finer and finer details, including tiny correlations captured by Eq.\,\eqref{Auto}, become relevant on later stages.
For $t$ exceeding the limit (sometimes called the Heisenberg time) that on the energy scale corresponds to the smallest spacing between the populated levels, the perpetually oscillating function $P(t)$ reaches a stabilized regime and shows no further qualitative modifications.  

\begin{figure}[t!]
\begin{flushright}
\includegraphics[width=\textwidth]{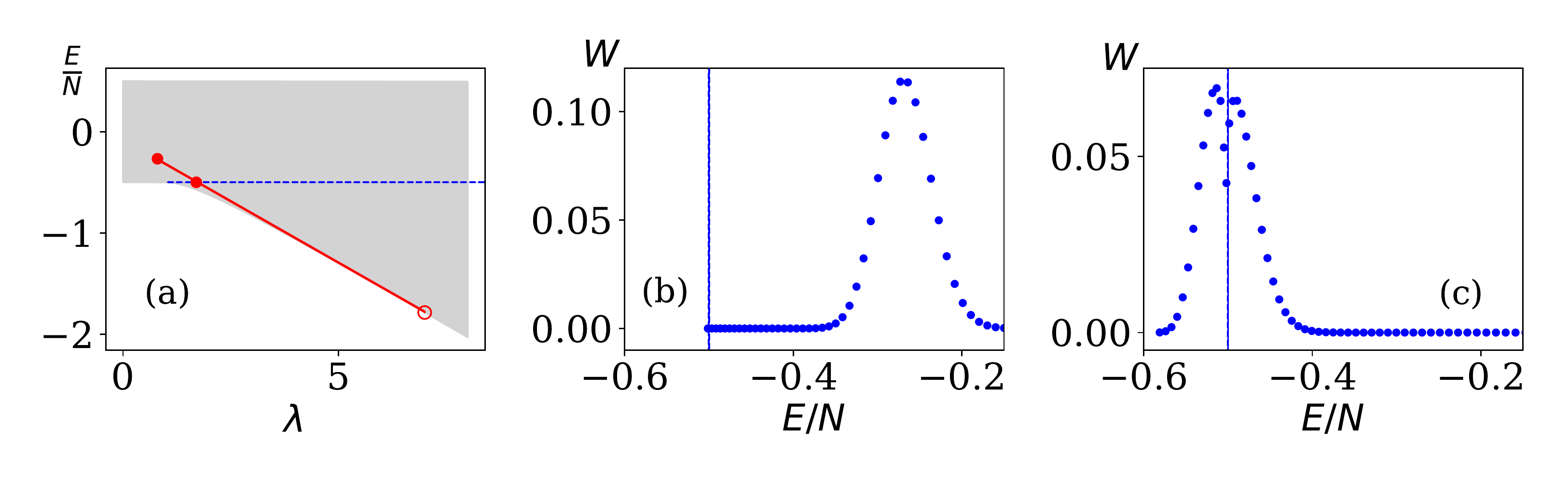}
\end{flushright}
\caption{Strength functions of two backward quenches in the Lipkin model. 
The second-order {\QPT} Hamiltonian \eqref{HLipkin} with $\chi=0$ is used with $N=2j=200$.
Panel (a) depicts the $\lambda\times E$ diagram, where the gray area represents the region where individual levels are located and the horizontal (blue) line marks an {\ESQPT} of type $(f,r)=(1,1)$.
The initial ground state at $\lamin=7$ (open circle) is \lq\lq quenched\rq\rq\ along the tilted (red) line to the excited domain at $\lamfi=0.8$ and $\lamfi\approx 1.75$; the full dots mark the respective final average energies.
The corresponding strength functions are shown in panels (b) and (c). 
The strength function of the critical quench in panel (c) exhibits a singularity at the {\ESQPT} energy (the vertical line).
Evolutions of the survival probability and Wigner functions are shown in Fig.\,\ref{Quo2}.
}
\label{Quo1}
\end{figure}

The average value $\ave{E}$ and width $\sqrt{\dis{E^2}}$ of the energy distribution \eqref{Stre} are determined from 
\begin{equation}
\hspace{-15mm}
\overbrace{\biave{\hat{H}(\lamfi)}_{\rm in}}^{\ave{E}}-\overbrace{\biave{\hat{H}(\lamin)}_{\rm in}}^{E_{\rm gs}(\lamin)}=\Delta\lambda\!\!\!\overbrace{\biave{\hat{H}'}_{\rm in}}^{\frac{d}{d\lambda}E_{\rm gs}(\lamin)},\quad\ 
\overbrace{\bidis{\hat{H}(\lamfi)^2}_{\rm in}}^{\dis{E^2}}=\Delta\lambda^2\,\bidis{\hat{H}'\,\!^2}_{\rm in},
\label{avevar}
\end{equation}
where ${\ave{\hat{A}}_{\rm in}=\matr{\psiin}{\hat{A}}{\psiin}}$ and ${\dis{\hat{A}^2}_{\rm in}=\ave{\hat{A}^2}_{\rm in}\I{-}\ave{\hat{A}}_{\rm in}^2}$ are the expectation value and variance of quantity $\hat{A}$ in the state $\ket{\psiin}$.
The first formula shows that the energy average after the quench is determined by the initial energy $E_{\rm gs}(\lamin)$ and its tangent $\frac{d}{d\lambda}E_{\rm gs}(\lamin)$.
This permits a simple geometrical prediction of the centroid of the final energy distribution, allowing us to design a quench focused to the middle of the selected final energy domain. 
The second formula implies that the width of the energy distribution is proportional to the width of distribution of $\hat{H}'$ in the initial state and grows linearly with $\Delta\lambda$.
This sets limits on the energy resolution with which the selected final energy can be tuned.
The geometrical visualization and strength functions of two quenches in the Lipkin model are shown in Fig.\,\ref{Quo1}.  
 
The decay of the survival probability at very short times is approximated by $P(t)\approx 1-\dis{E^2}t^2/\hbar^2$ and hence depends only on the energy width from the second formula of Eq.\,\eqref{avevar}.
On the other hand, elementary statistical measures of the survival probability fluctuations in infinite time, namely the average $\ave{P(t)}_{t\in[0,\infty)}\equiv\lim_{\tau\to\infty}\tau^{-1}\int_0^{\tau}dt\,P(t)$ and variance $\dis{P(t)^2}_{t\in[0,\infty)}$, 
follow from the fragmentation of the strength function among individual eigenstates of $\hat{H}(\lamfi)$.
It is quantified by the so-called participation ratio $\mathcal{P}=(\sum_i w_i^2)^{-1}$, which takes values between $\mathcal{P}=1$ for states fully localized in one of the selected basis states and $\mathcal{P}=d$ for states uniformly distributed among all $d$ basis states.
The following relations hold,   
\begin{equation}
\hspace{-15mm}
\biave{P(t)}_{t\in[0,\infty)}=\mathcal{P}^{-1},\qquad
\bidis{P(t)^2}_{t\in[0,\infty)}=\mathcal{P}^{-2}-\sum_iw_i^4,
\end{equation}
where the second term on the right-hand side of the second formula is of order ${\cal P}^{-3}$ for strongly delocalized states and can often be neglected.  

Since the pioneering studies, various issues related to quantum quench dynamics have been subject to extensive research.
Particularly the implications of ground-state {\QPTs} \cite{Sengu04} and quantum chaos (see, e.g., paper 19 in Ref.\,\cite{Binde19}) were studied in the evolution of the survival probability on the medium time scale (between the initial decay and the stabilized regime).
Moreover, new types of critical phenomena, unified under the name dynamical quantum phase transitions \cite{Sciol10,Heyl13,Heyl18,Zunko18}, were identified in connection with quantum quenches.
They represent non-analyticities appearing either in the long-time averages of physical observables as a function of quench parameters \cite{Sciol10}, or in the post-quench evolution of the survival probability as a function of the time variable \cite{Heyl13}.
A unified view on both these effects was presented in Ref.\,\cite{Zunko18}. 
Their robustness under the onset of short-range interactions in otherwise a fully connected system was studied in Ref.\,\cite{Leros18}.
Recently, an experimental realization of the first-type dynamical phase transition connected with a singularity in the spectrum of excited states was reported in the spinor condensate of sodium atoms \cite{Tian20}.

Impact of {\ESQPTs} on the quantum quench dynamics was discussed in several theoretical works.
An initial study of Ref.\,\cite{Peres11a} reported an {\ESQPT}-induced suppression of the survival probability at medium times for quenches within the interacting phase in the Dicke and Tavis-Cummings models (Sec.\,\ref{Open}) and in a related atom-molecule boson model (Sec.\,\ref{Algeb}).
An opposite effect, namely an {\ESQPT}-induced stabilization of the initial state for quenches from the non-interacting to the interacting phase in the Lipkin and ${n=3,4}$ vibron models (Secs.\,\ref{Quasi} and \ref{Ibms}), was described in Refs.\,\cite{Santo15,Santo16,Wang17,Perez17}.
In Ref.\,\cite{Kloc18}, both these complementary results were treated on a unified ground and qualitatively explained for the extended Dicke model.

It turns out that essential insight into the consequences of {\ESQPTs} in quantum quench dynamics follows from the semiclassical description based on the Wigner quasiprobability distribution in the phase space \cite{Hiller84}.
At time $t=0$, when the initial state is still intact, its Wigner distribution reads
\begin{equation}
\mathcal{W}(\vecb{q},\vecb{p},t\I{=}0)=\frac{1}{\pi\hbar}\int\! d\vecb{\delta}\ 
\bscal{\vecb{q}\I{+}\vecb{\delta}}{\psiin}\bscal{\psiin}{\vecb{q}\I{-}\vecb{\delta}}\,e^{-2\ii\vecb{\delta}\cdot\vecb{p}/\hbar}
\label{Wig},
\end{equation}
where $\ket{\vecb{q}\I{\pm}\vecb{\delta}}$ are generalized coordinate eigenstates of the system.
As the time increases, for ${t>0}$, the Wigner distribution evolves to $\mathcal{W}(\vecb{q},\vecb{p},t)$, which is given by the same formula as Eq.\,\eqref{Wig} but with $\ket{\psiin}$ replaced by ${\ket{\psi(t)}=e^{-i\hat{H}(\lamfi)t/\hbar}\ket{\psiin}}$.
The survival probability \eqref{Psur} is expressed as
\begin{equation}
P(t)=2\pi\hbar\int\! d\vecb{q}\,d\vecb{p}\ \mathcal{W}(\vecb{q},\vecb{p},t)\,\mathcal{W}(\vecb{q},\vecb{p},0)
\label{Psurcl}.
\end{equation}

The evolution of $\mathcal{W}(\vecb{q},\vecb{p},t)$ is strongly influenced by the existence of stationary points of the classical-limit Hamiltonian corresponding to $\hat{H}(\lamfi)$ at energies close to average ${\ave{E}\equiv\ave{\hat{H}(\lamfi)}_{\rm in}}$ of the distribution \eqref{Stre}.
This is what makes quantum quench dynamics near {\ESQPT} critical energies different from the same dynamics in non-critical energy domains \cite{Kloc20}.
If the stationary point ${(\vecb{q}^{(i)},\vecb{p}^{(i)})}$ is located inside the support of the initial Wigner distribution ${\mathcal{W}(\vecb{q},\vecb{p},t\I{=}0)}$, the semiclassical overlap \eqref{Psurcl} decays much slower and may keep increased even on long time scales.
Of course, an unstable stationary point cannot completely stop the decay as the Wigner distribution is never perfectly localized at this point and therefore must gradually deviate from the initial form. 
Nevertheless, the effect of stabilization can be very strong, as shown in the above-mentioned {\ESQPT} studies of Refs.\,\cite{Santo15,Santo16,Wang17,Perez17,Kloc18}.
This scenario applies particularly to \uvo{forward} quenches (${\lamfi>\lamin}$) of the ground state from the non-interacting ($\lamin\I{<}\lambda_{\rm c}$) to interacting phase ($\lamfi\I{>}\lambda_{\rm c}$), across the second-order {\QPT}, in which the global minimum of the classical-limit Hamiltonian becomes a local maximum or a saddle point with the same energy.

\begin{figure}[t!]
\begin{flushright}
\includegraphics[width=\textwidth]{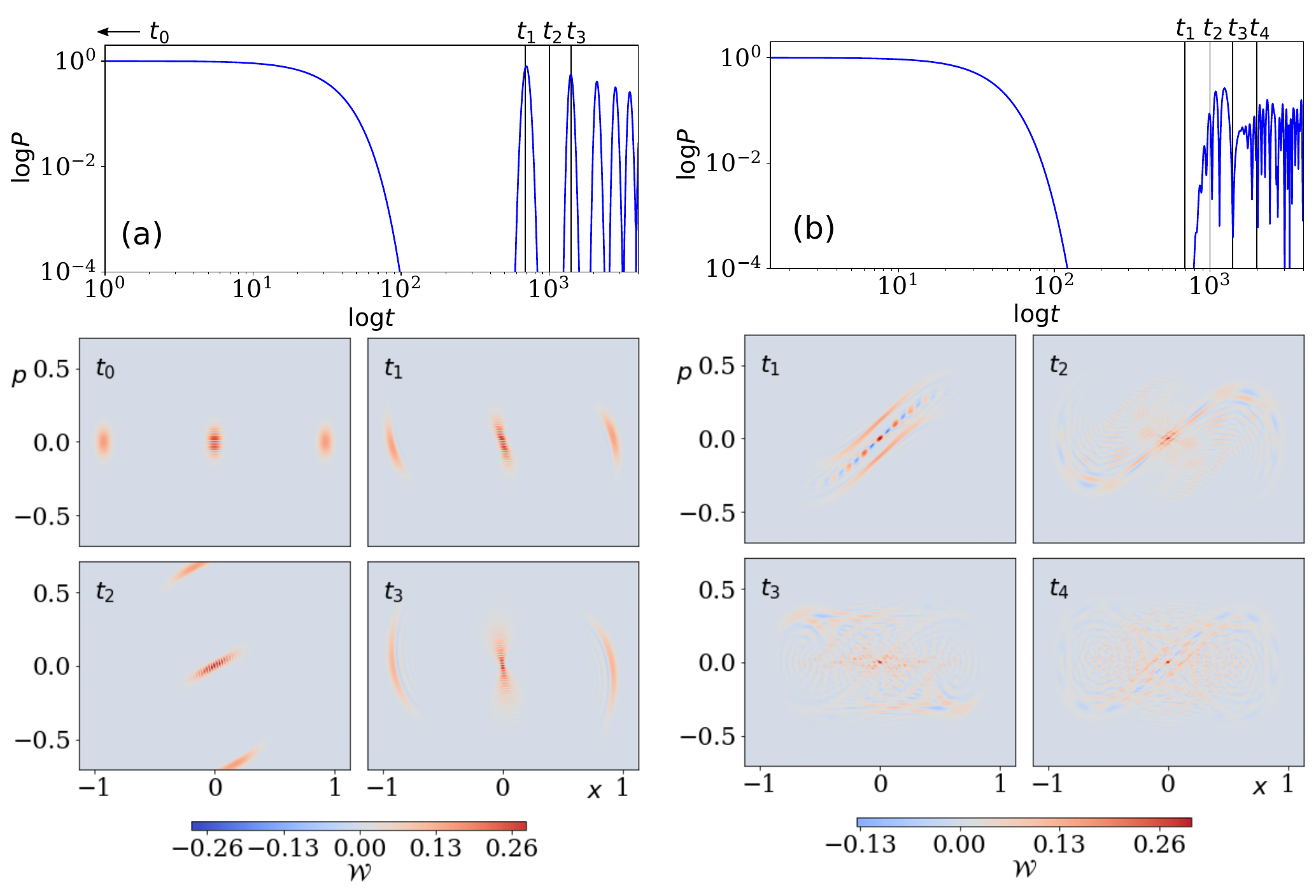}
\end{flushright}
\caption{
Evolutions of the survival probability (upper panels) and snapshots of the Wigner function (lower panels) for the two quenches from Fig.\,\ref{Quo1}.
The non-critical quench from Fig.\,\ref{Quo1}(b) is shown on the left [panel (a)], the critical quench from Fig.\,\ref{Quo1}(c) is on the right [panel (b)].
The non-critical quench shows repeated partial revivals of the initial state (with a power-law decrease of fidelity), which result from approximate returns of the Wigner distribution to the initial form.
The critical quench shows no revivals as the Wigner function gets captured at the stationary point $(q,p)=(0,0)$ associated with the {\ESQPT} of the final Hamiltonian.
The Wigner functions are taken at dimensionless times $(t_0,t_1,t_2,t_3,t_4)=(0,690,1000,1400,2000)$, see vertical lines in the upper panels. 
}
\label{Quo2}
\end{figure}

The degree of suppression of the post-quench decay of the initial state depends on the stability properties of classical dynamics in the vicinity of the corresponding stationary point.
More stable stationary points generate stronger suppression than less stable ones.
This was verified in the extended Dicke model (Sec.\,\ref{Open}) for quenches from the non-interacting phase to the {\ESQPT} region in the interacting phase  \cite{Kloc18}.
The decay of the initial ground state prepared at $\lambda_{\rm in}\I{=}0$ (energy $E=-j\omega_0$) was suppressed weakly for quenches to $\lambda_{\rm fi}\in(\lambda_{\rm c},\lambda_0)$, i.e., to the {\ESQPT} region of type $(f,r)=(2,1)$ [see Eq.\,\eqref{lac} and Fig.\,\ref{dede}].
In contrast, the decay of the same initial state was suppressed strongly for longer quenches to $\lambda_{\rm fi}\I{>}\lambda_0$, that is, to the {\ESQPT} region of type $(f,r)=(2,2)$. 
Indeed, the $(f,r)=(2,2)$ stationary point of the extended Dicke model turns out to be classically more stable than the (2,1) stationary point. 

What happens if the stationary point $(\vecb{q}^{(i)},\vecb{p}^{(i)})$ of the final Hamiltonian $\hat{H}(\lamfi)$ near the average energy $\ave{\hat{H}(\lamfi)}_{\rm in}$ is located away from the support of the initial Wigner distribution $\mathcal{W}(\vecb{q},\vecb{p},t\I{=}0)$? 
In this case, the early stage of the post-quench evolution is apparently not affected.
The existence of the stationary point may nevertheless show up in later stages of the evolution. 
For integrable systems (particularly for ${f=1}$ systems with closed classical orbits), the quasiclassical dynamics in absence of stationary points often leads to a regular sequence of partial revivals with an algebraically weakening amplitude (usually ${\propto 1/t}$) \cite{Kloc18}.
These are caused by periodic orbits that repeatedly drive the Wigner distribution through the initial phase space domain, as explicitly demonstrated in the ${\delta=0}$ version of the Dicke model \cite{Kloc20}.
The revivals are suppressed or completely avoided if a stationary point at any place in the energetically available phase space effectively captures a considerable part of the distribution.
This mechanism is responsible for the observed weakening of the power-law revivals in the $P(t)$ dependencies after \uvo{backward} (${\lamfi<\lamin}$) quenches of the ground state within the interacting regime of the Dicke and other models \cite{Perez17,Kloc18}.
Usually, the suppression of revivals is observed for quenches fine tuned right to the {\ESQPT} region, while for the neighboring shorter or longer quenches the revivals are present. 
This is illustrated for the Lipkin model (for both the critical and the non-critical quenches from Fig.\,\ref{Quo1}) in Fig.\,\ref{Quo2}.
For non-integrable ${f>1}$ models (e.g., the extended Dicke model with ${\delta>0}$), the effect strongly depends on the degree of chaos in the final energy region.
The study in Ref.\,\cite{Kloc18} demonstrates that in strongly chaotic cases the algebraic revivals are absent even for non-critical quenches.
The {\ESQPTs} then have no significant impact on post-quench dynamics.  
On the contrary, in less chaotic cases, if a part of the available phase space remains regular, the algebraic revivals are observed even for critical quenches to some {\ESQPTs}.

Formally the same analysis as described above in connection with quantum quench dynamics can be applied also in the description of {\em decoherence\/} in coupled quantum systems.
Consider a two-state system, a qubit, with basis vectors $\ket{0_{\rm Q}}$ and $\ket{1_{\rm Q}}$ defining the Hilbert space ${\Hi_{\rm Q}={\mathbb C}^2}$, which interacts with a bath system associated with an arbitrary (usually larger) Hilbert space~$\Hi_{\rm B}$.
A possible evolution over time $t$ in the full Hilbert space ${\Hi=\Hi_{\rm Q}\otimes\Hi_{\rm B}}$ can be expressed by a pair of formulas:
\begin{equation}
\hspace{-10mm}
\begin{array}{rcl}
\ket{0_{\rm Q}}\otimes\ket{\psi_{\rm B}(0)} & \stackrel{t}{\longrightarrow} & \ket{0_{\rm Q}}\otimes e^{-\ii\hat{H}_{\rm B0}t/\hbar}\ket{\psi_{\rm B}(0)}\equiv\ket{0_{\rm Q}}\otimes\ket{\psi_{\rm B0}(t)},\\
\ket{1_{\rm Q}}\otimes\ket{\psi_{\rm B}(0)} & \stackrel{t}{\longrightarrow} & \ket{1_{\rm Q}}\otimes e^{-\ii\hat{H}_{\rm B1}t/\hbar}\ket{\psi_{\rm B}(0)}\equiv\ket{0_{\rm Q}}\otimes\ket{\psi_{\rm B1}(t)},
\end{array}
\label{decohe}
\end{equation}
where $\ket{\psi_{\rm B}(0)}$ is an initial state of the bath, and $\hat{H}_{\rm B0}$ and $\hat{H}_{\rm B1}$ are two Hamiltonians describing conditional evolution of the bath for the qubit initial states $\ket{0_{\rm Q}}$ and~$\ket{1_{\rm Q}}$, respectively.
We note that the evolution \eqref{decohe} is special in the sense that it does not induce the ${\ket{0_{\rm Q}}\leftrightarrow\ket{1_{\rm Q}}}$ flips of the qubit.
It affects the qubit only through its decoherence, i.e., loss of purity of its quantum state.
If the evolution starts from a~factorized initial state ${\ket{\Psi(0)}=(\alpha_0\ket{0_{\rm Q}}\I{+}\alpha_1\ket{1_{\rm Q}})\otimes\ket{\psi_{\rm B}(0)}}$, the qubit (which at the beginning is in a pure superposition state with normalized amplitudes ${\alpha_0,\alpha_1\in{\mathbb C}}$) becomes entangled (in a~generic case) with states of the bath and its state gets mixed. 
The qubit state at any time is described by the density operator 
\begin{eqnarray}
\hspace{-20mm}
\hat{\varrho}_{\rm Q}(t)={\rm Tr}_{\rm B}\ket{\Psi(t)}\bra{\Psi(t)}&=&|\alpha_0|^2\ket{0_{\rm Q}}\bra{0_{\rm Q}}+|\alpha_1|^2\ket{1_{\rm Q}}\bra{1_{\rm Q}}
\label{denmat}\\
\hspace{-20mm}
&+&\alpha_0^*\alpha_1\underbrace{\scal{\psi_{\rm B0}(t)}{\psi_{\rm B1}(t)}}_{R(t)}\ket{1_{\rm Q}}\bra{0_{\rm Q}}+\alpha_0\alpha_1^*R^*(t)\ket{0_{\rm Q}}\bra{1_{\rm Q}}.
\nonumber
\end{eqnarray}
which yields ${{\rm Tr}\hat{\varrho}_{\rm Q}^2(t)=1\I{-}2(1\I{-}|R(t)|^2)|\alpha_0|^2|\alpha_1|^2}$.
We see that if ${|R(t)|=1}$, the qubit remains in a pure state for any coefficients ${\alpha_0,\alpha_1}$, while if ${|R(t)|<1}$, the qubit initiated in a pure state with ${\alpha_0,\alpha_1\neq 0}$ evolves into a mixed state with the same probabilities $|\alpha_0|^2$ and $|\alpha_1|^2$ to measure the basis states.
For ${R(t)=0}$, i.e., for orthogonal (fully distinguishable) evolved bath states on the right-hand side of Eq.\,\eqref{decohe}, the decoherence is maximal.
If the initial bath state coincides with an eigenstate of Hamiltonian $\hat{H}_{\rm B0}$, the function $R(t)$ is equivalent (modulo an unimportant phase factor) to the survival amplitude of the initial bath state after a quench from $\hat{H}_{\rm B0}$ to the new Hamiltonian~$\hat{H}_{\rm B1}$.
Indeed, the papers \cite{Relan08,Peres09}, which study a qubit coupled to a fully connected spin system described by a Lipkin Hamiltonian, report on a maximal enhancement of the qubit decoherence if the spin bath is excited to the {\ESQPT} region.
The corresponding dependence of $R(t)$ is consistent with the above described {\ESQPT} effects in the quantum quench dynamics (cf.\,Fig.\,\ref{Quo2}).

Quantum quench represents a general technique for exciting an isolated system to a selected domain of energy and control parameters. 
The dynamical response to the quench can be examined not only by the survival probability, as outlined above, but also by time dependencies of expectation values of various suitably selected observables, or even through some more complex techniques.
A lot of attention in recent years has been attracted by the concept of a so-called {\em out-of-time-order correlator}, see Refs.\,\cite{Hashi17,Swing18} and the references therein. 
This quantity is defined as
\begin{equation}
O(t)=\ave{\hat{B}(t)^\dag\hat{A}(0)^\dag\hat{B}(t)\hat{A}(0)}_{\hat{\varrho}}
\label{otoc},
\end{equation}
where $\hat{A}(0)$ and $\hat{B}(t)=\hat{U}(t)^\dag\hat{B}(0)\hat{U}(t)$ are Heisenberg pictures of two chosen \uvo{probe} operators (usually unitary or Hermitian) at times $0$ and $t$, and $\ave{\bullet}_{\hat{\varrho}}$ denotes an expectation value in a general state, which in the Heisenberg picture is described by a time-independent density operator $\hat{\varrho}$.
Note that the symbol $O(t)$ hides the dependence on $\hat{A}$, $\hat{B}$ and $\hat{\varrho}$. 
If evaluated in a pure state ${\hat{\varrho}=\ket{\psi}\bra{\psi}}$, the expression \eqref{otoc} yields a scalar product $\scal{\hat{A}(0)\hat{B}(t)\psi}{\hat{B}(t)\hat{A}(0)\psi}$.
Even if both operators commute at $t\I{=}0$, we assume that ${[\hat{A}(0),\hat{B}(t)]\neq 0}$ for ${t>0}$.
This reflects scrambling of information about the \uvo{action} $\hat{B}$, initially separated from $\hat{A}$, over the system (imagine, e.g., two impulses at distant sites of an interacting lattice system).
The quantity \eqref{otoc} is an efficient measure of this process. 
It was shown that the out-of-time-order correlator can also serve as a quantum counterpart of the classical Lyapunov exponent \cite{Malda16,Rozen17,Chave19} and has close links to the Renyi entropy \cite{Fan17}, giving therefore opportunity to measure properties related to quantum chaos and quantum entanglement. 
A recent proposal \cite{Lewis19} describes an experimental setup capable of measuring a particular out-of-time-order correlator in the Dicke model.

We assume that $\hat{\varrho}\I{=}\ket{\psi_i(\lamin)}\bra{\psi_i(\lamin)}$ is an eigenstate of the Hamiltonian with parameter $\lamin$, while the evolution ${\hat{U}(t)=e^{-\ii\hat{H}(\lamfi)t/\hbar}}$ of operator $\hat{B}(t)$ is realized by the Hamiltonian with parameter $\lamfi$.
This enables us to interpret Eq.\,\eqref{otoc} in the language developed for the description of quantum quench dynamics.
In particular, associating the ${t=0}$ operators $\hat{A}(0)$ and $\hat{B}(0)$ with the respective Schrödinger pictures $\hat{A}$ and $\hat{B}$, we identify the out-of-time-order correlator with the expectation value of a time-dependent operator ${\hat{O}(t)=\hat{B}^\dag\hat{U}(t)\hat{A}^\dag\hat{U}(t)^\dag\hat{B}\hat{U}(t)\hat{A}\hat{U}(t)^\dag}$ in the state ${\ket{\psi(t)}=\hat{U}(t)\ket{\psi_i(\lamin)}}$ evolving by the post-quench Hamiltonian.
If the initial state expressed in the final Hamiltonian eigenbasis is localized in the {\ESQPT} critical region, or on its either side, the time dependence \eqref{otoc} may exhibit some specific features.
The theoretical analysis of the Lipkin model with the second-order {\QPT} and the corresponding $(f,r)=(1,1)$ {\ESQPT} in Ref.\,\cite{Wang19b} confirms this anticipation.
It shows that using operators ${\hat{A}=\hat{B}\propto\hat{J}_x}$, one can distinguish quenches to the normal and parity breaking quantum phases by the long-time averages of the corresponding $O(t)$ dependence.
Another theoretical study \cite{Humme19} of a similar system with {\ESQPT} disclosed a tendency for long-time periodicity of a particular out-of-time-order correlator for quenches to the {\ESQPT} critical region, with the period determined by the local increase of the level density at the critical borderline.
The works in Refs.\,\cite{Pilat19} and \cite{Xu20} use the out-of-time-order correlator to detect {\ESQPT} based instabilities of semiclassical dynamics in the integrable version of the Dicke model and in the Lipkin model.

\subsection{Adiabatic and counterdiabatic driving}
\label{driven}

Let us now briefly outline effects of finite-speed driving with a general, externally prescribed time dependence of the Hamiltonian control parameters.
Such protocols are usually applied with the aim to prepare a desired quantum state, taken in general as the $i$th eigenstate (mostly the ground state) of the Hamiltonian at the final value of the control parameter.
The final Hamiltonian keeps the required state intact but it is implicitly assumed to be so complex that a direct preparation of its eigenstate is too difficult.
The most popular protocols make use of the adiabatic theorem. 
They achieve the task by initiating the system in the $i$th Hamiltonian eigenstate at another parameter value (where it is easier, for some reasons) and by driving it very slowly to the final parameter value.
To assure that this procedure is transitionless (i.e., that the system all the time remains in the $i$th instantaneous eigenstate), one needs to satisfy conditions for adiabaticity, which depend on the chosen trajectory in the parameter space \cite{Polko11,Kolod17}.
For the purposes of the present section we consider a multidimensional parameter space $\vecb{\lambda}\equiv(\lambda_1,\lambda_2,...)\equiv\{\lambda_{\alpha}\}$ with a general trajectory $\vecb{\lambda}(t)$ implying a time-dependent driving speed ${\frac{d}{dt}\vecb{\lambda}(t)\equiv\dot{\vecb{\lambda}}(t)\equiv\{\dot{\lambda}_{\alpha}(t)\}}$.
We introduce the gradient operator in the parameter space, ${\vecb{\nabla}\equiv\bigl\{\nabla_{\alpha}\equiv\frac{\partial}{\partial\lambda_{\alpha}}\bigr\}}$, which is applied to the Hamiltonian $\hat{H}(\vecb{\lambda})$ as well as to its individual eigenvectors $\ket{\psi_i(\vecb{\lambda})}$.

The exact evolution of the system obeys time dependent Schrödinger equation ${\ii\hbar\frac{d}{dt}\ket{\psi(t)}=\hat{H}(\vecb{\lambda}(t))\ket{\psi(t)}}$ with the initial state ${\ket{\psi(0)}=\ket{\psi_i(\vecb{\lambda}_{\rm in})}}$ coinciding with an eigenstate of $\hat{H}(\vecb{\lambda}_{\rm in})$.
Expanding the solution in the eigenbasis of the instantaneous Hamiltonian $H(\vecb{\lambda}(t))$, that is ${\ket{\psi(t)}=\sum_{i'} a_{i'}(t)\ket{\psi_{i'}(\vecb{\lambda}(t))}}$ with complex amplitudes $a_{i'}(t)$ satisfying the initial condition ${a_{i'}(0)=\delta_{i'i}}$, one derives a set of coupled differential equations for the amplitudes.
The equation for $a_i(t)$ reads
\begin{equation}
\hspace{-20mm}
\ii\hbar\frac{da_i(t)}{dt}
=E_i\bigl(\vecb{\lambda}(t)\bigr)\,a_i(t) 
-\dot{\vecb{\lambda}}(t)\I{\cdot}\sum_{i'}\bmatr{\psi_{i'}\bigl(\vecb{\lambda}(t)\bigr)}{\hat{\vecb{A}}\bigl(\vecb{\lambda}(t)\bigr)}{\psi_{i}\bigl(\vecb{\lambda}(t)\bigr)}\,a_{i'}(t)
\label{dridy},
\end{equation}
where the dot product denotes the sum of products of individual $\alpha$-components and
\begin{equation}
\hspace{-22mm}
\bmatr{\psi_{i'}(\vecb{\lambda})}{\hat{\vecb{A}}(\vecb{\lambda})}{\psi_{i}(\vecb{\lambda})}
\I{=}\ii\hbar\,\bscal{\psi_{i'}(\vecb{\lambda})}{\vecb{\nabla}\psi_i(\vecb{\lambda})}
\I{=}\left\{\begin{array}{ll}
\!\!\!\ii\hbar\scal{\psi_{i}(\vecb{\lambda})}{\vecb{\nabla}\psi_i(\vecb{\lambda})} & {\!\!\!\!\rm for\ }i'\I{=}i,\\
\!\!\!\frac{\matr{\psi_{i'}(\vecb{\lambda})}{\vecb{\nabla}\hat{H}(\vecb{\lambda})}{\psi_i(\vecb{\lambda})}}{E_{i}(\vecb{\lambda})-E_{i'}(\vecb{\lambda})} & {\!\!\!\!\rm for\ }i'\I{\neq}i.
\end{array}\right.
\label{Agauge}
\end{equation}
The first term on the right-hand side of Eq.\,\eqref{dridy} and the ${i'=i}$ term of the sum describe phase changes of the amplitude $a_i(t)$, preserving its absolute value.
The first term corresponds to the ordinary dynamical phase, while the diagonal term of the sum generates the geometric (Berry) phase \cite{Wilcz89}.
The ${i'\neq i}$ terms of the sum induce transitions to other states $\ket{\psi_{i'}(\vecb{\lambda})}$, changing therefore the absolute value $|a_i(t)|$.
To keep the driving adiabatic, these transitional effects must be suppressed by reducing the speed $\dot{\vecb{\lambda}}$ at the places $\vecb{\lambda}$ where the non-diagonal terms are large.

The form of Eq.\,\eqref{dridy} refers to a deep mathematical background for the description of driven quantum systems, the so-called {\em adiabatic gauge theory\/} \cite{Kolod17}.
It follows from a local \uvo{gauge} transformation ${\hat{U}(\vecb{\lambda})=\sum_{i'}\ket{\phi_{i'}}\bra{\psi_{i'}(\vecb{\lambda})}}$ of the system with variable $\vecb{\lambda}(t)$ to the \uvo{moving frame} with a fixed basis $\{\ket{\phi_{i'}}\}$.
The result of this transformation are the terms in Eq.\,\eqref{dridy} containing matrix elements of the gauge potential $\hat{\vecb{A}}(\vecb{\lambda})$ defined in Eq.\,\eqref{Agauge}.
The underlying geometry is deduced from the geometric tensor \cite{Provo80,Wilcz89}
\begin{eqnarray}
\hspace{-15mm}
\chi_{i\,\alpha\beta}(\vecb{\lambda})
=\frac{1}{\hbar^{2}}\left[
\biave{\hat{A}_{\alpha}(\vecb{\lambda})\hat{A}_{\beta}(\vecb{\lambda})}_{i}\I{-}
\biave{\hat{A}_{\alpha}(\vecb{\lambda})}_{i}\biave{\hat{A}_{\beta}(\vecb{\lambda})}_{i}\right]
\nonumber\\
=\sum_{i'(\neq i)}\frac{\matr{\psi_{i'}(\vecb{\lambda})}{\nabla_{\alpha}\hat{H}(\vecb{\lambda})}{\psi_{i}(\vecb{\lambda})}\matr{\psi_{i'}(\vecb{\lambda})}{\nabla_{\beta}\hat{H}(\vecb{\lambda})}{\psi_{i}(\vecb{\lambda})}^*}{\bigl(E_{i'}(\vecb{\lambda})\I{-}E_{i}(\vecb{\lambda})\bigr)^2}
\label{geo},
\end{eqnarray}
defined for an arbitrary eigenstate $\ket{\psi_i(\vecb{\lambda})}$ through the expectation values $\ave{\bullet}_i$ of the gauge potential components and their products in the state $\ket{\psi_i(\vecb{\lambda})}$.
The symmetric (real) part of the geometric tensor forms the metric tensor, and the antisymmetric (imaginary) part determines the Berry curvature tensor:
\begin{equation}
\hspace{-15mm}
g_{i\,\alpha\beta}(\vecb{\lambda})=\frac{1}{2}\,[\chi_{i\,\alpha\beta}(\vecb{\lambda})\I{+}\chi_{i\,\beta\alpha}(\vecb{\lambda})]
\,,\quad
F_{i\,\alpha\beta}(\vecb{\lambda})=\ii\ [\chi_{i\,\alpha\beta}(\vecb{\lambda})\I{-}\chi_{i\,\beta\alpha}(\vecb{\lambda})]
\label{metcurv}\,.
\end{equation} 
We stress that all these definitions are apparently invariant under a local phase transformation ${\ket{\psi_{i'}(\vecb{\lambda})}\mapsto e^{\ii\varphi(\vecb{\lambda})}\ket{\psi_{i'}(\vecb{\lambda})}}$ of all eigenvectors.

Both metric and curvature tensors describe the response of the system to nearly adiabatic driving.
It is well known that the integral of the Berry curvature tensor over an area inside a closed curve in the parameter space determines the geometric phase acquired by the adiabatically evolving system during a single loop along the curve \cite{Wilcz89}.
The variation of the phase of the $i$th eigenstate by adiabatic driving can be understood as a parallel transport of a vector on a curved manifold. 
The metric tensor defines an infinitesimal distance on this manifold.
In particular, we have
\begin{equation}
ds_i^2(\vecb{\lambda})=\sum_{\alpha,\beta}g_{i\,\alpha\beta}(\vecb{\lambda})\,d\lambda_{\alpha}d\lambda_{\beta}=1-\left|\scal{\psi_i(\vecb{\lambda}\I{+}d\vecb{\lambda})}{\psi_i(\vecb{\lambda})}\right|^2
\label{metric},
\end{equation}
where the second expression shows that the squared distance $ds^2_i(\vecb{\lambda})$ represents a probability of the departure from the $i$th eigenstate after an infinitesimal quench from $\vecb{\lambda}$ to ${\vecb{\lambda}+d\vecb{\lambda}}$.
Moreover, the quantity $\hbar^2\dot{s_i}(t)^2$, proportional to the squared metric velocity ${\dot{s}_i=\sum_{\alpha,\beta}g_{i\,\alpha\beta}\dot{\lambda}_\alpha\dot{\lambda}_\beta}$, can be shown to coincide with the variance of energy ${\Delta E_i(t)^2=\dis{\hat{H}(\vecb{\lambda}(t))^2}_{\ket{\psi(t)}}}$ in the instantaneous state $\ket{\psi(t)}$ in the leading order of the adiabatic perturbation theory \cite{Kolod17,Bukov19}.
While the evolving level energy $E_i(\vecb{\lambda}(t))$ is analogous to a potential energy that measures the reversible energy costs of adiabatic driving of the $i$th eigenstate, the dispersion $\Delta E_i(t)$ determines sort of friction induced by this driving.
Maximizing the fidelity of the system delivery into the $i$th eigenstate of the final Hamiltonian is equivalent to minimizing the friction induced by driving.

Consider a set of trajectories $\vecb{\lambda}(t)$ in the parameter space leading from point $\vecb{\lambda}_{\rm in}$ at time ${t=0}$ to point $\vecb{\lambda}_{\rm fi}$ at time ${t=\tau}$.
The final time must be sufficiently large for the first-order adiabatic perturbation formulas be valid.
The minimization of the energy variance averaged along the whole trajectory, 
\begin{equation}
\hspace{-15mm}
\frac{1}{\tau}\int_{0}^{\tau}\!\!\! dt\ \Delta E_i(t)^2
=\frac{\hbar^2}{\tau}\int_{0}^{\tau}\!\!\! dt \sum_{\alpha,\beta}g_{i\,\alpha\beta}\bigl(\vecb{\lambda}(t)\bigr)\,\dot{\lambda}_{\alpha}(t)\dot{\lambda}_{\beta}(t)
=\frac{\hbar^2}{\tau}\int_{0}^{\tau}\!\!\! dt\ \dot{s_i}(t)^2
\label{Eflu},
\end{equation}
selects the geodesic path ${\vecb{\wp}\equiv\{\vecb{\lambda}(t)\}_{t=0}^\tau}$ minimizing the distance between both points $\vecb{\lambda}_{\rm in}$ and $\vecb{\lambda}_{\rm fi}$, with the time dependence such that the metric velocity $\dot{s}_i(t)$ remains constant \cite{Kolod17,Bukov19}.
This means that near the points $\vecb{\lambda}$ where the metric tensor is large, the speed $\dot{\vecb{\lambda}}$ is small, and vice versa.
As seen from Eq.\,\eqref{geo}, the metric tensor is large if the gaps between energy levels are reduced and/or when the corresponding matrix elements of the Hamiltonian gradient are enlarged.
Note that these conclusions hold also in systems with a single parameter $\lambda$.

\begin{figure}[t!]
\begin{flushright}
\includegraphics[width=\textwidth]{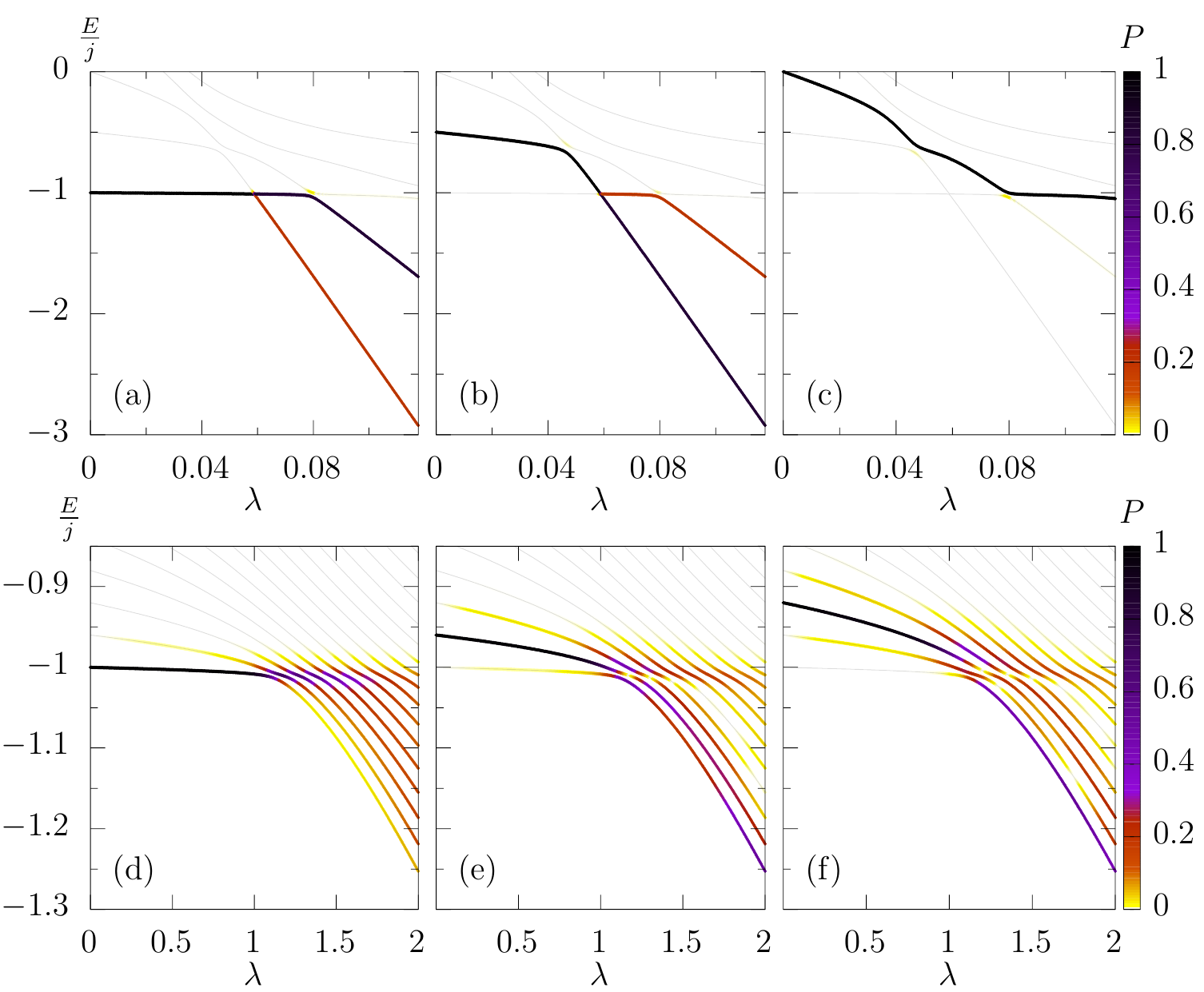}
\end{flushright}
\caption{
Evolution of eigenstate populations (encoded in the color of the curve corresponding to the respective energy level with running $\lambda$) for a system described by the first- and second-order {\QPT} Lipkin Hamiltonian (upper and lower rows, respectively) during a slow change of the control parameter $\lambda$ across {\QPT} and {\ESQPTs} spectral structures.
In panels from left to right in each row, the drive to the strongly interacting phase behind the {\QPT} is shown for the system initially prepared in the ground state, and in the first and second excited state of the non-interacting Hamiltonian.
Panels (a)--(c) corresponds to the passage through the first-order {\QPT} region [Hamiltonian \eqref{HLipkin} with ${\chi\I{=}4}$ and ${N\I{=}2j\I{=}4}$] at speed ${\dot{\lambda}\I{=}10^{-4.5}}$,  panels (d)--(f) to the passage through the second-order {\QPT} region [Hamiltonian \eqref{HLipkin} with ${\chi\I{=}0}$ and ${N\I{=}2j\I{=}100}$] at speed ${\dot{\lambda}\I{=}0.2}$.
We observe that non-adiabatic transitions to other states occur during the passage through avoided crossings of energy levels.
Courtesy of J.\,Dolej{\v s}{\'\i} \cite{Dolej20}.
}
\label{adia}
\end{figure}

The reduction of speed $\dot{\vecb{\lambda}}(t)$ must be performed particularly near quantum critical points.
This applies if the ground state of the system is driven through a {\QPT}.
If an excited state is involved, the slow down is needed during the passage through {\ESQPTs} characterized by a local increase of the level density or by a higher occurrence of sharp avoided crossing of levels.
These conclusions are schematically illustrated in Fig.\,\ref{adia} using the familiar Lipkin Hamiltonian \eqref{HLipkin}.
The {\ESQPTs} are relevant also for driving of thermal states, so they influence the ground-state results in realistic situations with low but non-zero temperatures.
The increased duration of quasiadiabatic driving in presence of {\QPTs} and {\ESQPTs} may represent a serious obstacle to the related quantum information protocols.
This was discussed in connection with adiabatic quantum computation in Ref.\,\cite{Schut06}, showing that for the first-order {\QPT} the time requirements grow exponentially with the system's size which kills any potential advantage of quantum computing.
In a more general context, excitations generated by a passage of an isolated quantum system through a second-order {\QPT} were presented in parallel to creation of topological defects in the early Universe \cite{Zurek85}.
Extensive literature devoted to this subject can be traced in Ref.\,\cite{Polko11}.

Though a general application of the adiabatic gauge theory in systems with {\ESQPTs} is still missing, several particular studies of (quasi)adiabatic driving in such systems already exist.
They are mostly based on the Lipkin model (Sec.\,\ref{Quasi}) with second-order ground-state {\QPT} and a subsequent $(f,r)\I{=}(1,1)$ {\ESQPT} separating the normal and parity breaking quantum phases.
The work \cite{Puebl15} reports on an irreversible mixing of states with different parity quantum numbers achieved in procedures involving cyclic adiabatic evolution across the {\ESQPT} borderline.
The work \cite{Kopyl17} analyzes various quantities characterizing the departure from adiabaticity during the passage of an excited system through the {\ESQPT} region.
In particular, the variance of energy induced by a non-adiabatic response to the driving at different speeds is shown to follow an algebraic scaling law with the exponent depending on the selected initial excited state.
This conclusion, which can be reproduced also on the semiclassical (mean-field) level, forms an {\ESQPT} analogy of the Kibble-Zurek mechanism known for the ground-state {\QPTs} \cite{Zurek85}.
The work \cite{Puebl20} analyzes modifications of this mechanism (so-called anti-Kibble-Zurek behavior) for the Lipkin (or Rabi, see Sec.\,\ref{EDM3}) system coupled within the Lindblad formalism to an external bath (cf.\,Sec.\,\ref{dissip}).
It is shown the the presence of {\ESQPT} modifies the non-adiabatic response of the open system to the external driving and leads to new algebraic scaling laws.
We stress that all these results are strictly valid only for the {\ESQPTs} in ${f=1}$ systems that show up as logarithmic divergences of the semiclassical level density.
Effects of less spectacular {\ESQPTs} in ${f>1}$ systems have not been studied and are probably weaker.

The Berry phase associated with a fully adiabatic cyclic driving of the Lipkin Hamiltonian was studied in Ref.\,\cite{Yuan12}.
Since the curvature tensor in Eq.\,\eqref{metcurv} feels the avoided crossing of energy levels, the Berry phase strongly reflects the {\QPT} and {\ESQPT} singularities. 
Indeed, the cited work examines the effect of the full rotation of the $\hat{J}_x$ operator around the $z$-direction (which activates the imaginary part of the geometric tensor), demonstrating that the Berry phase of the ground state as a function of the control parameter has a discontinuous derivative at the {\QPT} critical point, while the phases of excited states show diverging derivative at the {\ESQPT} critical borderline.
 
As indicated above, non-adiabatic effects represent a severe problem of quantum state preparation protocols based on slow driving.
A possible solution resorts to so-called {\em counterdiabatic driving} (also known as an adiabatic shortcut) \cite{Campo13,Sels17}.
The idea \cite{Demir03,Berry09} is simply to add to the Hamiltonian $\hat{H}(\vecb{\lambda}(t))$ a new component, which precisely compensates the action of the off-diagonal elements of the sum in Eq.\,\eqref{dridy} during the drive.
Thus we obtain the counterdiabatic Hamiltonian 
\begin{eqnarray}
\hspace{-11mm}
\hat{H}_{\rm CD}(t)=&&\hat{H}\bigl(\vecb{\lambda}(t)\bigr)+\hat{H}_{\rm cd}\bigl(\vecb{\lambda}(t),\dot{\vecb{\lambda}}(t)\bigr),
\label{Hcd}\\
&&\matr{\psi_{i'}(\vecb{\lambda})}{\hat{H}_{\rm cd}(\vecb{\lambda},\dot{\vecb{\lambda}})}{\psi_{i}(\vecb{\lambda})}=\dot{\vecb{\lambda}}\cdot
\matr{\psi_{i'}(\vecb{\lambda})}{\hat{\vecb{A}}(\vecb{\lambda})}{\psi_{i}(\vecb{\lambda})}.
\nonumber
\end{eqnarray}
The additional Hamiltonian $\hat{H}_{\rm cd}(\vecb{\lambda},\dot{\vecb{\lambda}})$ is assumed to have vanishing diagonal matrix elements in the instantaneous eigenbasis of $\hat{H}(\vecb{\lambda})$, while the off-diagonal elements are expressed through the gauge potential as written in second line of Eq.\,\eqref{Hcd}.

The counterdiabatic driving allows one, at least in theory, to shorten the time needed for a transitionless drive between $\vecb{\lambda}_{\rm in}$ and $\vecb{\lambda}_{\rm fi}$ to an arbitrarily small value.
The price to pay includes the costs of creating the counterbalancing field in the Hamiltonian.
With only limited resources there will exist a lower bound of the drive time, which brings us to the hot field of {\em quantum speed limits}.
The works \cite{Kolod17,Bukov19,Funo17} present very close links of quantum speed limits to geodesic paths and the theory outlined above.
As follows from these analyses, the minimal time $\tau_{i\,{\rm min}}$ needed to complete the counterdiabatic drive of the $i$th eigenstate between points $\vecb{\lambda}_{\rm in}$ and $\vecb{\lambda}_{\rm fi}$ is achieved for the geodesic path connecting these points.
The minimal time is given by
\begin{equation}
\tau_{i\,{\rm min}}(\vecb{\lambda}_{\rm in},\vecb{\lambda}_{\rm fi})=\frac{\hbar}{\ave{\Delta E_{i\,{\rm CD}}}_{\vecb{\wp}}}\
s_{i\,{\rm min}}(\vecb{\lambda}_{\rm in},\vecb{\lambda}_{\rm fi})
\label{speedli},
\end{equation}
where the quantity $\ave{\Delta E_{i\,{\rm CD}}}_{\vecb{\wp}}$ estimates the energy cost of the driving protocol.
It is the square root of the variance of the counterdiabatic Hamiltonian \eqref{Hcd} in the evolving state averaged along the whole path $\vecb{\wp}$.
The length $s_{i\,\rm min}(\vecb{\lambda}_{\rm in},\vecb{\lambda}_{\rm fi})$ of the geodesic path  has a crucial impact on the feasibility of the protocol at any selected scale of the energy cost.
In some cases, it can be just a factor of an order of unity, but for paths crossing {\QPT} or {\ESQPT} singularities it quickly increases with the size of the system. 
To keep the time $\tau_{i\,{\rm min}}$ in Eq.\,\eqref{speedli} short enough in the latter situations may put unrealistic energy requirements on the driving procedure.
This limits the performance of the counterdiabatic driving protocols in quantum critical systems.

An example of an {\ESQPT}-induced effect on the quantum speed limit (though not defined in the same way as outlined above) was presented in Ref.\,\cite{Wang19} within the generalized Lipkin model of a qubit coupled to a fully connected spin bath \cite{Relan08,Peres09}, see the text around Eq.\,\eqref{decohe}.
The {\ESQPT} linked to diverging semiclassical level density of the spin bath leads to a sharp increase of the time needed to emulate the evolution of the qubit if the bath is excited to the {\ESQPT} domain.
The problem is open for further studies using alternative state preparation protocols in various {\ESQPT} systems.

\subsection{Dissipative systems and feedback control}
\label{dissip}

An interesting question concerns the fate of {\ESQPT} effects in presence of dissipation.
So far, this question was addressed only in a rather limited number of studies.
The state of a quantum system interacting with a Markovian environment is commonly modeled by a time-dependent density operator $\hat{\varrho}(t)$ obeying a master equation in the Lindblad form \cite{Gardi14,Binde19,Manza20}.
For instance, for a Lipkin-type system (Sec.\,\ref{Quasi}), expressed solely in terms of collective quasispin operators \eqref{quasis}, the Lindblad master equation can read as follows \cite{Lee14,Kopyl15b},
\begin{eqnarray}
\hspace{-10mm}
\frac{d\hat{\varrho}(t)}{dt}=&&-\frac{\ii}{\hbar}\left[\hat{H}_{\rm NH}(\lambda,\kappa)\hat{\varrho}(t)-\hat{\varrho}(t)\hat{H}^{\dag}_{\rm NH}(\lambda,\kappa)\right]+\frac{\kappa}{N}\hat{J}_{+}\hat{\varrho}(t)\hat{J}_{-},
\label{lindbl}\\
&& \hat{H}_{\rm NH}(\lambda,\kappa)=\hat{H}(\lambda)-\ii\frac{\kappa}{2N}\hat{J}_{-}\hat{J}_{+}
\label{Heff},
\end{eqnarray}
where $\hat{H}_{\rm NH}(\lambda,\kappa)$ is an effective non-Hermitian Hamiltonian composed of a Hermitian term $\hat{H}(\lambda)$ describing the original conservative system depending on parameter $\lambda$, and an anti-Hermitian term weighted by coefficient $\kappa$ describing the coupling to the environment. 
The last terms in Eqs.\,\eqref{lindbl} and \eqref{Heff} express the action of the environment on the system.
It needs to be stressed, however, that it is implicit in form of these equations that the system and environment remain decoupled for all times.
More complex dynamics obtained if this assumption is lifted leads to non-Markovian master equations \cite{Gardi14,Binde19}, which are generally difficult to solve and have not been considered yet in connection with {\ESQPTs}. 

\begin{figure}[t!]
\begin{flushright}
\includegraphics[width=0.85\textwidth]{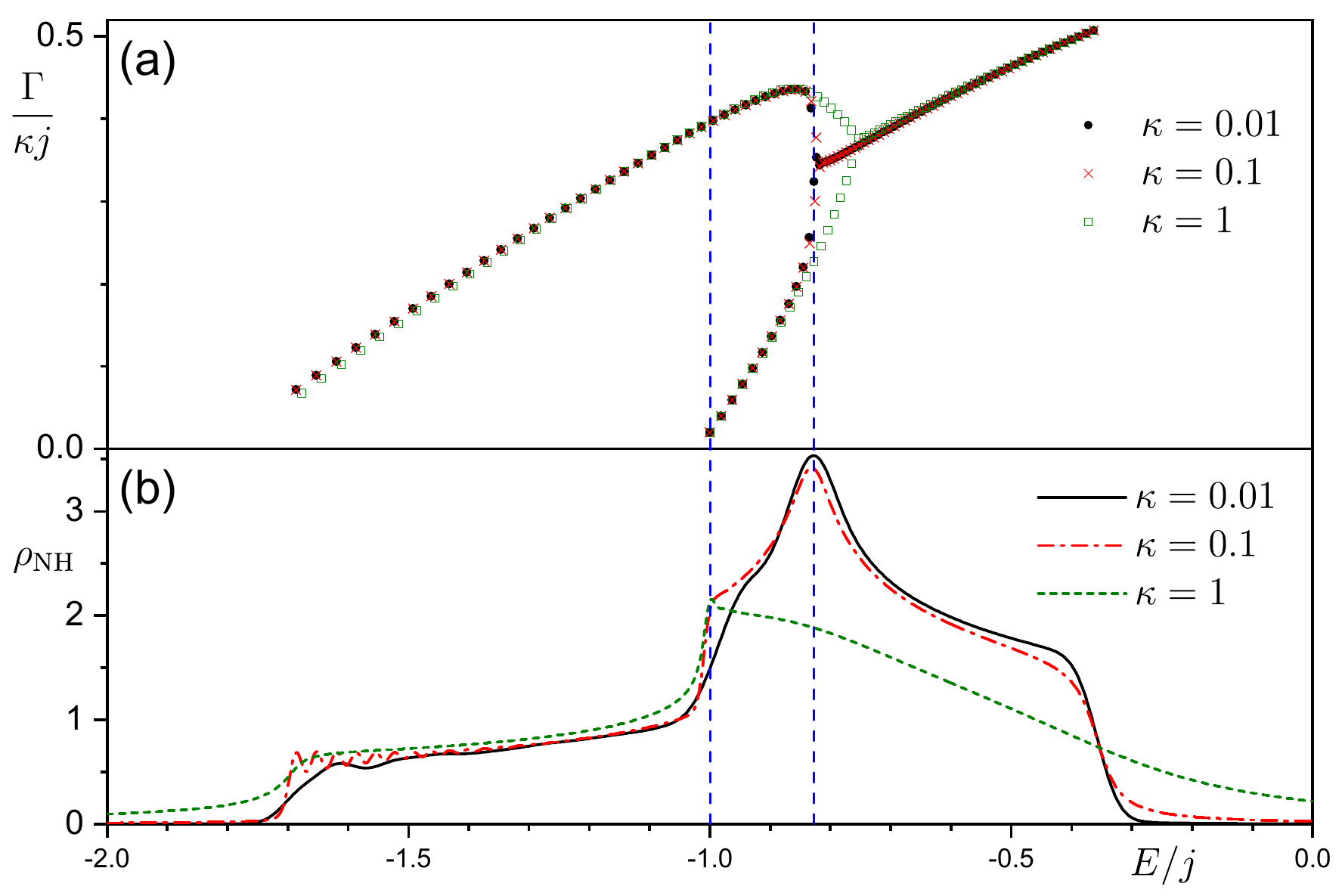}
\end{flushright}
\caption{The complex energy spectrum of the effective Hamiltonian \eqref{Heff} for various values of the dissipation strength $\kappa$.
The Hermitian term $\hat{H}(\lambda)$ coincides with the Lipkin Hamiltonian from Fig.\,\ref{lipkac} at $\lambda\I{=}0.08$ (the dashed vertical line in Fig.\,\ref{lipkac}) with $N\I{=}2j\I{=}100$.
Panel (a) shows the energy spectra in the complex plane and panel (b) depicts the corresponding \uvo{non-Hermitian} level density \eqref{contlev}.
The level density is smoothed by integration over an interval whose width decreases with increasing $\kappa$.
For the two lower values of $\kappa$ we observe remnants of the $(f,r)=(1,0)$ and $(1,1)$ {\ESQPTs} from the $\kappa\I{=}0$ system (cf.\,Fig.\,\ref{lipkac}). 
}
\label{neher}
\end{figure}

If the original Hamiltonian $\hat{H}(\lambda)$ shows an {\ESQPT}, one can look whether the complex-valued spectrum of the effective Hamiltonian $\hat{H}_{\rm NH}(\lambda,\kappa)$ carries some signatures of the original singularity.
The answer found in Ref.\,\cite{Kopyl15b} is positive.
An example for another Hamiltonian (the same as in Fig.\,\ref{lipkac}) can be seen in Fig.\,\ref{neher}, where we show, for several values of $\kappa$, individual complex energies ${\cal E}_i(\lambda,\kappa)={E_i(\lambda,\kappa)-\frac{\ii}{2}\Gamma_i(\lambda,\kappa)}$ (with $E_i\I{\in}\mathbb{R}$ and $\Gamma_i\I{\in}\mathbb{R}$, respectively, standing for the real energy and decay width of the level) in the domain containing the {\ESQPTs} accompanying the first-order {\QPT} of the original Hamiltonian $\hat{H}(\lambda)$.
The complex spectra in Fig.\,\ref{neher} can be characterized by a \uvo{non-Hermitian} level density defined as
\begin{equation}
\hspace{-23mm}
\rho_{\rm NH}(\lambda,\kappa,E)=-\frac{1}{\pi}\,{\rm Im}\,{\rm Tr}\,\frac{1}{E\I{-}\hat{H}_{\rm NH}(\lambda,\kappa)}=
\sum_{i}\frac{1}{\pi}\frac{\frac{1}{2}\Gamma_i(\lambda,\kappa)}{\bigl(E\I{-}E_i(\lambda,\kappa)\bigr)^2+\frac{1}{4}\Gamma_i(\lambda,\kappa)^2}
\label{contlev},
\end{equation}
cf.\,the right-hand side of Eq.\,\eqref{led}. 
Note that in the limit $\kappa\I{\to}0$ (no dissipation) we get $\Gamma_i\I{\to}0$, so the level density \eqref{contlev} becomes a chain of $\delta$-functions defining the ordinary level density \eqref{led} of a closed system.
In Fig.\,\ref{neher} we use some $\kappa$-dependent smoothing of $\rho_{\rm NH}(\lambda,\kappa,E)$, hence the density showed there represents an extension of the smoothed level density \eqref{leds} to an open system. 
It is clear from this figure that the singularities present in the smoothed level density at $\kappa\I{=}0$ survive in the density \eqref{contlev} up to relatively large values of $\kappa$.  

However, a different conclusion concerning the {\ESQPT} effects in presence of dissipation is drawn from the behavior of expectation values of various observables.
Using the classical Hamiltonian dynamics for a closed system with $\kappa\I{=}0$, one can determine relations between time averages $\ave{J_{\alpha}}$ of individual quasispin operators along individual trajectories \cite{Engel15}.
These relations capture mutual dependencies of the corresponding quantum expectation values $\ave{\hat{J}_{\alpha}}_i$ in individual eigenstates for large values of the size parameter $N$, and reveal singularities related to the {\ESQPTs} (cf.\,Sec.\,\ref{Leflo}).
A similar problem can be solved also for a dissipative system.
The time-dependent expectation value of quantity $\hat{A}$ in density matrix $\hat{\varrho}(t)$ following the master equation \eqref{lindbl} is calculated as ${A(t)={\rm Tr}[\hat{A}\hat{\varrho}(t)]}$.
Using the mean-field approach, in which one assumes ${{\rm Tr}[\hat{A}\hat{B}\hat{\varrho}(t)]\approx A(t)B(t)}$, it is possible to derive from Eq.\,\eqref{lindbl} a~set of coupled first-order differential equations for quasispin components $J_{\alpha}(t)$ \cite{Kopyl15b}.
Since these equations conserve ${J^2(t)=\sum_{\alpha=x,y,z}J^2_{\alpha}(t)}$, the dynamics can be visualized on the Bloch sphere, which allows us to analyze various dynamical phases of the dissipative Lipkin model.
We can also calculate time averages $\ave{J_{\alpha}}$ and compare them with the situation at $\kappa\I{=}0$.
The outcome of this analysis is that in presence of dissipation, the {\ESQPT} singularities of the $\kappa\I{=}0$ case get smoothed \cite{Kopyl15b}.  

A specific way to restore the \uvo{{\ESQPT} signal} in dissipative quantum systems was studied in Ref.\,\cite{Kopyl15b}.
The method was based on the so-called {\em feedback control\/} of the system.
This type of control is achieved by variations of the Hamiltonian parameter $\lambda$ with respect to the values of some dynamical quantities characterizing the system's actual evolution.
For example, consider the Lipkin Hamiltonian \eqref{HLipkin} with $\chi\I{=}0$, which shows the second-order {\QPT} from the non-interacting to interacting phase at $\lambda\I{=}\lambda_{\rm c}$.
In this case, the interaction strength can be controlled via the formula ${\lambda(t)=\lambda_0+\gamma[J_z^2(t)\I{-}J_z^2(t\I{-}\tau)]/N^2}$, where $\lambda_0\I{>}\lambda_{\rm c}$ and $\gamma\I{>}0$ are constants and ${J_z(t')={\rm Tr}[\hat{J}_z\hat{\rho}(t')]}$ is the average of $\hat{J}_z$ at time $t'$ \cite{Kopyl15b}.
The value $\lambda(t)$ reflects the difference between $J_z^2(t')$ at the current time ${t'=t}$ and a past time ${t'=t-\tau}$, where $\tau$ is a selected time delay.
This means that an increase (decrease) of $|J_z(t)|$ over the period $\tau$ induces an increase (decrease) of the interaction strength.
The description of the controlled system is a non-trivial task involving the stability analysis of the resulting equations of motions in various regimes and the determination of the system's dynamic phases.

Here we focus only on a single output of Ref.\,\cite{Kopyl15b}, namely the fact that the $\kappa\I{>}0$ system controlled according to the above formula with a suitable choice of parameters restores the non-analytic dependencies of observables characterizing the $\lambda\I{>}\lambda_{\rm c}$ {\ESQPT} in the $\kappa\I{=}0$ system.
This seemingly surprising result may be understood from the anomalous localization of the Hamiltonian eigenstates close to the {\ESQPT} in the $\ket{m\I{=}-j}$ basis state, which is associated with the $J_z\I{=}-j$ saddle point of the $\lambda\I{>}\lambda_{\rm c}$ classical-limit Hamiltonian.
This unstable stationary point is a remnant of the global minimum of the $\lambda\I{<}\lambda_{\rm c}$ classical Hamiltonian and its stability decreases with $\lambda$ increasing above the critical point.
Hence the above feedback control works so that any departure of the system from the stationary point induces decrease of $\lambda(t)$, which enforces stabilization of the system at the stationary point.
An analogous technique can be used also in the Tavis-Cummings regime of the Dicke model (Sec.\,\ref{Open}) with similar types of {\QPT} and {\ESQPT} singularities \cite{Kopyl15a}.
Stabilization of other {\ESQPT} types would require different feedback control schemes.
Note also that an alternative control scheme for the same Lipkin Hamiltonian was studied in Ref.\,\cite{Zimme18}.

An entirely different perspective on quantum dissipative dynamics was presented in Ref.\,\cite{Furtm17}, where the Lindblad formalism was used to determine properties of the Hamiltonian eigenstates in non-dissipative systems. 
The method was applied in the analysis of the {\ESQPT} scaling behavior in the Tavis-Cummings model.

\section{Extensions to other systems}
\label{Exten}

So far, the signatures of {\ESQPTs} have been studied in energy conserving spatially localized bound quantum systems with discrete energy spectra.
Now we will transcend these limits, showing that close analogues of {\ESQPTs} exist also in systems of other types.
At first we will consider spatially extended periodic systems, then we will look at periodically driven systems, and finally we will discuss unbound scattering systems.

\subsection{Spatially extended periodic systems}
\label{Spati}

As stressed on many places in this review, noticeability of the {\ESQPT} singularities in quantum spectra requires, at least in typical situations, that the many-body system comprising of $N\I{\to}\infty$ constituents activates only a moderate number $f$ of effective {\DoFs}.
That is why we have so far discussed spatially localized systems (hereafter abbreviated as $D\I{=}0$ systems) with strictly collective behavior.
The spatially extended systems, like lattices in $D\I{=}1,2,...$ spatial dimensions (1D, 2D,\,... crystals), seem excluded from these considerations since each of their $N$ constituents (lattice sites) brings in principle one or more independent {\DoFs}.
However, it turns out that singularities of very similar nature as {\ESQPTs} can appear even in such $D\I{>}0$ systems, though exclusively under the condition of somehow restricted dynamics activating only few-particle or, on the contrary, only some collective modes of motions.

First, let us consider motion of a single particle in a periodic lattice with $N\I{\to}\infty$ sites in $D\I{>}0$ spatial dimensions.
The dynamics is described by the Bloch theory, which explains the onset of the band structure of the energy spectrum. 
An essential object of Bloch theory is the {\em dispersion relation} $E(\vecb{\mathfrak{p}})$ defining the dependence of the particle energy on the quasimomentum $\vecb{\mathfrak{p}}\I{=}\hbar\vecb{k}$, where $\vecb{k}$ is the wave vector in the reciprocal lattice.
The smoothed density of states at energy $E$ in the given band is determined from the size of the $E(\vecb{\mathfrak{p}})\I{=}E$ manifold in the quasimomentum space, namely by the formula  
\begin{equation}
\overline{\rho}(E) =\frac{1}{(2\pi\hbar)^f} \int_{\mathscr{B}} d^f\vecb{\mathfrak{p}}\ \delta\bigl(E-E(\vecb{\mathfrak{p}})\bigr)
\label{DoS},
\end{equation}
where the integration is confined to the domain $\mathscr{B}$ of the quasimomentum space associated with the selected Brillouin zone. 
For a single particle, the dimension of the quasimomentum space is $f\I{=}D$.
However, the same approach can be used for $n\I{\geq}1$ mutually non-interacting particles moving in the lattice.
In this case we define the total $n$-particle energy $\sum_{k=1}^n E(\vecb{\mathfrak{p}}_k)=E(\vecb{\mathfrak{p}}_1,...,\vecb{\mathfrak{p}}_n)\equiv E(\vecb{\mathfrak{p}})$ depending on the total quasimomentum vector $\vecb{\mathfrak{p}}\equiv(\vecb{\mathfrak{p}}_1,...,\vecb{\mathfrak{p}}_n)$. 
So in general $f=nD$ in Eq.\,\eqref{DoS}. 

Formula \eqref{DoS} strongly resembles the familiar phase-space relation \eqref{smole} used in $D\I{=}0$ systems. 
It allows for a similar analysis as that presented in Sec.\,\ref{Leden}, disclosing that stationary points of the $E(\vecb{\mathfrak{p}})$ dependence generate singularities of the state density $\overline{\rho}(E)$.
An important difference is the lower dimension $f$ of integration in Eq.\,\eqref{DoS} compared to $2f$ in Eq.\,\eqref{smole}, which shifts the level-density singularities to lower derivatives.  
Another news is that the dimension of integration can be odd, which for degenerate stationary points makes all types of singularities sketched in Fig.\,\ref{Typ} potentially relevant.
The singularity typically appears in the $k$th derivative of the level density, where $k=(f\I{-}2)/2$ for $f$ even and  $k=(f\I{-}1)/2$ for $f$ odd.

Non-analyticities of the density of states of periodic lattice systems related to the saddle points of $E(\vecb{\mathfrak{p}})$ were first described already in 1953 by van Hove \cite{Hove53}.
Since then, these {\em van Hove singularities\/} became relevant in various condensed-matter settings. 
They appear in electronic spectra of solids \cite{Bassa75} and in this context they are often mentioned in connection with the high-temperature superconductivity \cite{Newns92} and electronic properties of 2D materials such as graphene \cite{Yuan19}.
In particular, as first described by Lifshits \cite{Lifshi60,Blant94}, the instability related to the crossing of the van Hove singularity with the Fermi energy may trigger phase changes which substantially alter the electronic conductance properties of the system.
The connection of van Hove singularities to {\ESQPTs} was recognized in Ref.\,\cite{Dietz13} in the context of so-called photonic crystals, realized experimentally in superconducting microwave resonators (\uvo{billiards}) \cite{Dietz15}.
A support of the {\ESQPT}-like interpretation of the observed singularity was gained from the verified logarithmic scaling of the singularity with the system's size, which was related to the results of Ref.\,\cite{Capri08}.

Despite these applications in electronic spectroscopy, the original van Hove idea \cite{Hove53} was related to the spectra of crystal vibrations.
In these cases, the particle moving through the $D$-dimensional periodic structure is not an electron but a phonon.
This takes us closer to the original {\ESQPT} framework of collective many-body systems. 
The number $N$ of atoms in the crystal (sites in the lattice) represents the size parameter, but the {\DoFs} connected with these atoms are frozen except $nD$ collective {\DoFs} describing vibrational modes of the whole crystal, where $n$ is the total number of phonons considered.
The $N\I{\to}\infty$ limit triggers non-analyticities that, as follows from Eq.\,\eqref{DoS}, typically appear in the zeroth (for $nD\I{=}1,2$), first (for $nD\I{=}3,4$) or higher (for $nD\I{\geq}5$) derivatives of the state density (Fig.\,\ref{Typ}). 
An algebraic theory of collective vibrations in $D\I{=}1$ and 2 lattices was given in Refs.\,\cite{Iache15,Dietz17}.

Here we give an example of {\ESQPT}-related effects in the $D\I{=}1$ {\it Bose-Hubbard model\/} with $N$ lattice sites \cite{Fishe89}.
We have already mentioned the $N\I{=}2$ version of this model in Sec.\,\ref{Ibms}, but now we consider $N\I{\gg}1$. 
We assume an ordered (equidistant) arrangement of sites in the linear geometry.
A simple Hamiltonian reads [cf.\,Eq.\,\eqref{HBH}]
\begin{equation}
\hat{H}=\varepsilon\sum_{i=1}^N\hat{b}^\dag_i\hat{b}_i-\tau\!\!\sum_{\langle ii'\rangle=1}^N (\hat{b}^\dag_i\hat{b}_{i'}+\hat{b}^\dag_{i'}\hat{b}_i)  
+\frac{U}{2}\sum_{i=1}^N\hat{b}^\dag_i\hat{b}^\dag_i\hat{b}_i\hat{b}_i
\label{BHam},
\end{equation}
where $\hat{b}^\dag_i$ and $\hat{b}_i$ create and annihilate the boson at the $i$th site.
The first term sums equal one-body energies $\varepsilon$ of individual bosons at all sites.
In applications related to cold atoms in optical lattices $\varepsilon$ is interpreted as the chemical potential and usually takes a negative value \cite{Gardi14}, while in the theory of crystal vibrations a natural choice of $\varepsilon$ is positive \cite{Iache15}. 
The second term of Eq.\,\eqref{BHam}, scaled by the parameter $\tau\I{\geq}0$, describes hopping of bosons between all neighboring sites $i$ and ${i'=i\I{+}1}$ (for ${i=N}$ the hopping term is zero), which is written as the sum over $\langle ii'\rangle$.
We set the Dirichlet boundary conditions for wave functions.
Note that in a way generalizable to higher dimension $D$ we can say that the lattice is bi-partite, i.e., consists of two sublattices between which the hopping takes place. 
The third term represents two-body interactions of bosons within the same site.
Here the interaction is taken as repulsive, hence weighted by a constant $U\I{\geq}0$.
Interactions between bosons from the neighboring (or even more distant) sites are neglected. 

The total number of bosons ${\hat{n}=\sum_i\hat{b}^\dag_i\hat{b}_i}$ is an integral of motion for the Hamiltonian \eqref{BHam}, so the energy spectrum consists of mutually non-interacting bands of states with ${n=0,1,2,...}$.
In the absence of hopping, i.e., for $\tau\I{=}0$, the system trivially conserves also the numbers of bosons on all individual sites, ${\hat{n}_i=\hat{b}^\dag_i\hat{b}_i}$, so the eigenstates of the Hamiltonian coincide with states $\ket{n_1,n_2,...,n_N}$.
In this case, the ground state has a localized, so-called Mott insulator form \cite{Sachd99}.
The populations of all sites are equal, given by values $n_i\I{=}0$ for $\varepsilon\I{>}0$ and $n_i\I{=}1,2,...$ for $\varepsilon\in(-U,0)$, $(-2U,-U)$, ..., respectively.
At the boundary values of $\varepsilon$ the configurations with the corresponding pair of the $n_i$ numbers become degenerate.
If hoping is switched on by setting $\tau\I{>}0$, the bosons get delocalized and numbers $n_i$ can in general show quantum fluctuations.
As $-\varepsilon$ increases, the Mott-insulator phases with sharp values $n_i\I{=}0,1,2,...$ alternate with intervals in which $n_i$ fluctuates around an expectation value $\ave{\hat{n}_i}$ that continuously interpolates the respective pairs of Mott-insulator values. 
In these intervals, the ground state belongs to the superfluid phase.
The transition between the superfluid phase and each of the Mott-insulator phases is a continuous {\QPT}.
A detailed quantum analysis of the ground-state phases of the Bose-Hubbard model with bi-partite $D$-dimensional lattices and possible inclusion of disorder was performed in Ref.\,\cite{Freer96}.

\begin{figure}
\begin{flushright}
\includegraphics[width=0.8\textwidth]{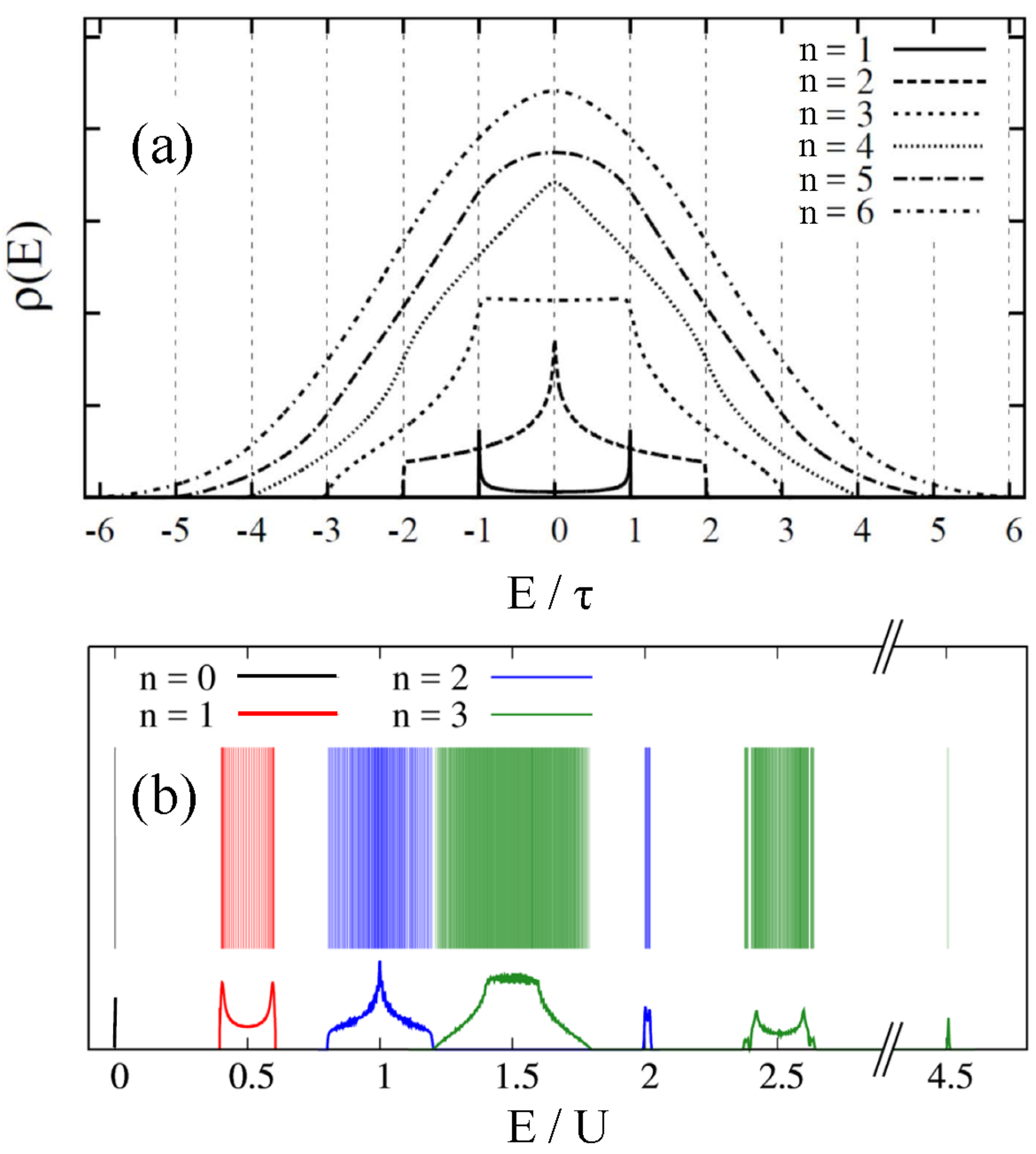}
\end{flushright}
\caption{Energy spectra of the Bose-Hubbard model \eqref{BHam} on a 1D linear chain.
Panel~(a): Level densities of the $N\I{\to}\infty$ limit of the Hamiltonian with $\varepsilon\I{=}U\I{=}0$ and $\tau\I{=}0.5$ for various boson numbers (adapted from Ref.\,\cite{Macek19b}). 
Panel~(b): Actual $N\I{=}30$ spectrum of the model with $\varepsilon\I{=}0.5$, $\tau\I{=}0.05$ and $U\I{=}1$ including the bands with $n\I{=}0,1,2,3$. Energy levels are marked by thin vertical lines and the corresponding smoothed level densities by curves at the bottom.
The components of the spectrum with different $n$ are distinguished by color.
The smoothed level densities were obtained by convolution of the real spectrum with Gaussian functions whose widths and normalization factors were chosen separately for each $n$ to make the curves comparable.}
\label{Huba}
\end{figure}

In addition to the rich {\QPT} results, careful examination of excited spectra of the Bose-Hubbard model shows also numerous {\ESQPT}-like singularities of the van Hove type \cite{Iache15,Dietz17,Macek19b}.
These appear in level densities describing the bands of states with various values of $n$.
The dispersion relation $E(\mathfrak{p}_1,...,\mathfrak{p}_n)$ for an arbitrary fixed number $n$ of bosons with quasimomenta $\mathfrak{p}_{l}$ can be calculated analytically in the free-hopping limit $U\I{=}0$.
The level densities of the corresponding bands in the infinite-size limit $N\I{\to}\infty$ are depicted in panel (a) of Fig.\,\ref{Huba}.
In agreement with the above explanations, the singularities of the level density connected with saddles of the dispersion relation are apparent in the zeroth derivative for the $n\I{=}1$ and 2 bands.
In particular, we observe inverse square-root singularities at the edges of the $n\I{=}1$ band (see the right-hand panel of Fig.\,\ref{Typ}) and a logarithmic singularity in the middle of the $n\I{=}2$ band (the left-hand panel of Fig.\,\ref{Typ}). 
For $n\I{=}3$ and 4, the singularities affect analogously the first energy derivative of the level density, and for $n\I{\geq}5$ they appear in even higher derivatives.
As $n$ increases, the level density of the band tends to a smooth Gaussian-like profile.

The actual spectrum of the Bose-Hubbard Hamiltonian with a non-vanishing repulsive interaction of bosons within each site is exemplified in panel (b) of Fig.\,\ref{Huba}.
There we see bands with the lowest values $n\I{=}0,1,2$ and 3 calculated numerically for the system in the $n_i\I{=}0$ Mott-insulator ground-state phase for $N\I{=}30$.
In this $\varepsilon\I{>}0$ setting, the low-$n$ bands correspond to low-energy part of the spectrum.
While the bands with $n\I{=}0$ and 1 keep the same level-density profile as in the $U\I{=}0$ case, those with $n\I{=}2,3,...$ become modified.
In particular, the intra-site repulsion induces detachment of a certain fraction of states to higher energies, which leads to splitting of the $n\I{\geq}2$ bands to several \uvo{sub-bands}.
The detached levels correspond to so-called overtone states with some of the site occupation numbers taking values $n_i\I{\geq}2$.
For $n\I{=}2$ there is only one such an overtone sub-band (at $E/U\I{\approx}2$ in Fig.\,\ref{Huba}), for $n\I{=}3$ there are two (at $E/U\I{\approx}2.5$ and 4.5), and so on.
The level density profile of the main sub-band (with all $n_i\I{\leq}1$) keeps the features of the $U\I{=}0$ limit with the corresponding $n$. 
Some of the overtone bands also exhibit precursors of van Hove singularities, but these are compatible with ${f=(n\I{-}1)D}$ (see, e.g., the $n\I{=}2$ case in Fig.\,\ref{Huba}).
We note, however, that in the limit $N\I{\to}\infty$ the number of overtone states becomes negligible in comparison with the number of states in the main sub-band.  

Level densities of bands with low values of $n$ for $D\I{=}2$ lattices of square and hexagonal shapes were determined analytically in Ref.\,\cite{Iache15}.
The resulting theoretical results were also compared with experimental data from microwave photonic crystals.
The structure of the Hamiltonian eigenfunctions for $n\I{=}1$ bands in these lattices was presented in Ref.\,\cite{Dietz17}.
There, the $k$th eigenfunction $\ket{\psi_k}$ was expanded in the $\ket{n_1,n_2,...,n_N}$ basis as ${\ket{\psi_k}=\sum_{i=1}^N\psi_k(x_i,y_i)\ket{0_1,...0_{i-1},1_i,0_{i+1},...0_N}}$, where the amplitudes $\psi_k(x_i,y_i)$ represent the wave function in the discrete coordinate space of lattice site positions $(x_i,y_i)$.
For the square lattice, the $n\I{=}1$ band exhibits a logarithmic divergence of the level density, while the $n\I{=}2$ band shows a non-analyticity in the first derivative of the level density.   
For the hexagonal (honeycomb) lattice, the level densities take more complicated forms, which are also known from the vast literature on graphene.
In particular, the $n\I{=}1$ band contains two logarithmic divergences of the level density separated by the so-called Dirac point with vanishing level density \cite{Iache15}.
In both lattice types, the wave functions $\psi_k(x_i,y_i)$ of the eigenstates close to the {\ESQPT} energies manifest some anomalous localization properties \cite{Dietz17,Macek19c}. 
This is reminiscent of ${D\I{=}0}$ systems with ${f\I{=}1}$, for which anomalous spatial localization of eigenstates is also observed near {\ESQPT} energies, but a possible deeper connection of these phenomena needs to be clarified. 

\subsection{Periodically driven systems}

In this section we will consider time-dependent Hamiltonians $\hat{H}(t)$ satisfying $\hat{H}(t)\I{=}\hat{H}(t\I{+}T)$, where $T=2\pi/\omega$ is a time period and $\omega$ the corresponding frequency.
The periodicity is ensured by an external driving of the system.
We stress that the present case is principally different from the examples of driven dynamics discussed in Sec.\,\ref{Dynam}. 
Here, we will not just vary parameters of a Hamiltonian which in the stationary case has an {\ESQPT}.
Phase transitions of the static version of the Hamiltonian (if any) are not important. 
Instead, we describe the evolution induced by the periodically driven Hamiltonian via a {\em stationary effective Hamiltonian\/} and study emergent critical properties (including potential {\ESQPTs}) of this Hamiltonian.

A general $T$-periodic Hamiltonian can be expressed as 
\begin{equation}
\hat{H}(t)=\hat{H}_0+\!\!\sum_{k\in{\mathbb Z}\setminus\{0\}}\hat{V}_k e^{\ii k\omega t}
\quad{\rm with}\ 
\hat{V}_{-k}=\hat{V}^{\dag}_{+k}
\label{Haper},
\end{equation}
where ${\hat{H}_0=\hat{H}_0^\dag}$ and $\hat{V}_k$ are arbitrary time-independent operators satisfying the given constraint.
The special cases, which are most often considered in the literature, are:
(i) monochromatically driven (sometimes called ac-driven) Hamiltonians 
\begin{equation}
\hspace{-12mm}
\hat{H}(t)=\hat{H}_0+\hat{V}_0+\hat{V}\cos(\omega t)+\hat{W}\sin(\omega t)
\quad\ {\rm with}\,
\left\{\begin{array}{l}
\hat{V}\I{=}\hat{V}_{1}\I{+}\hat{V}_{1}^{\dag},\\
\hat{W}\I{=}\ii\bigl(\hat{V}_{1}\I{-}\hat{V}_{1}^{\dag}\bigr),\\
\hat{V}_k\I{=}0\ {\rm for}\ |k|\I{\geq}2
\end{array}\right.
\label{Hamon}
\end{equation}
(where, without loss of generality, we may set either one of the operators $\hat{V}$ and $\hat{W}$ to zero), and (ii) periodically kicked Hamiltonians 
\begin{equation}
\hspace{-12mm}
\hat{H}(t)=\hat{H}_0+\hat{V}T\sum_{l=-\infty}^{+\infty}\delta(t\I{-}lT)
\quad\ {\rm with}\ \,
\hat{V}_k=\hat{V}\ \,\forall\, k
\label{Hakic},
\end{equation}
also known from the so-called quantum maps \cite{Haake10,Berry79} and time crystals \cite{Sacha15}.

The periodically driven systems are in some aspects similar to the spatially periodic stationary systems discussed in Sec.\,\ref{Spati}.
They can be described in terms of the {\em Floquet theory}, which is analogous to the Bloch theory of lattice systems.
In particular, the unitary evolution operator of the system over one period, the so-called Floquet operator, is expressed as \cite{Rahav03}
\begin{equation}
\hspace{-12mm}
\hat{U}(t_0,t_0\I{+}T)={\rm T}\exp\left[{-\frac{\ii}{\hbar}\int_{t_0}^{t_0+T}\!\!\!\!\! dt'\,\hat{H}(t')}\right]=
e^{-\ii\hat{K}(t_0)}e^{-\ii\hat{H}_{\rm F}T/\hbar}e^{\ii\hat{K}(t_0)}
\label{Floq},
\end{equation}
where the symbol ${\rm T}$ denotes the time ordering in the exponential expansion.
The Hermitian operator $\hat{H}_{\rm F}$ on the right-hand side of Eq.\,\eqref{Floq} is a stationary effective (Floquet) Hamiltonian (\uvo{quasi-Hamiltonian}) which expresses the evolution of the system in an appropriate time-dependent basis ${\{\ket{\phi_i(t)}=e^{\ii\hat{K}(t)}\ket{\phi_i}\}}$, with $\{\ket{\phi_i}\}$ denoting an arbitrary fixed basis.
The \uvo{gauge} transformation to and from this basis is generated by a time-dependent Hermitian operator $\hat{K}(t)$ satisfying ${\hat{K}(t)=\hat{K}(t+T)}$, and is applied in Eq.\,\eqref{Floq} at the initial and final times $t_0$ and ${t_0+T}$, respectively \cite{Rahav03,Goldm14}.
Therefore, if expressed in the basis $\ket{\phi_i(t_0)}$, the evolution of the system prepared at ${t=t_0}$ in an initial state $\ket{\psi_{\rm in}}$ (e.g., an eigenstate of $\hat{H}_0$) over the time interval ${\Delta t=t-t_0}$ equal to an integer multiple of the period $T$ is determined solely by the effective Hamiltonian $\hat{H}_{\rm F}$.
The situation is similar to the study of quantum quench dynamics (Sec.\,\ref{quench}), when we also analyzed the evolution of a state $\ket{\psi_{\rm in}}$ [an eigenstate of $\hat{H}(\lamin)$] by an \uvo{alien} stationary Hamiltonian $\hat{H}(\lamfi)$.
Eventual critical properties of the effective Hamiltonian $\hat{H}_{\rm F}$, and particularly the {\ESQPTs} in the spectrum of its eigenvalues ${\mathfrak E}_i$ (the so-called quasienergies), have similar consequences in periodically driven dynamics as the critical properties of $\hat{H}(\lamfi)$ in the quantum quench dynamics.

Methods for evaluation of the effective Hamiltonian are based on the Baker-Campbell-Hausdorff expansion.
In particular, for kicked systems \cite{Haake10,Schar88} the initial time $t_0$ can be naturally taken just before (or just after) the kick and the transformations with $\hat{K}(t_0)$ may be disregarded.
Hence $e^{-\ii\hat{V}T/\hbar}e^{-\ii\hat{H}_0T/\hbar}=e^{-\ii\hat{H}_{\rm F}T/\hbar}$.
This leads to an expansion of $\hat{H}_{\rm F}$ in powers of $T$, each term of the resulting series being expressed via the operators $\hat{H}_0$ and $\hat{V}$ \cite{Schar88,Grozd88}.
For large frequencies of the driving the expansion can be truncated at some low order.
Similar approximation techniques can be applied also to monochromatically driven Hamiltonians \cite{Engel13}.

It is clear that the choice of the initial time $t_0$, which alters the resulting evolution operator over one period, is ambiguous.
This holds already for kicked systems and becomes particularly problematic for general periodic systems with no privileged value of $t_0$.
The change of $t_0$ induces just an unitary transformation of the full effective Hamiltonian, so it does not affect exact quasienergies, but the choice of $t_0$ can play a more important role if $\hat{H}_{\rm F}$ is determined only approximately (e.g., in a given order of the expansion in the driving period).
The more sophisticated approach captured in Eq.\,\eqref{Floq}, which was developed in Refs.\,\cite{Rahav03,Goldm14,Eckar15}, solves this problem by the above-mentioned transformation to the gauge in which the Hamiltonian becomes truly time-independent.
The evaluation of both operators $\hat{H}_{\rm F}$ and $\hat{K}(t)$ can be again obtained in the form of an expansion in powers of $T$ \cite{Goldm14}.

The time-independent effective Hamiltonian resulting from the Floquet description of a periodically driven system can be subject to the same phase-transitional analysis as any stationary Hamiltonian.
In connection with {\ESQPTs}, such analyses were performed within the Lipkin model (Sec.\,\ref{Quasi}), based on the familiar $f\I{=}1$ collective quasispin algebra, for several specific cases of driving.
In Ref.\,\cite{Engel13}, the monochromatic driving of the form \eqref{Hamon} with ${\hat{H}_0\propto\hat{J}_z-(\gamma_x\hat{J}_x^2+\gamma_y\hat{J}_y^2)/N}$, and ${\hat{V}\propto\gamma_x^{\prime}\hat{J}_x^2}$ (while $\hat{W}\I{=}0$), where $\gamma_x,\gamma_y,\gamma_x^{\prime}$ are free parameters, was studied yet without explicit reference to the {\ESQPTs}.
An aproximate effective Hamiltonian was determined and analyzed to obtain the corresponding quasienergy surface in the phase space, which is in fact equivalent to the semiclassical {\ESQPT} analysis.
In Ref.\,\cite{Basti14a}, the kicked system of the form \eqref{Hakic} with ${\hat{H}_0\propto\hat{J}_z^2}$ and ${\hat{V}\propto\hat{J}_x}$ was considered and its approximate effective Hamiltonian was analyzed.
This led to a clear identification of the {\ESQPT}-like singularities in the zeroth derivative of the density of quasienergy levels.
In Ref.\,\cite{Basti14b}, both monochromatic and kicked driving protocols were revisited on a common basis, using slightly different forms of driving Hamiltonians with ${\hat{H}_0\propto\hat{J}_x}$ and ${\hat{V}\propto\hat{J}_z^2}$ in both cases.
The resulting effective Hamiltonians were shown to manifest {\ESQPT} singularities of all the types $(f,r)\I{=}(1,0)$, (1,1) and (1,2), both driving methods yielding of course different quasienergy level densities and {\ESQPT} critical borderlines.
Note that the singularities caused by the global minimum and maximum of the classical-limit effective Hamiltonian appear in the spectrum of quasienergies because of their mapping ${{\mathfrak E}_i\mapsto{\mathfrak E}_i\,\mathrm{mod}\,(\hbar\omega)}$ into the first Brillouin zone.
Finally, in Ref.\,\cite{Bondy15}, the same kicked system as in Ref.\,\cite{Basti14a} was reanalyzed in terms of the more advanced method based on Eq.\,\eqref{Floq} and Refs.\,\cite{Rahav03,Goldm14}.
The results on {\ESQPT} were fully verified and extended.

We emphasize that in recent years, the periodically driven quantum systems have become an important playground for various theoretical concepts and advanced experimental techniques.
The first experimental realization of the kicked top system was reported in Ref.\,\cite{Chaud09}.
In the meantime it was proven that periodic driving can be used to create and control new dynamical phases of matter in the context of various many-body systems---see Refs.\,\cite{Bukov15,Leros19}, the references therein, and also those in Refs.\,\cite{Basti14b,Bondy15,Eckar15}.
Therefore, the identification of {\ESQPT}-like singularities in such systems has potentially far reaching consequences.

\subsection{Unbound systems}

So far, our attention was focused solely on bound quantum systems with discrete energy spectra.
Only for such systems the level density in Eq.\,\eqref{led} makes sense.
However, even unbound quantum systems with continuous energy---known in the context of various scattering problems---can be described via a suitably defined \uvo{level density} \cite{Levin69,Krupp98,Krupp99}.
This so-called {\em continuum level density\/} is given by
\begin{equation}
\hspace{-10mm}
\delta\rho(E)=-\frac{1}{\pi}\,\lim\limits_{\epsilon\to 0}\,{\rm Im}\,
\left({\rm Tr}\,\frac{1}{E\I{+}\ii\epsilon\I{-}\hat{H}}-{\rm Tr}\,\frac{1}{E\I{+}\ii\epsilon\I{-}\hat{H}_0}\right)
=\frac{1}{\pi}\,\frac{\partial\varphi(E)}{\partial E}
\label{coled},
\end{equation}
where $\hat{H}=\hat{H}_0\I{+}\hat{V}$ is the full Hamiltonian of the scattering system composed of a~free Hamiltonian $\hat{H}_0$ and a finite-range interaction $\hat{V}$. 
Both trace terms in the middle part of Eq.\,\eqref{coled} have the same form as the right-hand side of Eq.\,\eqref{led}, i.e., they are expressed through the energy representation of the Green operator associated with the respective Hamiltonian, but here the traces are performed over the continuous set of energy states.
We immediately note that the continuum level density $\delta\rho(E)$, being defined as a difference of two positive terms, is not a positive definite quantity.
The last equality in Eq.\,\eqref{coled} captures the relation (derived, e.g., in Ref. \cite{Levin69}) of the continuous level density to the relative phase shift $\varphi(E)$ (its variation with energy) between the elastically scattered systems in presence and in absence of the interaction~$\hat{V}$.
This will be important in the forthcoming considerations.

The most important contribution to the continuum level density, which results from the subtraction of both terms in Eq.\,\eqref{coled}, comes from the states whose wave functions exhibit an increased spatial localization in the interaction region with non-vanishing ${\hat{V}=\hat{H}-\hat{H}_0}$.
These resonance states can be identified via the so-called {\em complex scaling method\/} \cite{Suzuk05,Suzuk08}.
In brief, the procedure consists of two steps \cite{Moise11}: 
First, the system is put to a finite box, so the traces in Eq.\,\eqref{coled} become sums over discrete eigenvalues.
Second, the coordinates and momenta in Hamiltonians $\hat{H}$ and $\hat{H}_0$ are rotated in the complex plane in such a way that the resonance states associated with the full Hamiltonian $\hat{H}$ become normalizable eigenstates of the non-Hermitian Hamiltonian $\hat{H}_{\rm NH}$ resulting from the complex rotation.
These eigenstates have complex energies ${\cal E}_k=E_k\I{-}\frac{\ii}{2}\Gamma_k$ (enumerated by index $k$), where $E_k$ represents the resonance real energy and $\Gamma_k$ the width.
The remaining discrete eigenstates ${\cal E}_{l}=E_{l}\I{-}\frac{\ii}{2}\Gamma_{l}$ (enumerated by index $l$) of $\hat{H}_{\rm NH}$, as well as all discrete eigenstates ${\cal E}_{0l}=E_{0l}\I{-}\frac{\ii}{2}\Gamma_{0l}$ of the complex-rotated free Hamiltonian $\hat{H}_{\rm 0NH}$, represent discretized continuum states.
These form a background that would become continuous in the infinite-box limit.
Note that contributions of the background states associated with $\hat{H}_{\rm NH}$ and $\hat{H}_{\rm 0NH}$ to the traces in Eq.\,\eqref{coled} tend to cancel each other \cite{Suzuk05,Suzuk08}, so the resulting continuum level density is formed dominantly by the contributions of resonances. 
These have the same form as the right-hand side of Eq.\,\eqref{contlev}.

A generalization of this complex non-Hermitian approach was reported in Ref.\,\cite{Stran20}.
The real continuum level density \eqref{coled} was replaced by a complex density 
\begin{equation}
\Delta\rho({\cal E})=\frac{\ii}{\pi}\,
\left({\rm Tr}\,\frac{1}{{\cal E}\I{-}\hat{H}_{\rm NH}}-{\rm Tr}\,\frac{1}{{\cal E}\I{-}\hat{H}_{0{\rm NH}}}\right)
\label{Coled},
\end{equation}
which is defined in the domain of complex energy ${\cal E}$, naturally associated with the resonance and continuum states obtained from the complex scaling method.
It is clear that individual discrete eigenvalues  ${\cal E}_k$, ${\cal E}_l$ and ${\cal E}_{0l}$ represent poles of this function.
However, we are interested in the behavior of $\Delta\rho({\cal E})$ for ${{\cal E}=E+\ii 0}$, i.e., on the real energy axis.
The values of $\Delta\rho(E)$ are complex even for real energies, with ${\rm Re}\,\Delta\rho(E)$ coinciding with $\delta\rho(E)$ and ${\rm Im}\,\Delta\rho(E)$ being expressed by an easily derivable formula analogous to Eq.\,\eqref{contlev}.
What is the meaning of the imaginary density?
We anticipate, in analogy with the last equality of Eq.\,\eqref{coled}, that the complex level density $\Delta\rho(E)$ on the real energy axis is related to a complex phase shift $\Phi(E)$, which contains the ordinary phase ${\varphi(E)={\rm Re}\,\Phi(E)}$ as well as an imaginary phase $\ii\,{\rm Im}\,\Phi(E)$ expressing the attenuation of the wave in the selected channel.
Hence 
\begin{equation} 
\Delta\rho(E)=\frac{1}{\pi}\,\frac{\partial\Phi(E)}{\partial E} 
\label{Coledre}.
\end{equation}
This relation can indeed be verified on both the analytical and numerical levels for the 1D scattering processes mentioned below. 
Note that in practical evaluations we use a smoothed dependence $\Delta\overline{\rho}(E)$ of the complex continuum level density, which can be obtained by setting ${{\cal E}=E+\ii\epsilon}$ with a small $\epsilon\I{>}0$ in Eq.\,\eqref{Coled}, and the corresponding smoothed complex phase shift $\overline{\Phi}(E)$.

We apply these ideas to the scattering problems for systems with $f\I{=}1$, namely to the tunneling of a particle with mass $m$ through a 1D potential $V(x)$ \cite{Razav14}.
We assume that the potential is of a finite range, i.e., that $\hat{V}(x)$ vanishes outside a certain finite interval $(a,b)$ delimited by points ${a<b}$, while $V(x)$  inside the interval takes an arbitrary shape.
In particular, multibarrier potentials generating resonances in transmission and reflection probabilities can be treated in this way.
The usual asymptotics of wave functions is required, namely ${\psi(x)=e^{+\ii px/\hbar}+\alpha(E)e^{-\ii px/\hbar}}$ for $x\I{<}a$ and ${\psi(x)=\beta(E)e^{+\ii px/\hbar}}$ for $x\I{>}b$, where $p\I{=}\sqrt{2mE}$, and $\alpha(E)$ and $\beta(E)$ stand for the reflection and transmission amplitudes, respectively.
The transmission amplitude can be written as
\begin{equation}
\beta(E)=|\beta(E)|e^{\ii\varphi(E)}=e^{\ii\left[\varphi(E)-\ii\ln|\beta(E)|\right]}\equiv e^{\ii\Phi(E)}
\label{tra},
\end{equation}
where $\varphi(E)$ is a real phase shift of the transmitted wave and $\Phi(E)$ is a complex phase including also information on the attenuation of the transmitted wave.

Closely related to the phase shift of the tunneling particle is its assumed \uvo{time delay} behind the barrier \cite{Landa94,Carva02}.
Time relations in general scattering processes, and particularly those in the 1D scattering, represent a rather hot topic of present-day experiments in attosecond metrology (for more information see, e.g., Ref.\,\cite{Sokol18} and the references therein).
The simplest definition of a time delay in quantum tunneling is the so-called Eisenbud-Wigner time, which is expressed via the energy variation of the real phase shift: $\delta t(E)\equiv\hbar\frac{\partial}{\partial E}\varphi(E)$ \cite{Wigne55,Smith60}.
From the right-hand side of Eq.\,\eqref{coled} we see that this time is related to the continuous level density through a simple formula $\delta\rho(E)\I{=}\delta t(E)/(\pi\hbar)$.
There is a direct connection of this formula to the smoothed level density of $f\I{=}1$ bound systems, for which Eq.\,\eqref{smole} yields $\overline{\rho}(E)\I{=}T(E)/(2\pi\hbar)$, where $T(E)$ is a period of the closed orbit at energy $E$ (sum of periods if several orbits coexist).
The missing factor 2 in the denominator of the formula for $\delta\rho(E)$ can be linked to the fact that in that here we consider only a one-way passage of the particle through the interacting region instead of a closed orbit.

In the spirit of the above complex generalizations, we introduce \cite{Stran20}, besides the complex level density $\Delta\rho(E)$ and complex phase shift $\Phi(E)$, also a {\em complex time delay\/} ${\Delta{t}(E)=\hbar\frac{\partial}{\partial E}\Phi(E)}$.
This definition parallels the Eisenbud-Wigner time given above and its smoothed form reads ${\Delta\overline{t}(E)=\hbar\frac{\partial}{\partial E}{\overline\Phi}(E)}$. 
The use of Eq.\,\eqref{Coledre} (with smoothed quantities) leads to a simple relation
\begin{equation}
\Delta{\overline\rho}(E)=\frac{\Delta\overline{t}(E)}{\pi\hbar}
\label{cot}.
\end{equation}

The meaning of the formula \eqref{cot} for 1D scattering processes can be deduced from the following considerations.
From the semiclassical expression of the smoothed complex transmission amplitude 
$\overline{\beta}(E)$,
\begin{equation}
\overline{\beta}(E)e^{\ii pb/\hbar}= e^{\ii pa/\hbar}\ e^{\ii\left[\int_{a}^{b}dx' \sqrt{2m[E-V(x')]}+C\right]/\hbar}
\label{wkb}
\end{equation}
(where the constant $C$ includes phase shifts at the boundaries between the allowed and forbidden regions \cite{Berry72}), we can immediately determine the smoothed complex phase $\overline{\Phi}(E)$ and, by its differentiation, also the smoothed complex time shift $\Delta\overline{t}(E)$.
In this way we find that
\begin{eqnarray}
\hspace{-18mm}
{\rm Re}\,\Delta\overline{{\rho}}(E)&\I{=}&\frac{1}{\pi\hbar}
\left[\int_{E\geq V(x)}\!\!\!\!\! dx' \sqrt{\frac{m}{2[E\I{-}V(x')]}}\I{-}\sqrt{\frac{m}{2E}}(b\I{-}a)\right]
\I{=}\frac{t_{+}(E)\I{-}t_0(E)}{\pi\hbar},
\quad\label{ret}\\
\hspace{-18mm}
{\rm Im}\,\Delta\overline{{\rho}}(E)&\I{=}&\frac{1}{\pi\hbar}
\int_{E<V(x)}\!\!\!\!\! dx' \sqrt{\frac{m}{2[V(x')\I{-}E]}}
=\frac{t_{-}(E)}{\pi\hbar}
\label{imt}.
\end{eqnarray}
In the first formula, the integration within the interval $(a,b)$ includes only the classically allowed regions with $E\I{\geq}V(x)$, and $t_{\rm +}(E)$ is the time that a classical particle with energy $E$ needs to travel through these regions.
The time $t_0(E)$ corresponds to the passage of a free particle across the whole interval $(a,b)$.
The integration in the second formula includes, on the contrary, only the classically forbidden regions with $E\I{<}V(x)$, and the time $t_{-}(E)$ corresponds to a classical time which a particle with energy $-E$ would need to cross these regions in the inverted potential $-V(x)$.
Therefore we arrive at a remarkable finding that the real part of the complex level density \eqref{Coled} on the real energy axis expresses the time shift between the interacting and free particles in classically allowed regions, while the imaginary part reflects time relations in the classically forbidden regions after the $E\I{\to}-E$ transformation.
This represents a generalization of the celebrated instanton-like solutions of the tunneling problem \cite{Colem85,Takat99,Deunf10,Deunf13}.

\begin{figure}[t!]
\begin{flushright}
\includegraphics[width=0.85\textwidth]{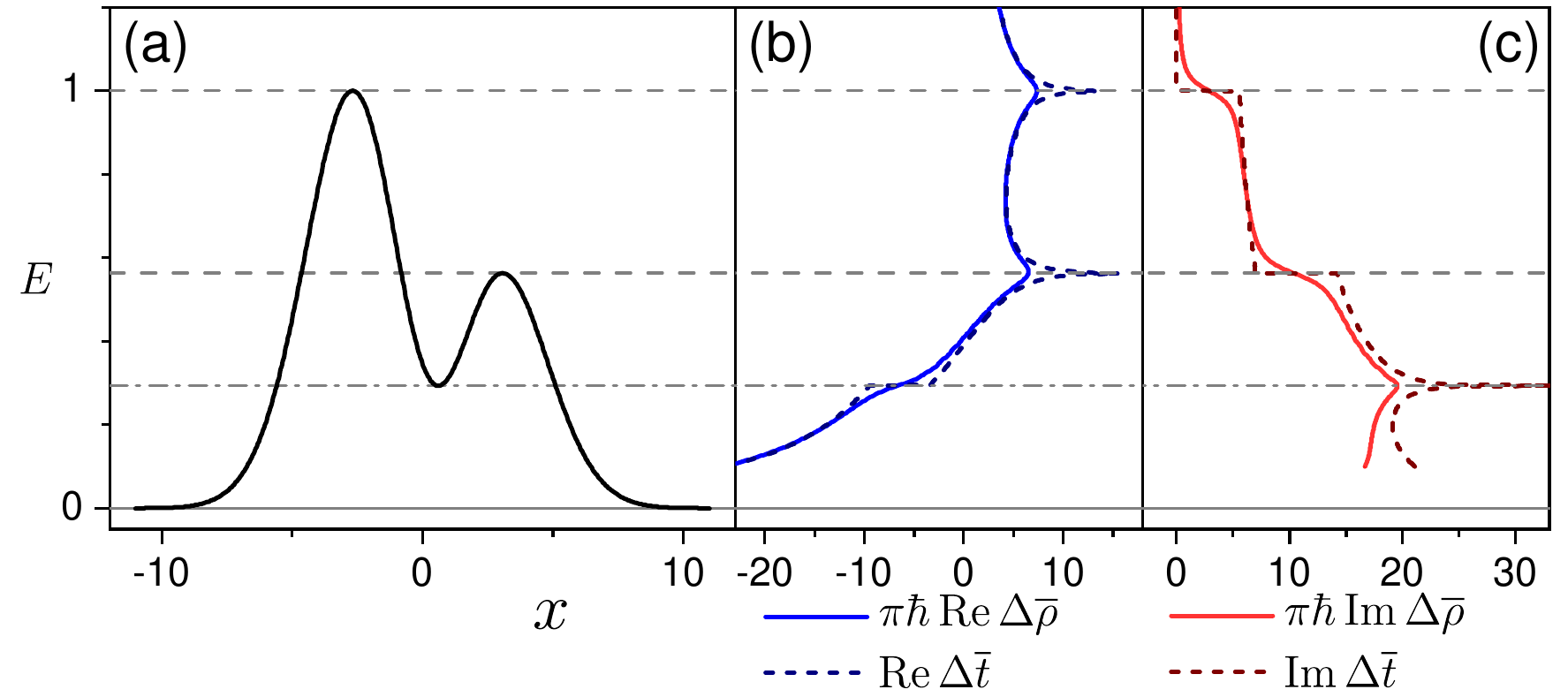}
\end{flushright}
\caption{The smoothed complex-extended continuum level density $\pi\hbar\Delta{\overline\rho}(E)$ and its semiclassical estimate $\Delta\overline{t}(E)$ for a double-barrier potential.
The exact density from Eq.\,\eqref{Coled}, determined by the complex scaling method, is smoothed by choosing ${\cal E}=E\I{+}i\epsilon$ with $\epsilon\I{=}0.05$. 
The estimate based on the complex time shifts is calculated from Eqs.\,\eqref{ret} and \eqref{imt}.
Panel (a): the potential of the form $V(x)=P(x)e^{-x^2/10}$, where $P(x)$ is a quadratic function. 
Panel (b): The real part of the smoothed level density (full curve) and the time shift from Eq.\,\eqref{ret} (dashed curve).
Panel (c): The imaginary part of the smoothed level density (full curve) and the imaginary time from Eq.\,\eqref{imt} (dashed curve).
The three stationary points of $V(x)$ and $-V(x)$ at energies marked by horizontal lines cause the corresponding {\ESQPT}-like singularities of both the real and imaginary parts of the continuum level density.
Based on Ref.\,\cite{Stran20}.
}
\label{reson}
\end{figure}

Formulas \eqref{ret} and \eqref{imt} have an immediate consequence:
They show that the real and imaginary parts of the smoothed continuum level density $\Delta\overline{{\rho}}(E)$ manifest {\ESQPT}-like singularities at the energies corresponding to stationary points in the classically allowed and forbidden regions, respectively.
In particular, ${\rm Re}\,\Delta\overline{{\rho}}(E)$ has singularities connected with stationary points of potential $+V(x)$ and ${\rm Im}\,\Delta\overline{{\rho}}(E)$ has singularities connected with stationary points of $-V(x)$.
This was verified by numerical calculations performed for several 1D tunneling potentials \cite{Stran20}, in which 
the times $t_{+}(E)$, $t_{-}(E)$ and $t_{0}(E)$ on the right-hand side of Eqs.\,\eqref{ret} and \eqref{imt} were calculated independently from ${\rm Re}\Delta\overline{{\rho}}(E)$ and ${\rm Im}\Delta\overline{{\rho}}(E)$ on the left-hand side. 
One of the results is illustrated in Fig.\,\ref{reson} for a sample potential with two maxima and one minimum, see panel~(a).
Panels~(b) and~(c), respectively, depict the real and imaginary parts of the level density $\pi\hbar\Delta\overline{{\rho}}(E)$ (full curves) and the real and imaginary parts of the time shift $\Delta\overline{t}(E)$ (dashed curves).
Both curves in each panels are in a reasonable match and exhibit $f\I{=}1$ {\ESQPT}-like singularities (or their precursors) at energies of the three stationary points of the potential.
The singularities in panel (b) follow the typology from Eq.\,\eqref{clasfull} for potential $+V(x)$ (two maxima, one minimum), while the singularities in panel (c) follow the typology for potential $-V(x)$ (two minima, one maximum).
Note that in the forbidden regions, the kinetic term must also be inverted, so that the indices of the stationary points are ${r=1}$ and $2$, respectively, which leads to the upward logarithmic divergence and two downward jumps of the imaginary density in panel (c).

We note that some particular cases of singularities in the density of tunneling resonances have been previously studied in molecular physics; see, e.g., Refs.\,\cite{Conno68,Roy78}.
The present formulation creates a unified framework for all such phenomena and connects them to the physics of {\ESQPTs}.
Since the density of continuum states $\Delta{\rho}(E)$ fully determines the transmission amplitude $\beta(E)$, its singularities have measurable consequences.
This is challenging since the present technologies make it possible to realize customized resonant tunneling potentials via fabricating suitable nanostructures, and therefore enable us to test the predictions of Eqs.\,\eqref{ret} and \eqref{imt}.
An open question concerns possible extension of these effects to $f\I{>}1$ systems.

\section{Outlook}
\label{Conc}


In summary, we have seen---in great majority of the above-studied examples---that the {\ESQPT} represents criticality affecting non-thermal excitations of a bound quantum system composed of a large number of interacting constituents, which are nevertheless described by a few collective {\DoFs}.
The infinite-size limit of such systems is anchored in the classical limit, and this holds also for their {\ESQPT} singularities.
Analogues of {\ESQPTs} exist also in other types of systems, including periodic lattice systems and scattering systems with continuous energy spectra.
The {\ESQPTs} are constituted by non-analyticities of the level density and level flow in energy and control parameters, and show up as non-analyticities of microcanonical averages of various quantities over the energy shell.
Signatures of {\ESQPTs} can be found in thermal properties and in the response of the system to various kinds of driving. 
We stress potential relevance of the {\ESQPT}-related effects in the design of various quantum information protocols, which are often realized in quantum systems satisfying the above criteria.

Rather than trying to summarize diversity of the above-discussed results, we close this review by attempting to foresee near developments of the {\ESQPT} field.
On the theoretical side, more {\ESQPT} singularities will most probably be disclosed in quantum spectra of new or so far unexamined many-body models.
However, the most needed advance in the foundations of the {\ESQPT} concept concerns its possible extension to non-collective systems with unlimited {\DoF} numbers.
These systems also manifest abrupt changes of the structure of excited states  (e.g., the eigenstate phase transitions \cite{Pal10,Huse13}), although the  nature of these changes oversteps the semiclassical description.
We believe that a part of the {\ESQPT} methodology can be applied also to these genuinely quantum non-analyticities.

We also anticipate substantial progress in the identification and description of {\ESQPT}-related thermodynamic and dynamic effects.
The present knowledge on these effects is just an initiation of the field.
We know how {\ESQPTs} affect the microcanonically thermalized systems, but so far recognized links to canonical thermodynamics are only fragmentary.
Progress in this field is important for the analysis of various thermal devices based on finite quantum systems \cite{Kloc19}.
Diverse dynamic effects probably represent the most promising field of the {\ESQPT} study.
In particular, the driven evolution of quantum systems through the {\ESQPT} region (activated, e.g., due to a non-zero temperature of the system) modifies the efficiency of quantum state preparation protocols.
Similarly, consequences of {\ESQPTs} affect quantum-state manipulation techniques that employ quantum quenches. 
Understanding of the {\ESQPT}-like singularities in quasienergy spectra of periodically driven systems is essential for the correct design of purpose-built driving procedures.

It is beyond any doubt that more experimental verification of the {\ESQPT} effects is needed.
So far, the signatures of {\ESQPTs} have been found in the spectra of some simple molecules \cite{Winne05,Zobov05,Lares13} and those of 2D photonic crystals \cite{Dietz13}.
Detection of dynamic {\ESQPT} signatures connected with quantum quench dynamics or some types of out-of-time-order correlators may be on the way in the interacting atom-field systems described by the Dicke-like models \cite{Lewis19,Li20}.
Recently, the spinor Bose-Einstein condensates were proposed as another experimental platform for the {\ESQPT}-related studies \cite{Feldm20}.
Various still unrecognized dynamic signatures of {\ESQPT} may be present in the vast body of experimental data---either already existing (see, e.g., Ref.\,\cite{Zhao14}), or in principle available---on various kinds of artificial quantum systems.
One may anticipate experimental evidence of the {\ESQPT} analogues in kicked systems based on suitable quantum simulators, and in 1D scattering systems fabricated via customized resonance tunneling nanostructures.
We believe that some of these evidences will be brought in a near future.

\section*{Acknowledgments}

This review is dedicated to Tobias Brandes, whose intended involvement in its creation was thwarted by a cruel arrow of outrageous fortune. 
We thank Curro P{\'e}rez-Bernal for useful comments on the preprint.
The work was supported by the Czech Science Foundation under the grant no.\,20-09998S and by the Charles University in Prague under the project UNCE/SCI/013.

\vspace{1cm}

\end{document}